\documentclass[acmtog, nonacm]{acmart}

\usepackage{wrapfig} 
\usepackage{sistyle} 

\SIthousandsep{,}
\SIgroupfourfalse

\usepackage[ruled]{algorithm2e} 

\SetAlFnt{\small}
\SetAlCapFnt{\small}
\SetAlCapNameFnt{\small}
\SetAlCapHSkip{0pt}

\begin{document}
\title{Intrinisic Mesh Simplification}

\author{Randy Shoemaker}
\email{rwshoemaker@wm.edu}
\orcid{0009-0009-7589-4356}
\affiliation{%
 \institution{College of William \& Mary}
 \city{Williamsburg, VA}
 \country{USA}
}
\author{Sam Sartor}
\email{slsartor@wm.edu}
\orcid{0009-0001-1915-6887}
\affiliation{%
 \institution{College of William \& Mary}
 \city{Williamsburg, VA}
 \country{USA}
}
\author{Pieter Peers}
\email{ppeers@siggraph.org}
\orcid{0000-0001-7621-9808}
\affiliation{%
 \institution{College of William \& Mary}
 \city{Williamsburg, VA}
 \country{USA}
}

\renewcommand\shortauthors{Shoemaker et al.}

\def\sectionautorefname{Section}
\def\etal{{et al.}}

\begin{abstract}
  This paper presents a novel simplification method for removing vertices
  from an intrinsic triangulation corresponding to extrinsic vertices lying
  on near-developable (i.e., with limited Gaussian curvature) and general surfaces. We
  greedily process all intrinsic vertices with an absolute Gaussian curvature below a
  user selected threshold.  For each vertex, we repeatedly perform local
  intrinsic edge flips until the vertex reaches the desired valence (three for
  internal vertices or two for boundary vertices) such that removal of the
  vertex and incident edges can be locally performed in the intrinsic
  triangulation. Each removed vertex's intrinsic location is tracked via
  (intrinsic) barycentric coordinates that are updated to reflect changes in
  the intrinsic triangulation.  We demonstrate the robustness and
  effectiveness of our method on the \emph{Thingi10k} dataset and analyze the effect of the curvature threshold on the solutions of PDEs.
\end{abstract}



\maketitle

\newcommand{\R}[1]{\mathbb{R}^#1}
\newcommand{\ceil}[1]{\lceil#1\rceil}
\def\M{\mathsf{M}}
\def\V{\mathsf{V}}
\def\E{\mathsf{E}}
\def\F{\mathsf{F}}
\def\h{h}
\def\l{\ell}
\def\g{\kappa}
\def\Q{\mathsf{Q}}
\def\P{\mathsf{P}}
\def\A{A}
\begin{figure}
  \includegraphics[width=\columnwidth]{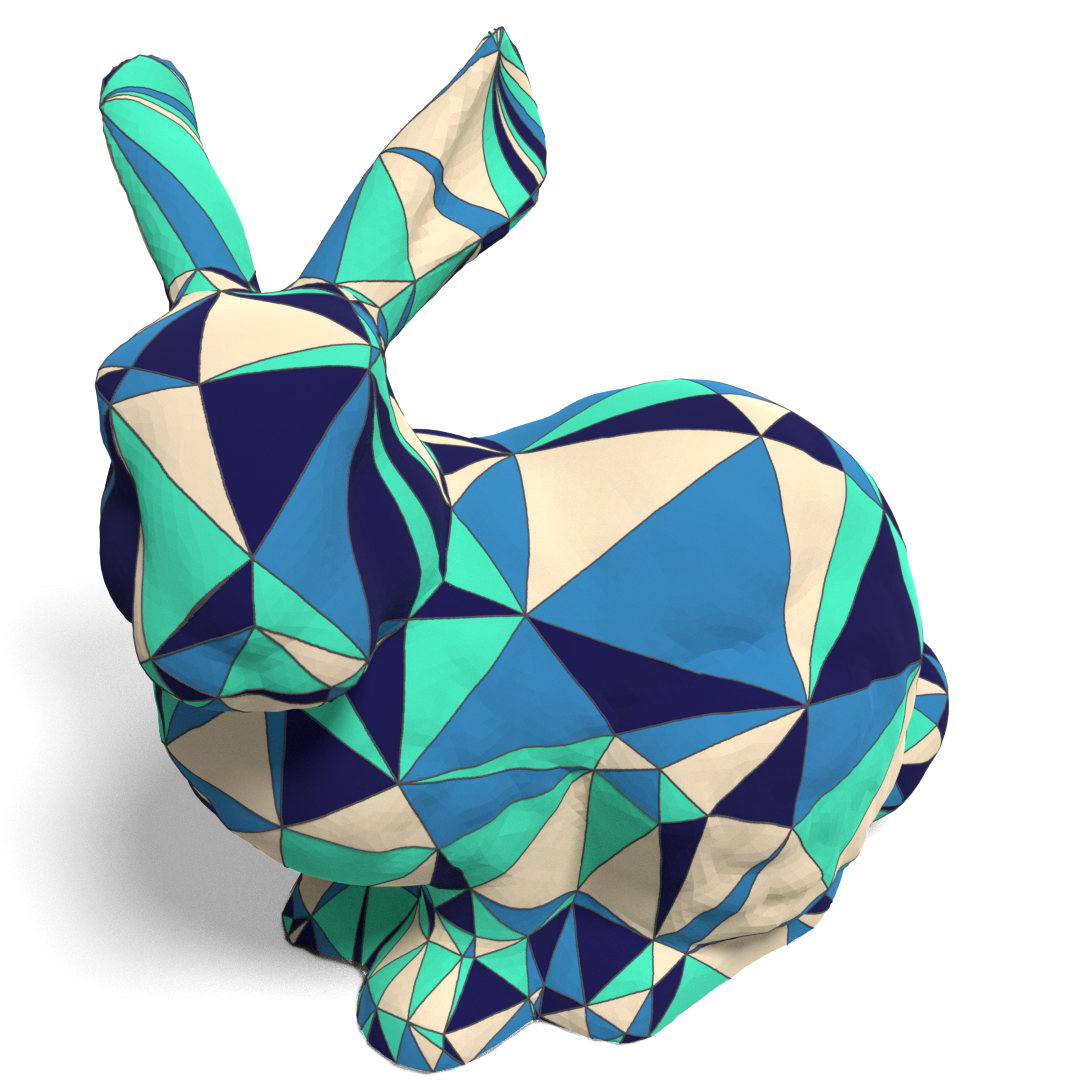}
  \caption{Our Intrinsic Simplification algorithm simplifies the metric of a
  surface by approximating the intrinsic geometry with developable patches.}
  \label{fig:bunny}
\end{figure}

\section{Introduction}
\label{sec:intro}

Geometric data encountered in the wild are often \emph{``messy''} from a
geometry processing perspective, necessitating the need for robustified
processing
methods~\cite{Zhou:2016:MAS,Hu:2018:TMW,Sellan:2019:SGP,Sawhney:2020:MCG,Qi:2022:BFW}. Intrinsic
triangulation
frameworks~\cite{Fisher:2007:AAC,Sharp:2019:NIT,Gillespie:2021:ICI} have been
proposed as an alternative strategy for robust geometry processing in the
wild.  Intrinsic geometry processing approaches geometry processing from an intrinsic
view where all operations are expressed as combinations of atomic operations
(e.g., edge flipping, face splitting, inserting edges, etc.)  defined on
distances between points over the 2D manifold.  While existing intrinsic
triangulation frameworks differ in data-structure and efficiency of certain
atomic operations, they all share that each ``extrinsic'' vertex has an
immutable counterpart in the intrinsic triangulation, and consequently, vertex
removal of initial extrinsic vertices is not supported.  Many in-the-wild
triangle meshes are finely triangulated in order to faithfully approximate
curved surfaces in $\R{3}$ by piecewise planar surfaces.  
\setlength{\intextsep}{0pt}%
\setlength{\columnsep}{16pt}%
\begin{wrapfigure}{r}{0.3\linewidth}
    \centering
    \includegraphics[width=\linewidth]{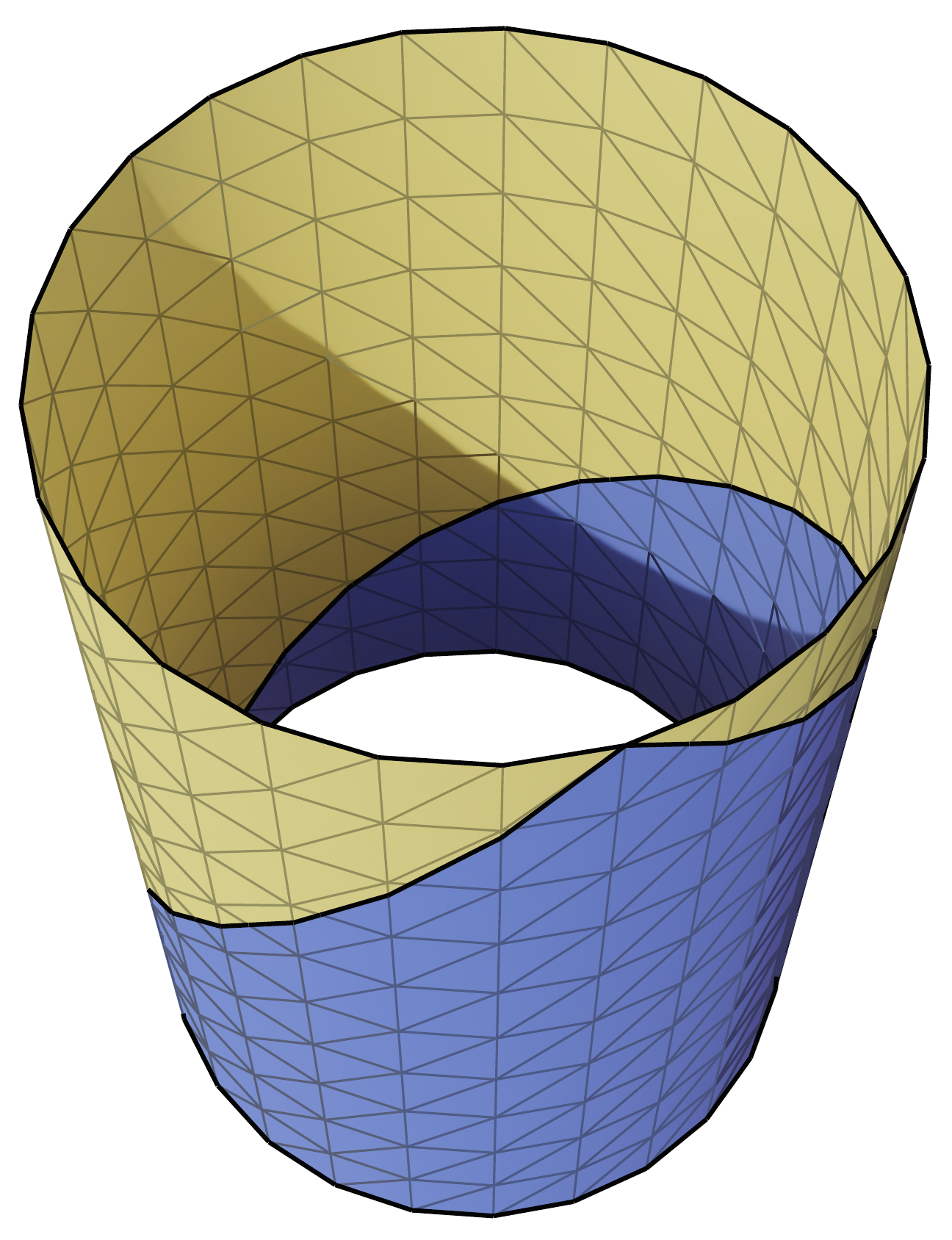}
\end{wrapfigure}
However, a
significant portion of in-the-wild triangle meshes are designed with CAD tools
or are the result of 3D scans of real-world surfaces formed by combining
developable patches.  Intrinsically, such developable parts are isometric to a
plane.  For example consider a cap-less cylinder which can be fully modeled
using just two intrinsic triangles (see inset).
However, to faithfully capture the curvature, thousands of extrinsic triangles
are needed in $\R{3}$. Not only does blindly embedding a heavily tessellated
developable surface incur significant overhead, it also impacts the efficiency
of many downstream processing algorithms such as optimal Delaunay
triangulations~\cite{Chen:2004:ODT}, adaptive intrinsic mesh
refinement~\cite{Sharp:2019:NIT}, computing geodesics~\cite{Sharp:2020:YCF},
geodesic distances, and other tasks.

In this paper we present a topology-preserving method for simplifying a
triangle mesh directly on the intrinsic manifold.  A key insight is that
vertices with zero Gaussian curvature can be removed without impacting the
accuracy of the metric defined on the embedding.  Moreover, a locally
developable approximation can be obtained by allowing vertices with small
Gaussian curvature to be removed as in~\autoref{fig:bunny}.  However, classic vertex merging and edge
collapse require updating of edge lengths in the intrinsic setting which can
be non-trivial when removing a vertex with non-zero curvature.  Instead, we
introduce a novel intrinsic simplification method based on edge flipping, a
stable atomic intrinsic operation.  For each vertex we would like to remove,
we perform edge flips until the vertex has valence 3 (or valence 2 for
vertices on the boundary).  We can then remove the vertex if the resulting
triangulation remains valid.  If the triangulation becomes invalid, we undo
the edge flips in reverse order and reschedule the vertex for later
evaluation. We process vertices in a greedy
\emph{``intrinsically-flattest-first''} order until no more vertices can be
removed.  In addition, we keep track of the intrinsic location of deleted
vertices by their intrinsic barycentric coordinates.  Compared to extrinsic
simplification which approximates the surface with fewer planar triangles,
intrinsic simplification can be seen as approximating the surface with
developable patches. Since the metric of an intrinsic triangulation is fully
 described by its edge lengths, intrinsic simplification can be seen as 
 simplification of the metric rather than the extrinsic mesh.

We demonstrate and validate the
robustness of our method on the \emph{Thingi10k} dataset~\cite{Zhou:2016:TTK},
and evaluate the impact of relaxing the Gaussian curvature threshold on PDE
solutions.
\section{Related Work}
\label{sec:related}

\paragraph{Mesh Simplification}
There exists a vast body of work on \emph{extrinsic} mesh simplification and
an exhaustive enumeration is beyond the scope of this paper.  Instead we will
focus on seminal papers in this area and contrast them against our intrinsic
mesh simplification method; we refer the reader to \cite{Khan:2020:SRS}
(Sec. 4.1) for an in-depth systematic review. Extrinsic mesh simplification
methods aim to reduce the number of vertices in the mesh such that some
quality metric is best preserved. The most commonly preserved quality is the
visual appearance of a
mesh~\cite{Schroeder:1992:DTM,Garland:1997:SSQ,Rossignac:1993:MR3,Popovic:1997:PSC,CohenSteiner:2004:VSA}. Vertex
decimation~\cite{Schroeder:1992:DTM} iteratively deletes vertices according to
an extrinsic criterion and the resulting hole is carefully re-triangulated.
Our method of vertex deletion is similar to vertex decimation, except that we
employ intrinsic edge flips until the ring of a vertex is a triangle which can
be removed without retriangulation.  In their seminal work, Garland and
Heckbert~\shortcite{Garland:1997:SSQ} greedily merge vertices to minimize a
quadric error metric (QEM) via edge contraction.  We also follow a greedy
approach, but instead of QEM, we use Gaussian curvature to drive the
simplification.

Recently, Lescoat~\etal~\shortcite{Lescoat:2020:SMS} proposed spectral mesh
simplification, a greedy extrinsic mesh simplification strategy, that aims to
preserve the intrinsic geometry (i.e., minimize the change in the first $k$
eigenvectors of the Laplace-Beltrami operator). Spectral mesh simplification
is able to reduce the number of extrinsic triangles while minimizing errors
when computing spectral distances.  However, spectral mesh simplification is
relatively computationally expensive and limited to reducing extrinsic
surfaces. In contrast, intrinsic mesh simplification is computationally light
weight and more efficient in reducing the number of elements in
developable patches while preserving the intrinsic geometry.

In concurrent work, Liu~\etal~\shortcite{Liu:2023:SSI} propose a similar method for
intrinsic mesh simplification via intrinsic error metrics. Their method tracks an approximation of the accumulated error during simplification,
informing the order of vertex deletion. Our method uses the (dynamic) magnitude of the Gaussian curvature of a vertex to decide deletion order, which is easily updated after deletion.
Our method removes vertices by reducing them to a desired valence whereby they can
be safely removed. Once the valence of a vertex has been reduced, our method implicitly 
intrinsically flattens the neighborhood of a vertex via a simple update whereas Liu~\etal~\shortcite{Liu:2023:SSI} flattens the neighborhood of a vertex prior to valence reduction, requiring iterative optimization.

\paragraph{Intrinsic Triangulations}
Intrinsic triangulation
frameworks~\cite{Fisher:2007:AAC,Sharp:2019:NIT,Gillespie:2021:ICI} provide
tools and atomic operations to perform geometry processing algorithms that
only rely on intrinsic information directly on the intrinsic mesh (e.g.,
geodesic distance~\cite{Sharp:2020:YCF}, computing distortion minimizing
homeomorphisms~\cite{Takayama:2022:CIT}, or algorithms that rely on the
Laplace-Beltrami operator~\cite{Botch:2021:PMP}).  All existing intrinsic
triangulation frameworks support edge flipping, and the
Signpost~\cite{Sharp:2019:NIT} and Integer Coordinate
frameworks~\cite{Gillespie:2021:ICI} support additional atomic operations such
as adding vertices, repositioning (added) vertices, and computing common
subdivisions.  However, none of the existing intrinsic triangulation frameworks
currently support the removal of an extrinsic vertex's intrinsic
counterpart.  Our method only relies on edge
flipping to remove vertices, opening the door to possible adaptation to other
current and future intrinsic triangulation frameworks.
                  
\paragraph{Global Parameterization}
Global mesh parameterization is similar to intrinsic mesh simplification in that
both produce a base domain with a mapping to the original mesh. Various in depth
surveys discuss the variety of parameterization  
methods~\cite{Sheffer:2006:MPM, Hormann:2007:MPTP, Floater:2005:SPT}. In our case 
the base domain is an intrinsic triangulation and the mapping is specified by the 
barycentric coordinates of removed extrinsic vertices with respect to the 
intrinsic triangulation. Global parameterization methods utilizing simplicial and 
quadrilateral complexes as the base domain that apply iterative decimation 
schemes are most relevant to this work, for example~\cite{Khodakovsky:2003:GSP, Lee:1998:MAPS, Bommes:2013:IGMQ}. 
Such methods typically require some kind of embedding of mesh vertices as the 
algorithm progresses (usually in two dimensions) and are therefore unable take 
advantage of the additional degrees of freedom offered by intrinsic triangulations~\cite{Sharp:2019:NIT}
which are unencumbered by the requirement to maintain an embedding. Furthermore 
global parameterization methods, such as those producing integer grid maps (IGM) 
are computationally intensive whereas our edge flipping procedure is fast. 
Ebke ~\etal~\shortcite{Ebke:2016:IQR} offer a framework for computing an IGM on large meshes 
via a decimation objective based on the change in Gaussian curvature between the 
coarse and fine mesh. While our vertex ordering criteria are similar, their 
method is not designed to support intrinsic triangulations since both the 
parameterization domain and simplified meshes must be embedded in $\R{2}$ and $\R{3}$ 
respectively.

\section{Background}
\label{sec:background}

We briefly review the data structure required to support intrinsic triangulations. For a more detailed discussion of intrinsic triangulations we refer the reader to~\cite{Sharp:2019:NIT}.

Starting from a mesh $\M = \{\V, \E, \F\}$ encoded as a $\Delta$-complex, an intrinsic 
triangulation requires the lengths $\l_{ij}$ of the $ij$-th edge in $\E$ between vertices 
$i$ and $j \in \V$. Consistent with previous approaches, we represent the intrinsic triangulation as a $\Delta$-complex since it supports phenomena such as self-edges (edges that connect a vertex to itself) and degree one vertices. The lengths $\l_{ij}$ describe the shape of the triangles. 
It can be easily seen that the edge lengths fully define the intrinsic geometry. Other relevant intrinsic information can be directly computed from the edge lengths:
\begin{eqnarray}
  \theta_{jk}^{i} & = & \arccos \left( \frac{\l^2_{ij} + \l^2_{ik} -
                      \l^2_{jk}}{2\l_{ij}\l_{ik}} \right), \\
  \A_{ijk} & = & \sqrt{s(s-\l_{ij})(s-\l_{jk})(s-\l_{ki})}, \\
  s & = & (\l_{ij} + \l_{jk} + \l_{ki}) / 2
\end{eqnarray}
where $\theta_{jk}^{i}$ is the interior angle at $i \in \V$ in the triangle
$ijk \in \F$ and $\A_{ijk}$ is the area of the triangle $ijk \in \F$.

\section{Intrinsic Simplification by Edge Flipping}
\label{sec:flip}

Our goal is to remove intrinsic vertices that are part of a (near) developable
patch in the corresponding extrinsic mesh.  Removing such vertices will not
alter the intrinsic geometry since a developable patch is isometric to a
planar neighborhood.  A surface is developable around a vertex $i$ if it has
zero Gaussian curvature: $\g_i = 2\pi - \alpha_i$, where
$\alpha_i = \sum_{ijk} \theta_{jk}^{i}$ is the cone angle.  Ideally, only
vertices with zero Gaussian curvature should be removed such that the
intrinsic geometry is not changed.  However, developable surfaces are often
highly tessellated for accurate approximation in $\R{3}$. Depending on the
exact triangulation and/or numerical round-off errors, an exact zero Gaussian
curvature might not be reached.  Therefore, in practice we try to remove all
vertices with an absolute value Gaussian curvature less than some predetermined threshold
$\g_{max}$.  Setting $\g_{max}$ to a larger threshold, allows us to obtain an
intrinsic approximation where portions of the triangulation are replaced with
developable patches.

\begin{figure}
  \centering
  {\small
  \begin{tabular}{cccc}
        \def\svgwidth{0.2\linewidth}
        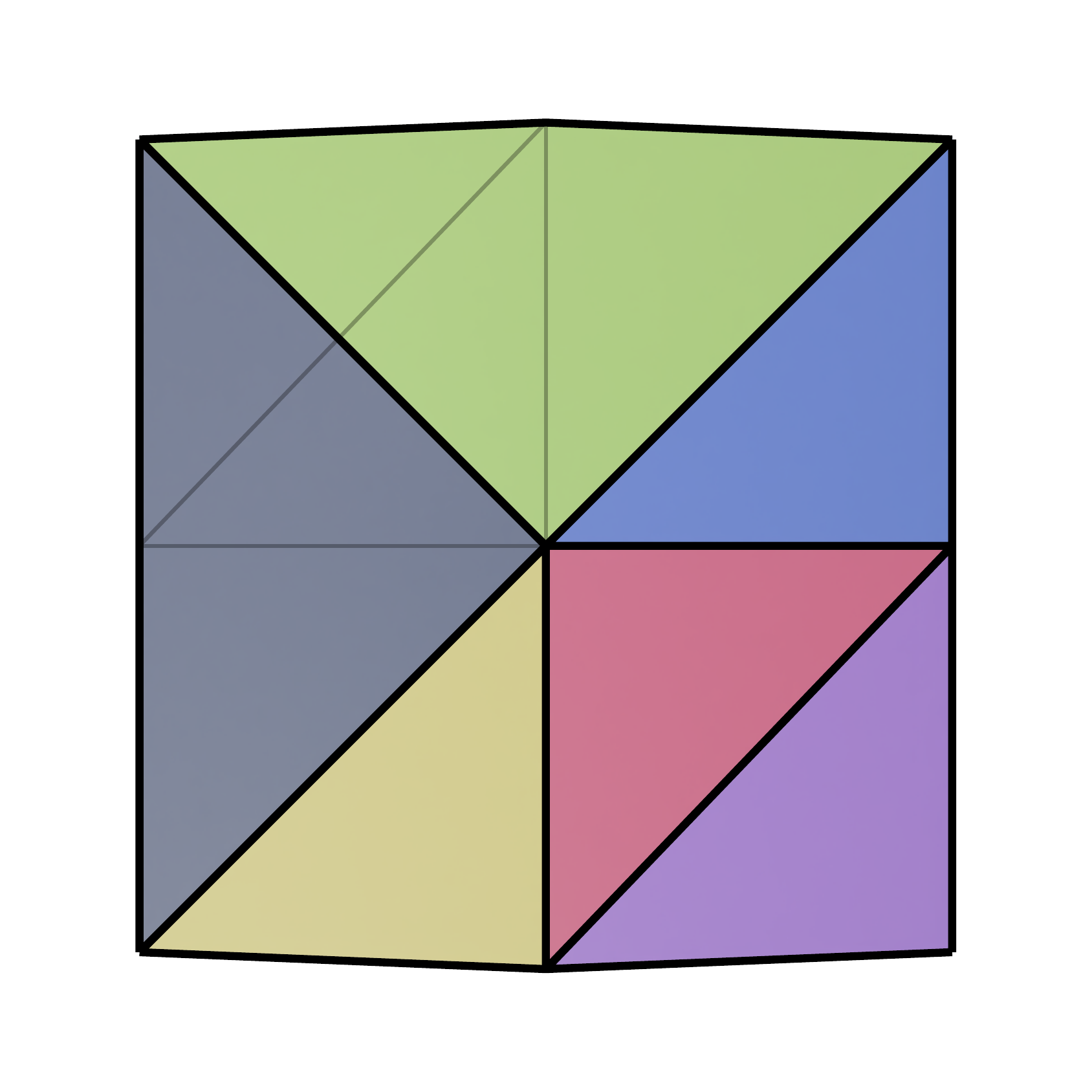 &
        \def\svgwidth{0.2\linewidth}
        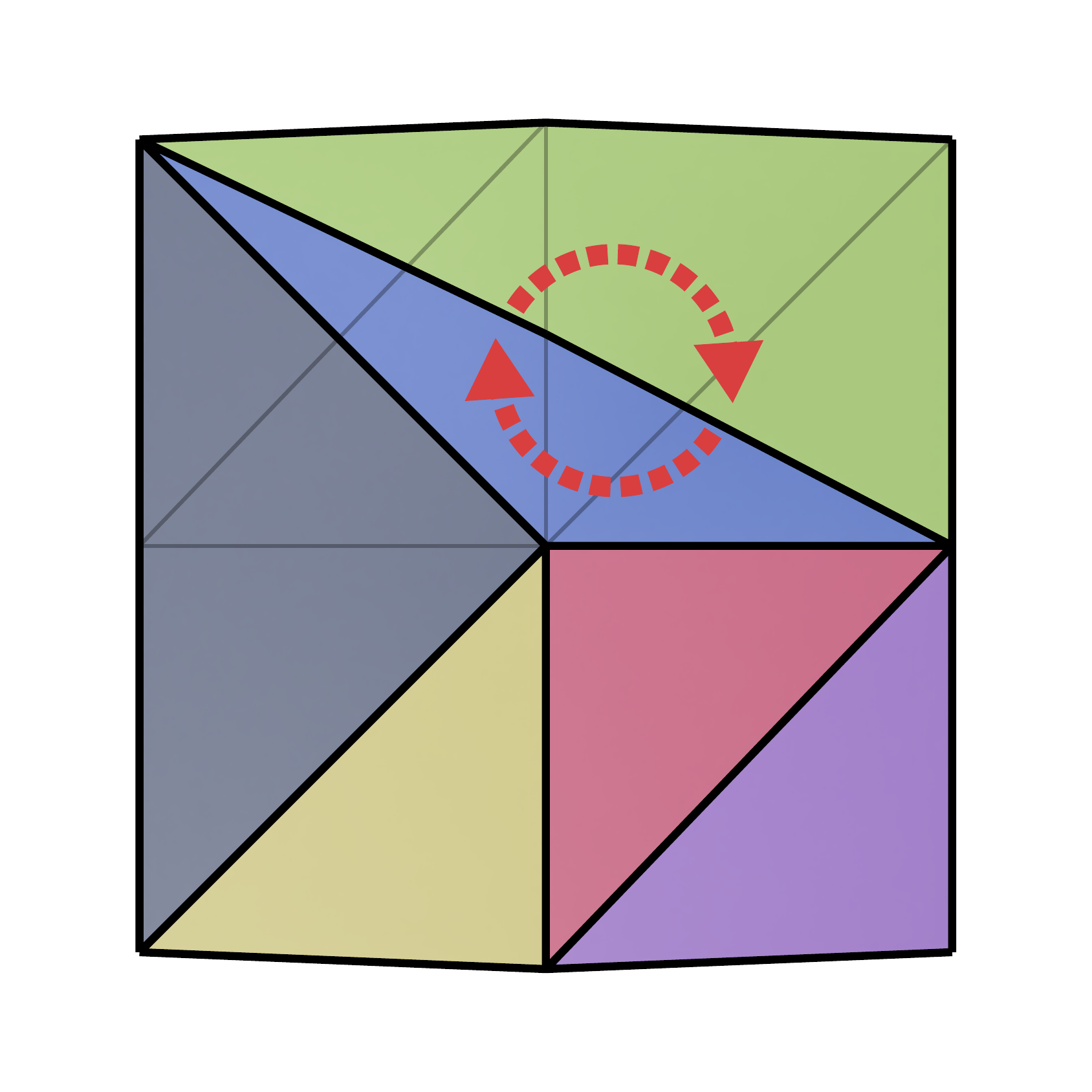 &
        \def\svgwidth{0.2\linewidth}
        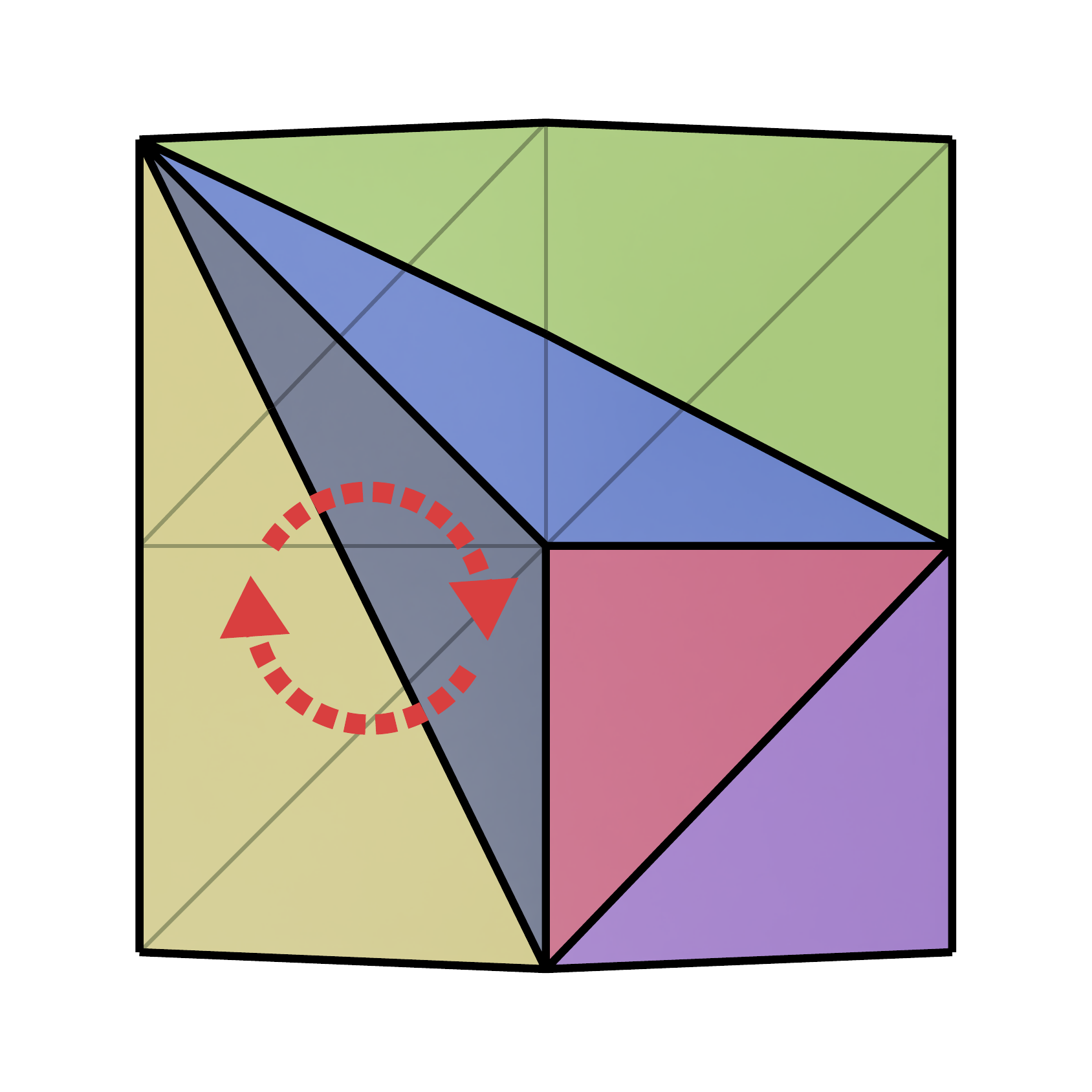 &
        \def\svgwidth{0.2\linewidth}
\begingroup%
  \makeatletter%
  \providecommand\color[2][]{%
    \errmessage{(Inkscape) Color is used for the text in Inkscape, but the package 'color.sty' is not loaded}%
    \renewcommand\color[2][]{}%
  }%
  \providecommand\transparent[1]{%
    \errmessage{(Inkscape) Transparency is used (non-zero) for the text in Inkscape, but the package 'transparent.sty' is not loaded}%
    \renewcommand\transparent[1]{}%
  }%
  \providecommand\rotatebox[2]{#2}%
  \newcommand*\fsize{\dimexpr\f@size pt\relax}%
  \newcommand*\lineheight[1]{\fontsize{\fsize}{#1\fsize}\selectfont}%
  \ifx\svgwidth\undefined%
    \setlength{\unitlength}{810bp}%
    \ifx\svgscale\undefined%
      \relax%
    \else%
      \setlength{\unitlength}{\unitlength * \real{\svgscale}}%
    \fi%
  \else%
    \setlength{\unitlength}{\svgwidth}%
  \fi%
  \global\let\svgwidth\undefined%
  \global\let\svgscale\undefined%
  \makeatother%
  \begin{picture}(1,1)%
    \lineheight{1}%
    \setlength\tabcolsep{0pt}%
    \put(0,0){\includegraphics[width=\unitlength,page=1]{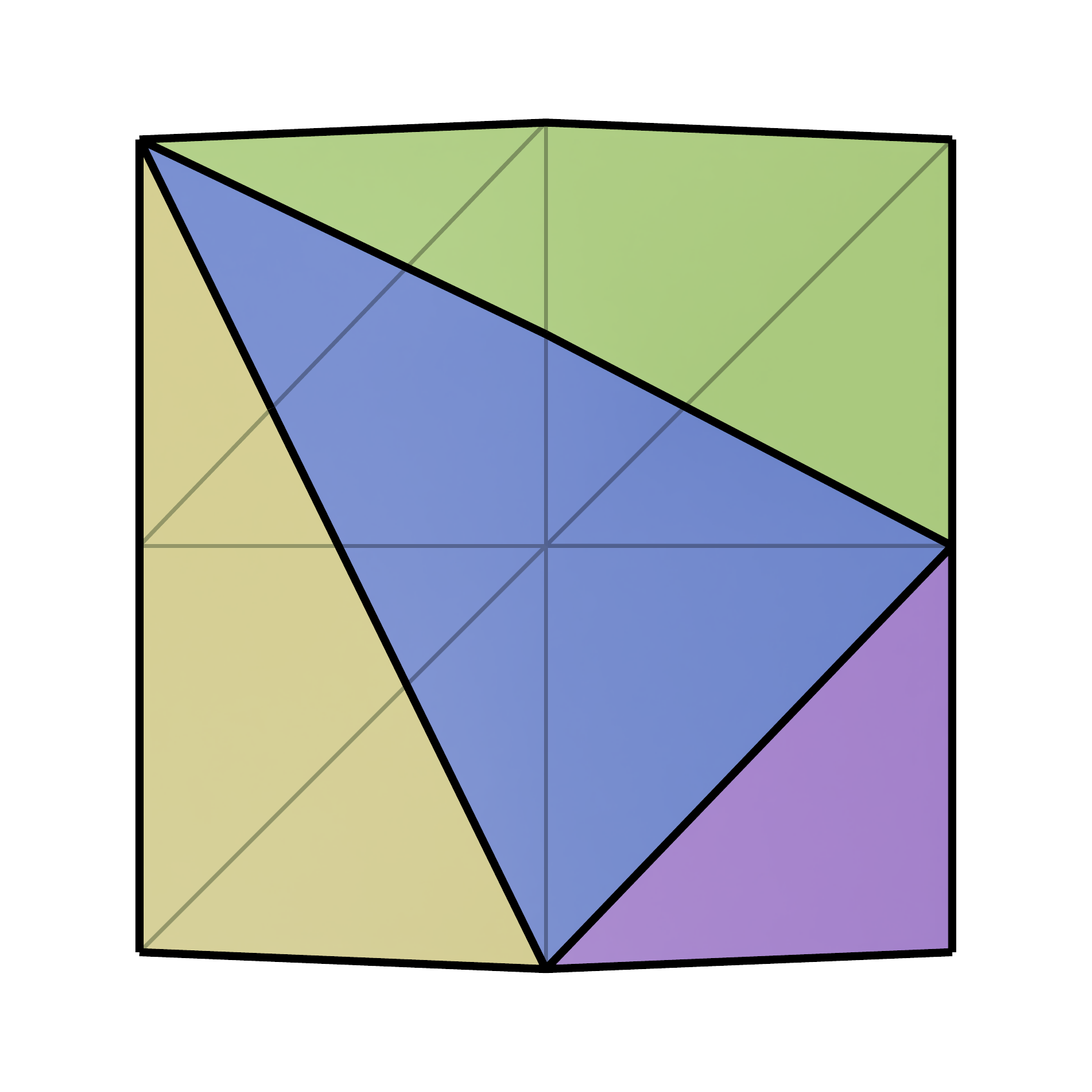}}%
    \put(0.06,0.84){\makebox(0,0)[lt]{\lineheight{1.25}\smash{\begin{tabular}[t]{l}l\end{tabular}}}}%
    \put(0.48,-0.02){\makebox(0,0)[lt]{\lineheight{1.25}\smash{\begin{tabular}[t]{l}j\end{tabular}}}}%
    \put(0.89,0.45){\makebox(0,0)[lt]{\lineheight{1.25}\smash{\begin{tabular}[t]{l}k\end{tabular}}}}%
    \put(0.48,0.45){\makebox(0,0)[lt]{\lineheight{1.25}\smash{\begin{tabular}[t]{l}i\end{tabular}}}}%
  \end{picture}%
\endgroup%

    \\
    initially & edge flip & edge flip & remove \\
    (valence 5) & (valence 4) & (valence 3) &
   \end{tabular}

    \begin{tabular}{cccc}

        \includegraphics[width=0.2\linewidth]{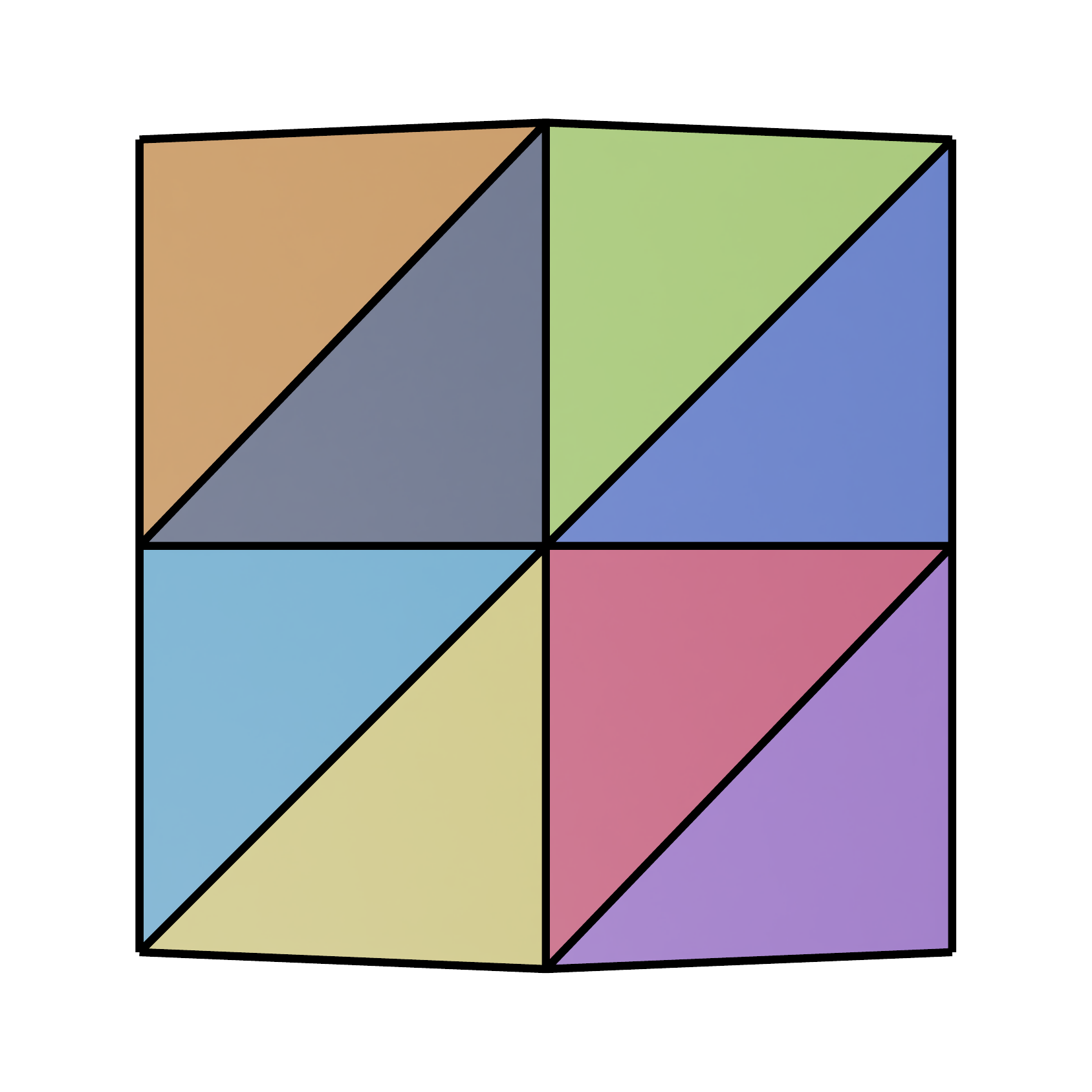} &
        \includegraphics[width=0.2\linewidth]{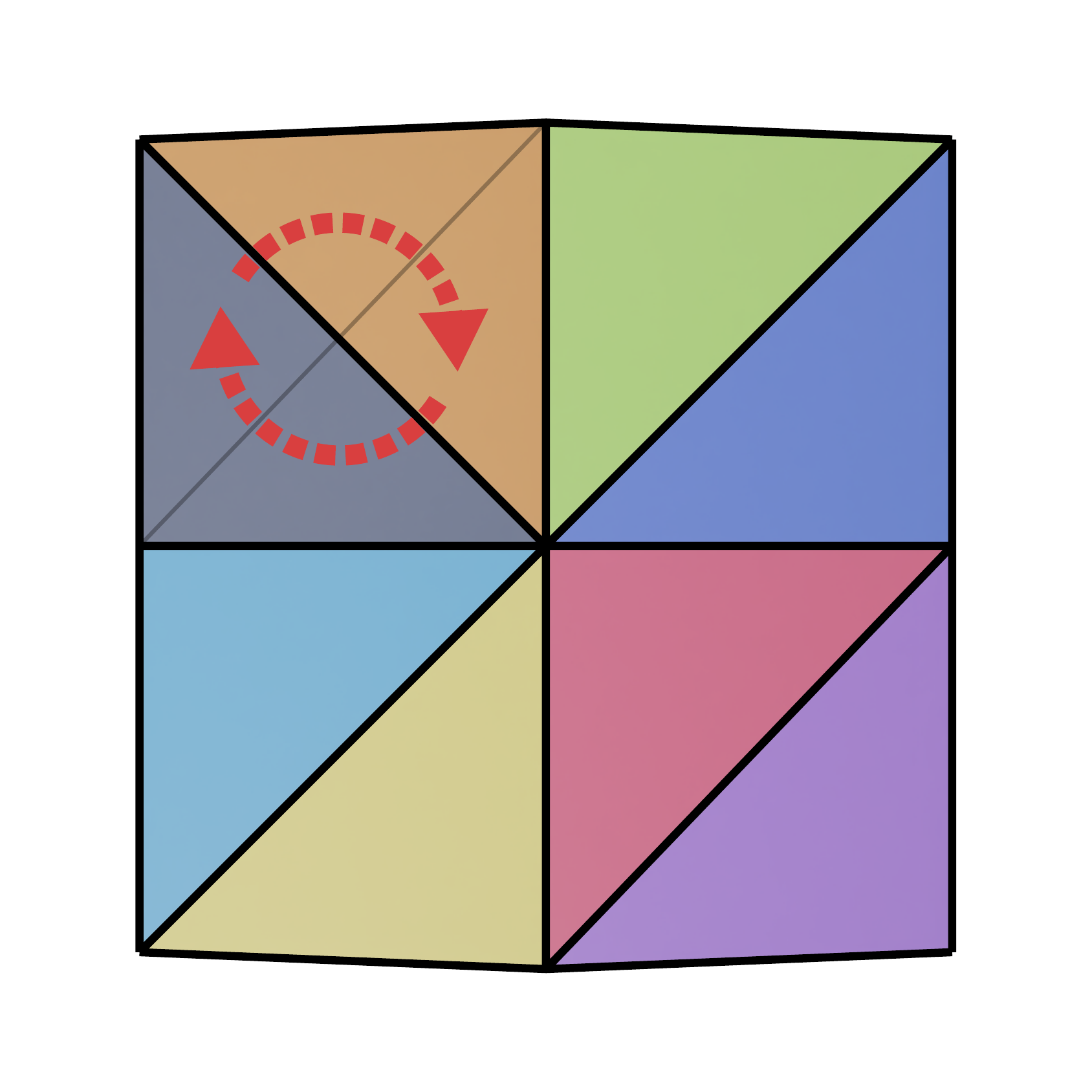} &
        \includegraphics[width=0.2\linewidth]{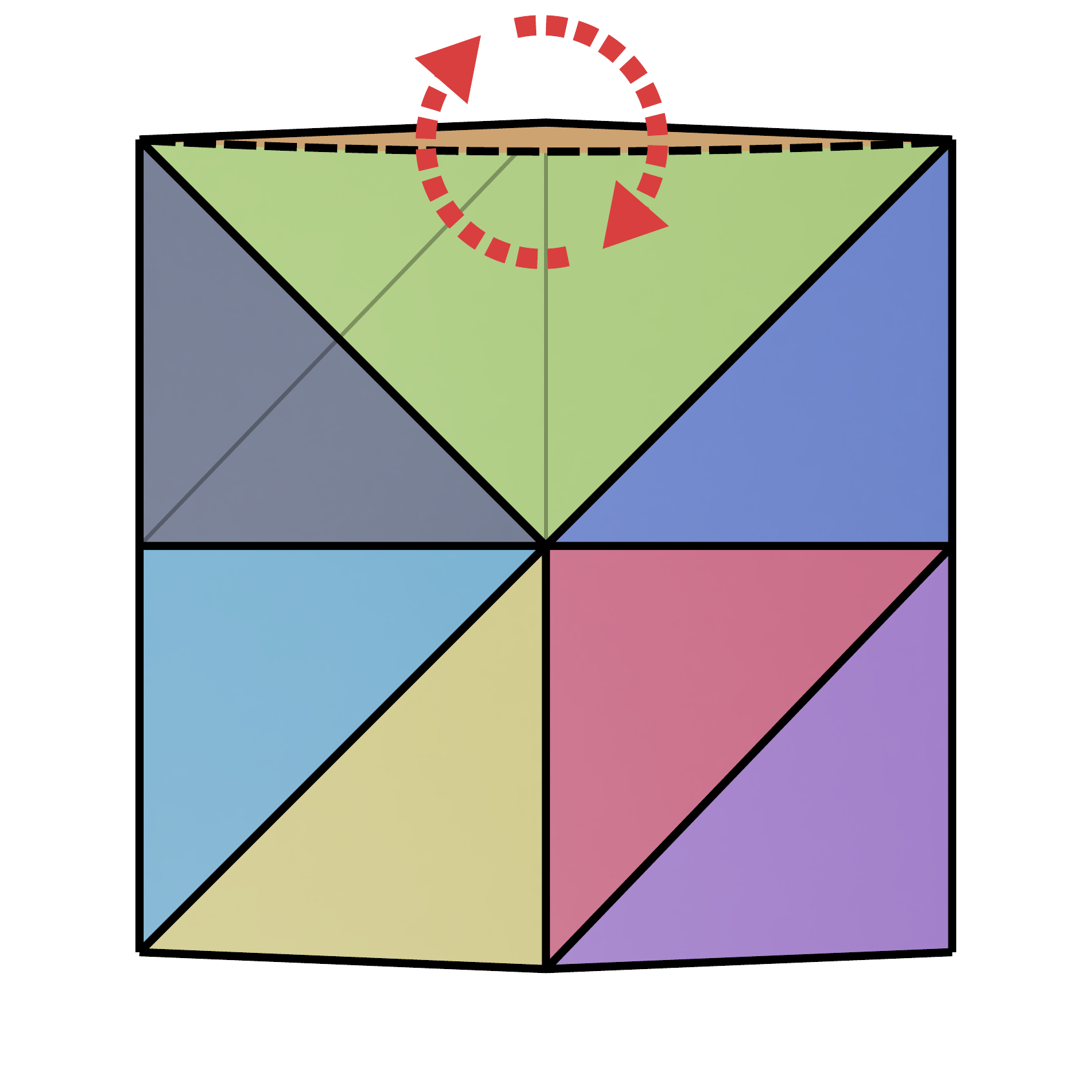} &
        \includegraphics[width=0.2\linewidth]{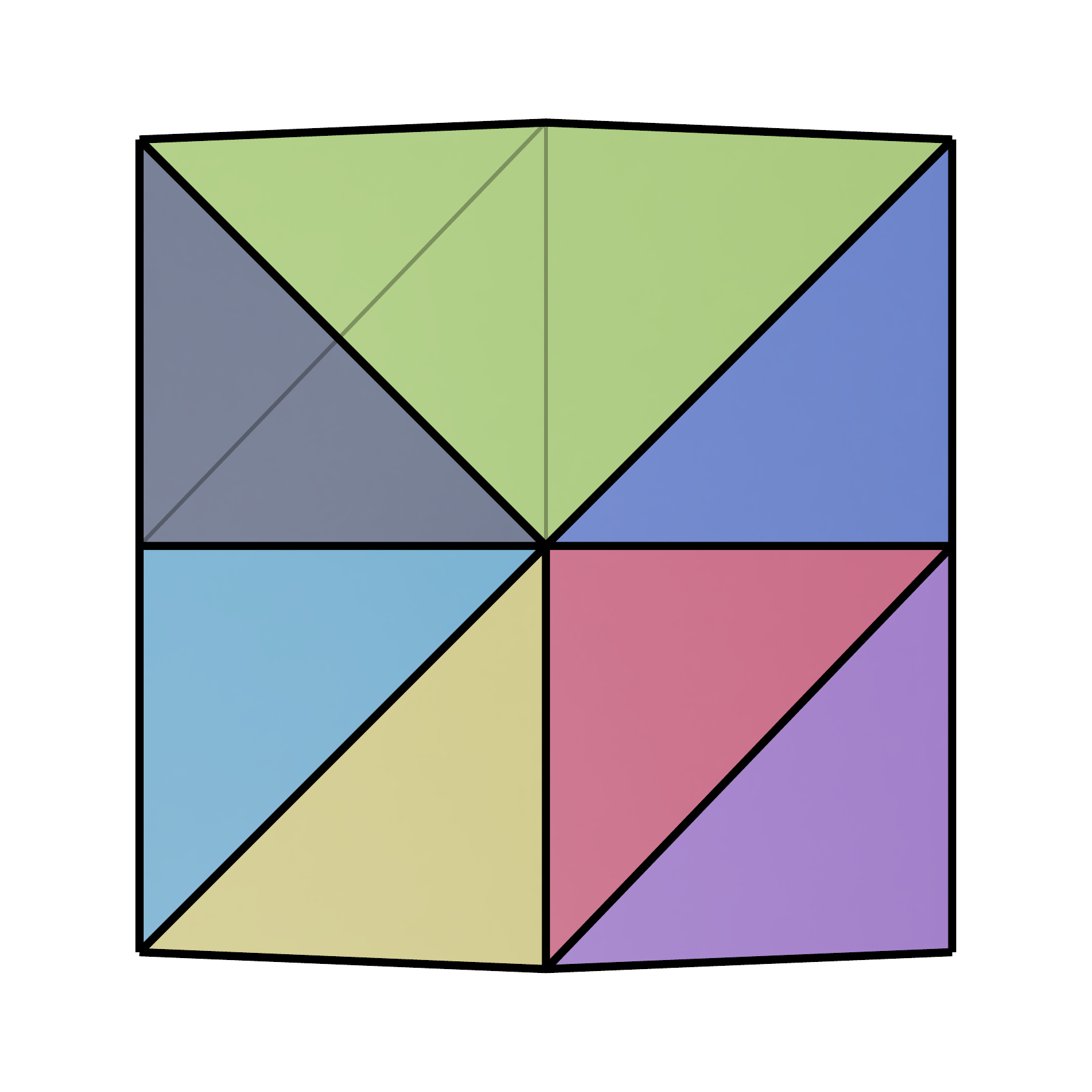}
      \\
      initially & edge flip & edge flip & remove \\
      (valence 4) & (valence 3) & (valence 2) &
    \end{tabular}
  }
    \caption{Illustration of intrinsic vertex removal by edge flipping for
      an interior (top) and a boundary (bottom) vertex.}
    \label{fig:simplification}
\end{figure}
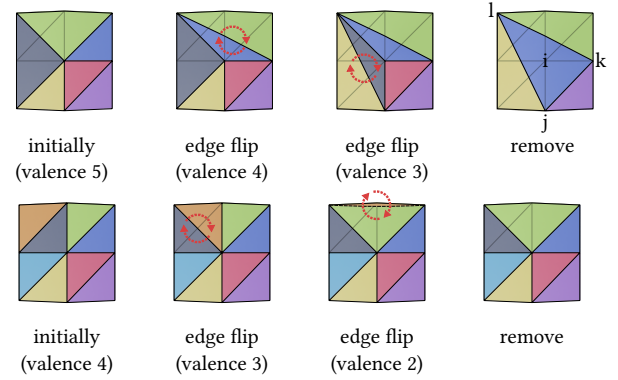

Although not strictly necessary, we start by performing an intrinsic Delaunay
retriangulation to ensure a well behaved mesh.  Next, we sort all vertices by the magnitude of
their Gaussian curvature in a queue $\P$ by smallest Gaussian curvature first,
and process the vertices in $\P$ until no vertices with a Gaussian curvature
less than $\g_{max}$ can be removed. For each vertex, we perform edge
flips on all incident edges until the desired valence (three for interior
vertices or two for vertices on boundaries) is reached. We avoid removing degree one vertices or those incident to self-edges and avoid flipping edges that would modify mesh topology.  We record each edge
flip in a FIFO queue $\Q$ for additional post-processing detailed below.  For
vertices with zero Gaussian curvature, we are guaranteed to reach the desired
valence~\cite{Sharp:2020:YCF, Gillespie:2021:ICI}. However this is not the case
for vertices with non-zero curvature. If the Gaussian curvature is extremely negative the valence can not be achieved. In practice models typically have few if any vertices with large negative curvature and we found that we can usually achieve
the desired valence.  When the desired
valence is reached, simplification can be easily achieved by removing the
vertex $i$ and all incident edges $ij$, $ik$ and $il$, such that the resulting
triangle $jkl$ forms a developable approximation
(\autoref{fig:simplification}). However, depending on the configuration of the
1-ring, removing the vertex can lead to an invalid triangulation.  We
therefore only remove vertices if $jkl$ strictly adheres to the triangle inequality.
When we are unable to reach the desired valence or if the triangle inequality condition is
not met, we undo the edge flips recorded in $\Q$ in reverse order to restore
the triangulation. It is possible that after removing more vertices, the
triangulation is more favorable for removing the vertex. Therefore, we
re-queue the vertex for re-processing after all outstanding vertices with a
Gaussian curvature less than $\g_{max}$ have been processed.

If a vertex $i$ can be successfully removed (i.e., it has the desired valence
after edge flipping, and it satisfies the triangle inequality condition), then we perform the
following steps:
\begin{itemize}
\item We update the Gaussian curvature of the vertices of $jkl$ in the
  processing queue $\P$ (either by updating the order if $j$, $k$, or $l$ was
  in the queue, or by adding the vertex if not yet queued).
\item We perform a post-removal edge flipping procedure. During edge flipping
  to achieve the desired valence, it is sometimes possible to create degenerate 
  triangles. Instead of trying to figure out a safe flipping order, we instead
  'repair' the triangulation after vertex removal. For each edge recorded in $\Q$, we 
  check (in reverse order) if the resulting edge is Delaunay.  If not, then we flip the 
  edge, and add the four edges of the resulting triangles to $\Q$.  We repeat this 
  process until $\Q$ is empty and the resulting triangulation meets the intrinsic Delaunay
  property again.
\end{itemize}

\paragraph{Discussion}
A benefit of our simplification algorithm is that it preserves the
Euler characteristic $\chi = V-E+F$ of the mesh.  For a closed mesh (without
boundary), each vertex removal step results in a deletion of $1$ vertex, $3$
edges, and $3$ faces, while adding $1$ new face ($\Delta\chi=-1+3-3+1=0$). In case there is a boundary, we remove $1$ vertex, $2$ edges
and $1$ face ($\Delta\chi=-1+2-1=0$) and boundary loops are unaffected. By virtue of the Gauss-Bonnet
theorem, we know that since the Euler characteristic is preserved, the sum
of the Gaussian curvature is unchanged.  Thus deleting a vertex $i$ implies
that its Gaussian curvature is redistributed to its neighbors. This also
justifies why after each vertex removal, we update the Gaussian curvature
sorted processing queue $\P$.

\begin{figure}
    \centering
    {\small
    \begin{tabular}{cccc}
    \includegraphics[width=0.2\linewidth]{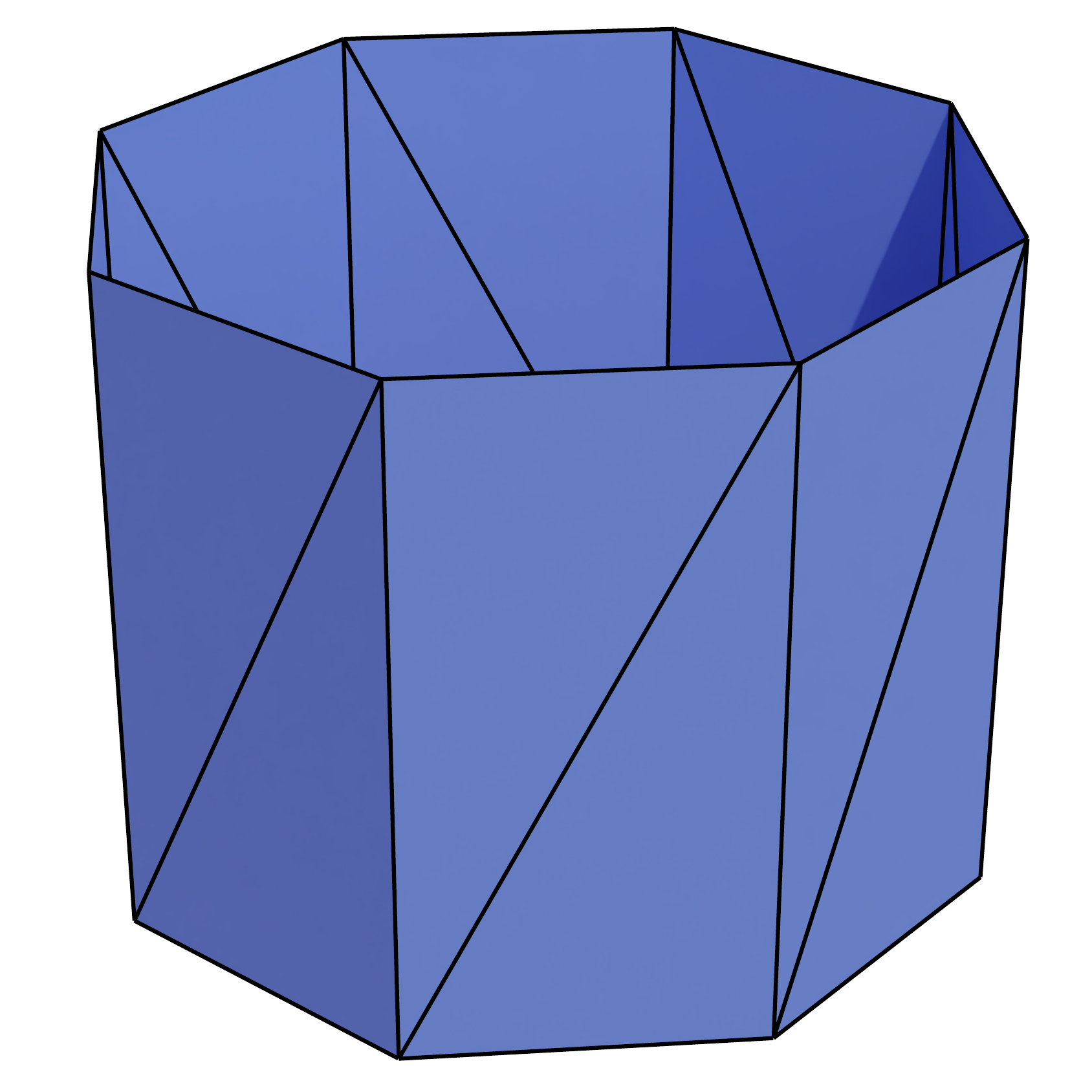} &
    \includegraphics[width=0.2\linewidth]{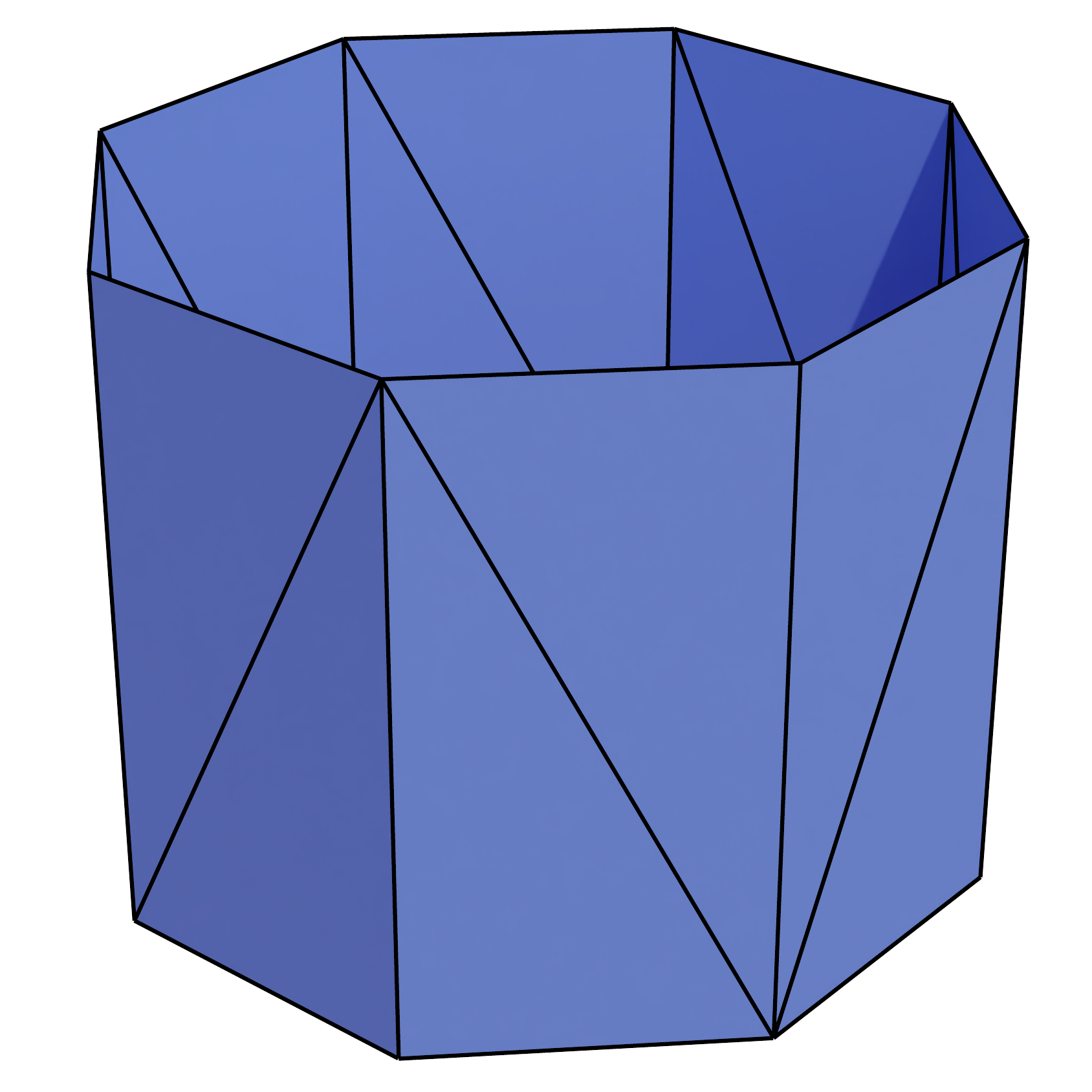} &
    \includegraphics[width=0.2\linewidth]{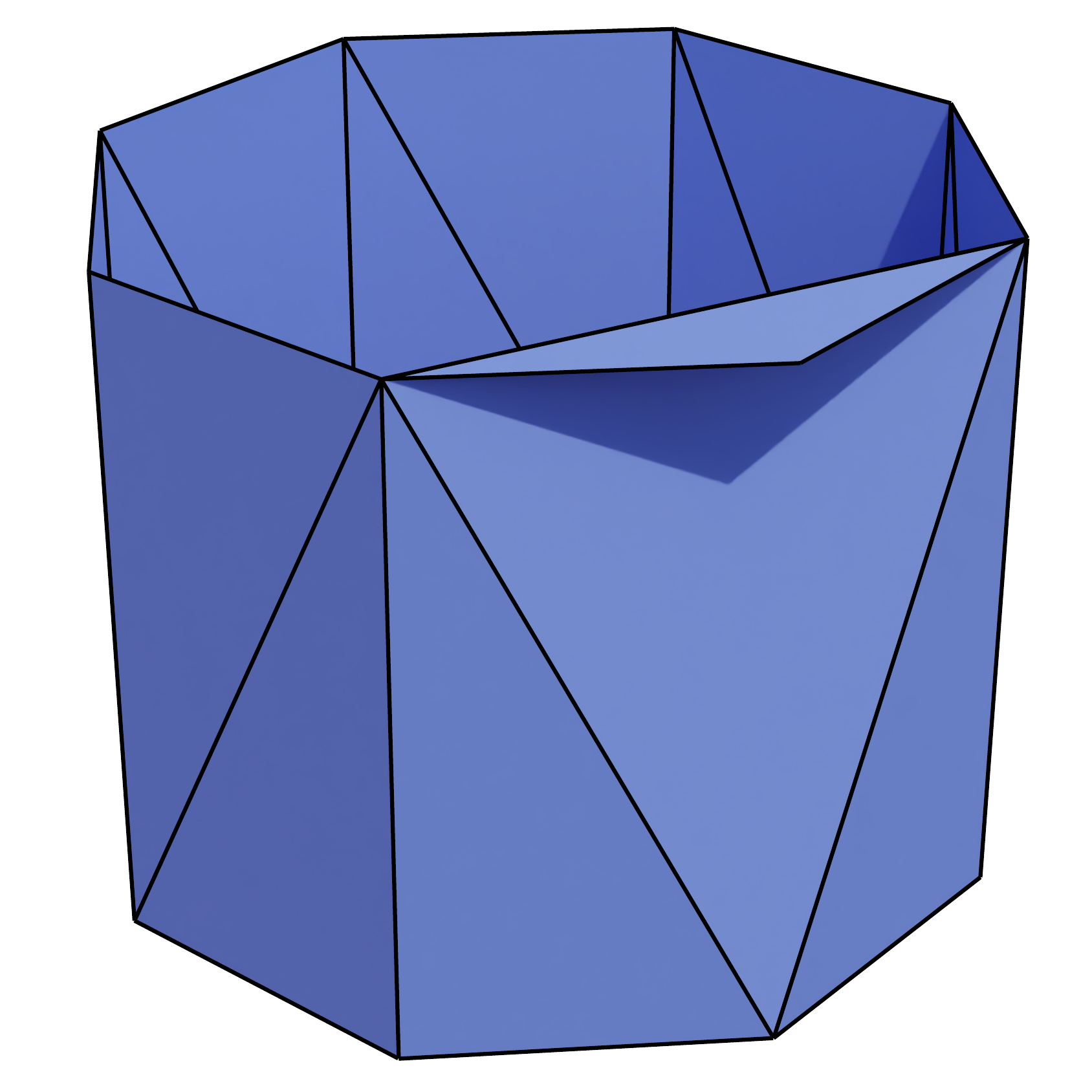} &
    \includegraphics[width=0.2\linewidth]{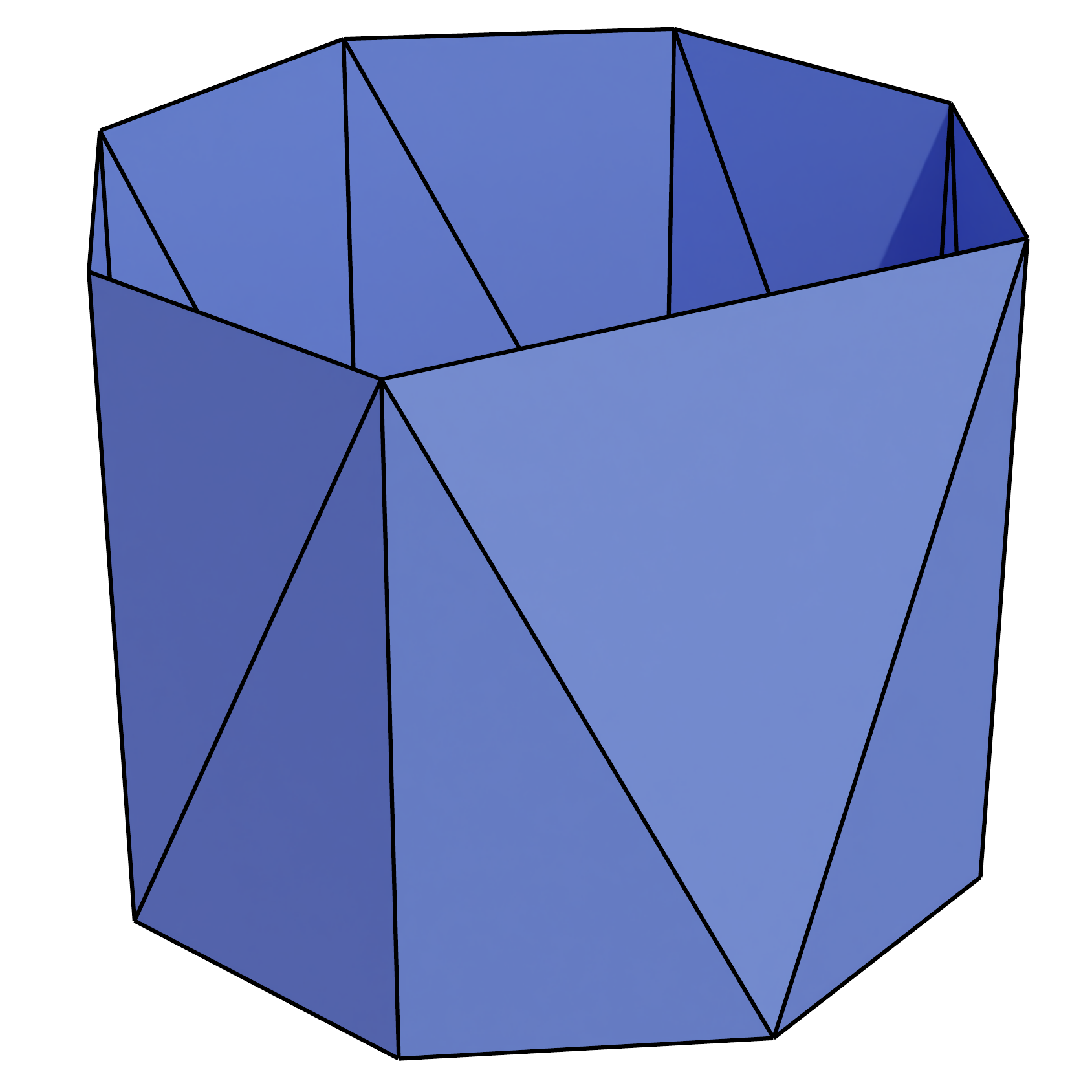}
      \\
      initially & edge flip & edge flip & remove
    \end{tabular}
    }
    \caption{Applying edge flipping for vertex removal in the extrinsic
      setting can significantly alter the metric.}
    \label{fig:badextrinsic}
\end{figure}

The above edge flipping strategy reduces the 1-ring polygon around each
candidate vertex for removal to the trivial re-triangulation case; i.e.,
such that no re-triangulation is needed.  We can apply this strategy only on
the intrinsic triangulation, not in an extrinsic setting where it can
significantly alter the metric. The metric can be altered when an edge flip 
is applied to the extrinsic mesh (\autoref{fig:badextrinsic}) whereas applying 
the same operation to an intrinsic edge preserves the metric.
Only the vertex removal step alters the surface metric, occurring only if 
the vertex has nonzero Gaussian curvature.

While vertex removal typically requires retriangulation in the extrinsic setting, it is unclear in the intrinsic domain how to update the edge lengths when the Gaussian curvature is not zero.  First, 
the 1-ring polygon needs to be flattened into to a developable surface, altering the global intrinsic geometry. Second, it is unclear what the updated edge 
lengths should be such that the candidate vertex is projected onto the 
developable approximation. By reducing the valence to three we do not need to
update any edge lengths in the intrinsic triangulations yielding a computationally light strategy that avoids both retriangulation and computing new edge lengths via optimization. While our approach avoids updating edge lengths
during vertex removal it requires us to carefully consider defining the barycentric coordinates of the removed vertex with respect to the remaining
triangle should they be required by downstream applications.  

\begin{table*}[htp]
  \centering
  \caption{Mesh simplification statistics over a subset of over $7,\!000$ manifold
    and oriented triangle meshes in the Thingi10K dataset comparing the percentage
    of vertices with a Gaussian curvature less then $\g_{max}$ versus the
    percentage removed after simplification.                                                            }
\label{tab:stats}
\begin{tabular}{cccccccc}
\toprule
& \multicolumn{2}{c}{Removable} & \multicolumn{2}{c}{Successfully Removed} & \multicolumn{3}{c}{Mean Time (s)} \\
\cmidrule(lr){2-3}
\cmidrule(lr){4-5}
\cmidrule(lr){6-8}
$\g_{max}$ &  mean & std. dev. &  mean & std. dev. & remove & track & total \\
\midrule
 $10^{-9}$ &  5.57\% &              13.84 & 99.56\% &               4.73 & 0.15s & 0.61s & 0.76s\\
 $10^{-6}$ &  7.48\% &              16.20 & 99.37\% &               5.15 & 0.16s & 0.87s & 1.03s\\
 $10^{-4}$ & 12.67\% &              20.23 & 95.58\% &               8.14 & 0.20s & 0.84s & 1.03s\\
 $10^{-2}$ & 33.96\% &              33.53 & 91.41\% &              10.12 & 0.33s & 1.24s & 1.57s\\
 $10^{-1}$ & 58.44\% &              37.81 & 89.18\% &              12.62 & 0.41s & 1.79s & 2.20s\\
 $1.00$    & 88.75\% &              22.34 & 94.87\% &               7.87 & 0.45s & 2.39s & 2.84s\\
 $\pi$     & 99.81\% &               2.21 & 94.57\% &               9.17 & 0.46s & 2.70s & 3.15s\\
\bottomrule
\end{tabular}
\end{table*}

\paragraph{Intrinsic-Extrinsic Correspondence}
Similar to the Signpost data structure\cite{Sharp:2019:NIT}, we keep track of the correspondence
between intrinsic and extrinsic triangulations.  Special care needs to be taken for
tracking the correspondence of deleted vertices.  This is achieved by
maintaining \emph{intrinsic} barycentric coordinates for each extrinsic vertex
deleted from the intrinsic triangulation.  There are three scenarios to
consider:
\begin{enumerate}
\item \emph{Defining the barycentric coordinate $(c_i^j, c_i^k, c_i^l)$ of a deleted
    vertex $i$} inside a triangle $jkl$. When the threshold $\g_{max}$ is not
  zero, the vertex is `projected' onto a developable approximation. Since this
  projection is an approximation, a number of viable options exist.  In our
  implementation, we opt for a conformal projection, which are known to have low
  distortion~\cite{Gillespie:2021:DCE, Springborn:2008:CET}. We fix
  the edge lengths of the bounding triangle $jk$, $kl$, $lj$ and seek a uniform
  scale of the edges $ij$, $ik$, and $il$ incident to $i$ that ensures the projection
  of $i$ is intrinsically flat (as in~\cite{Springborn:2008:CET} with fixed boundary).
  Since $i$ is degree three we can compute the uniform scale by ensuring that the corner
  angles of the projections of $ikl$, $ilj$, and $ijk$ agree with the corner angles of
  $jkl$. For example, for corner $j$ of $jkl$ we require that the scale $s$ satisfies: 
  $$\theta^{j}_{kl} = \theta^{j}_{ki} + \theta^{j}_{il}$$ where
  $\theta^{j}_{ki} = \arccos \left( \frac{l_{jk}^2 + s^2(l_{ij}^2 - l_{ki}^2)}{ 2l_{jk}s\l_{ij} }\right)$ 
  (similarly for $\theta^{j}_{il}$). In fact if the formula holds for any corner of $jkl$,
  say $j$ then the projection of edge $ji$ must lie in the (unfolded) plane of $ijk$
  and similarly for $i$ and the projections of edge $ki$ and $kl$, ensuring that the 
  new Gaussian curvature of $i$ is zero. Solving for $s$ in the corner angle 
  equation yields a quadratic in $s^2$. We then choose the solution for $s^2$ 
  (and hence $s$) that ensures the Gaussian curvature at the projection of $i$
  is zero and set the barycentric coordinate $(c_i^j, c_i^k, c_i^l)$ accordingly.
  If $s$ would cause the projections of any of $ikj$, $ijl$, and $ilk$ to violate
  the triangle inequality and $i$ (or any of its dependent vertices) is required (for downstream applications) we
  do not remove $i$ and instead restore the intrinsic mesh by undoing the edge flips
  recorded in $\Q$ in reverse order and reschedule it for deletion.
\item \emph{Updating the barycentric coordinates of a vertex $v$ that depends
    on a deleted vertex $i$}.  When a vertex is deleted, it is possible that a
  previously deleted vertex's barycentric coordinates depends on it.  Because
  our vertex deletion is only performed after the vertex has the desired
  valence, it follows that the dependent vertex must lie in one of the faces
  that will be merged. Hence, both the barycentric coordinates of the deleted
  and dependent vertex will depend on the same corner vertices after deletion.
  Thus, we can easily substitute the barycentric coordinates of the deleted
  vertex. For example, if $v \in ijk$, and $i$ is deleted, then the updated
  coordinates are: $(c_v^j + c_v^i c_i^j, c_v^k + c_v^i c_i^k, c_v^i c_i^l)$.
\item \emph{Updating the barycentric coordinate if the vertex is contained in
    a triangle whose edge is flipped.} In this case, we unfold both triangles isometrically into a local plane and recompute the barycentric coordinates
    with respect to the new face containing the vertex.
\end{enumerate}
\section{Results and Evaluation}
\label{sec:results}

We have implemented our intrinsic simplification method in $\mathsf{C\!+\!+}$
using a half-edge data structure annotated with edge lengths.
We have validated our method on a subset of the \emph{Thingi10k}~\cite{Zhou:2016:TTK} dataset containing valid manifold and oriented triangle meshes. Processing all $7k$ triangle meshes takes approximately \num{25}
hours to run the whole algorithm for the $7$ 
 threshold values for $\g_{max}$: $10^{-9}$, 
$10^{-6}$, $10^{-4}$, $10^{-2}$, $10^{-1}$, $1.0$, $\pi$ (in radians) on a
Intel i5-8265U (1.60GHz) CPU with $16$GB of memory (using a single core).

\begin{figure*}
    \centering
    {\small
    \def\reswidth{0.25\linewidth}
    \begin{tabular}{ccc}
    \includegraphics[width=\reswidth]{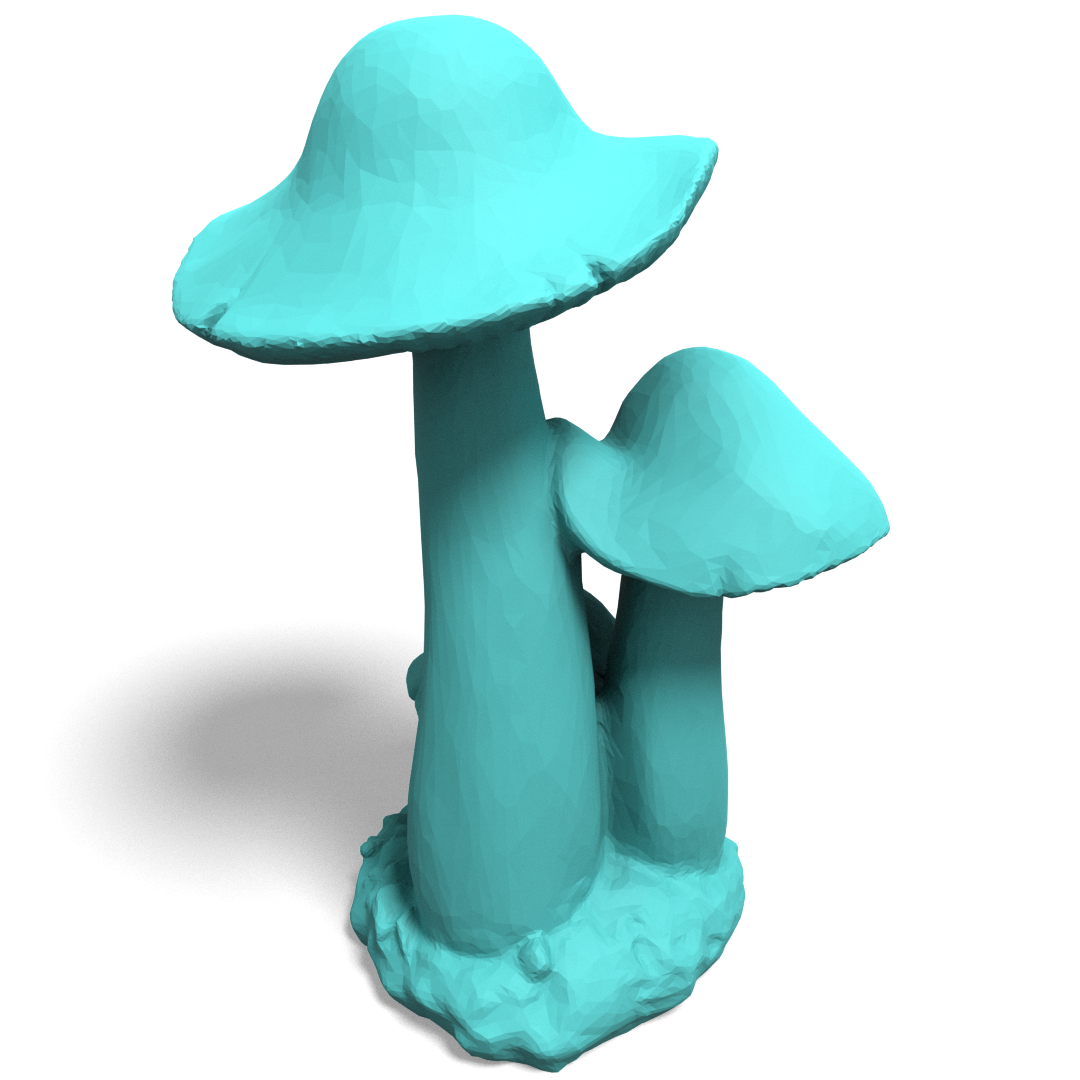} &
    \includegraphics[width=\reswidth]{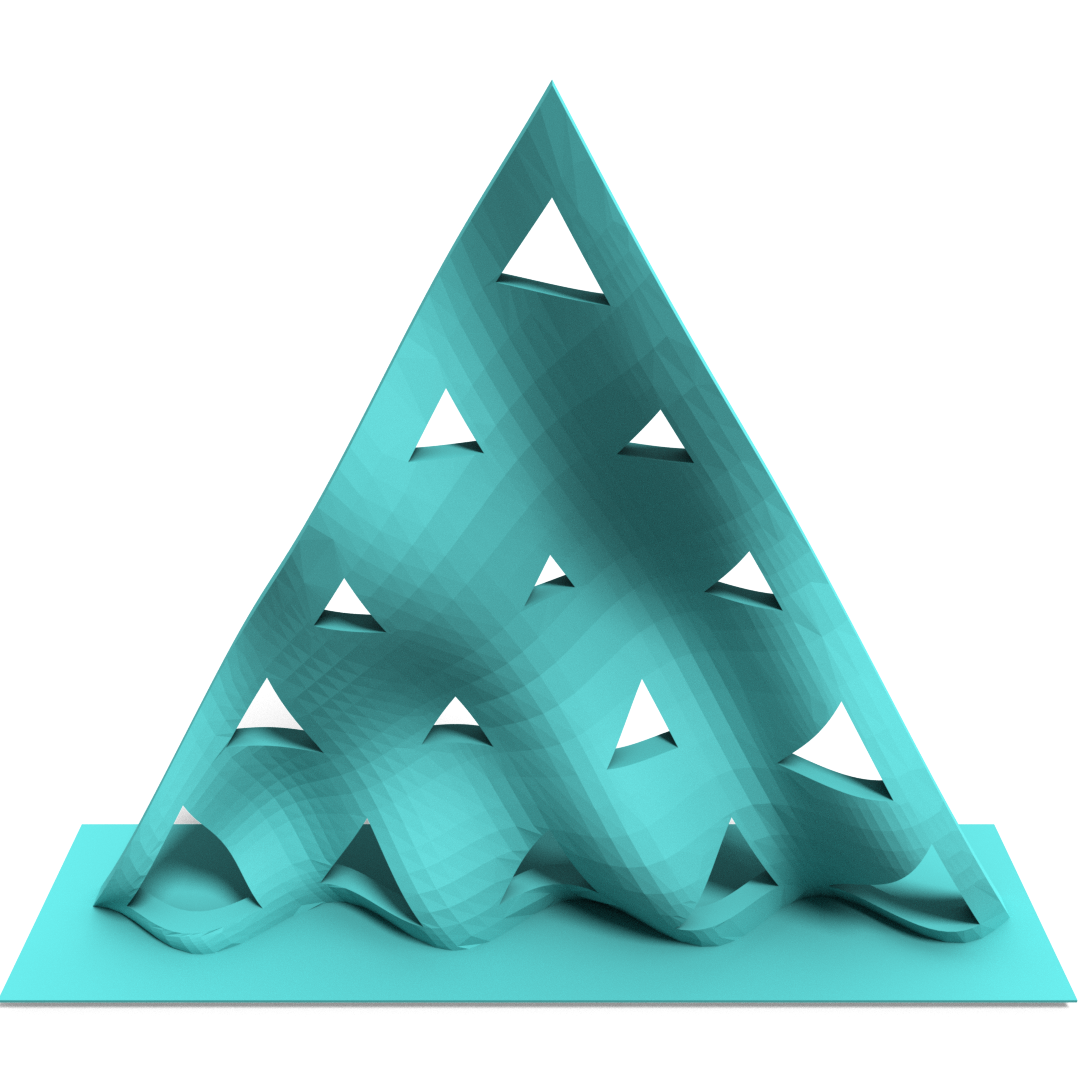} &
    \includegraphics[width=\reswidth]{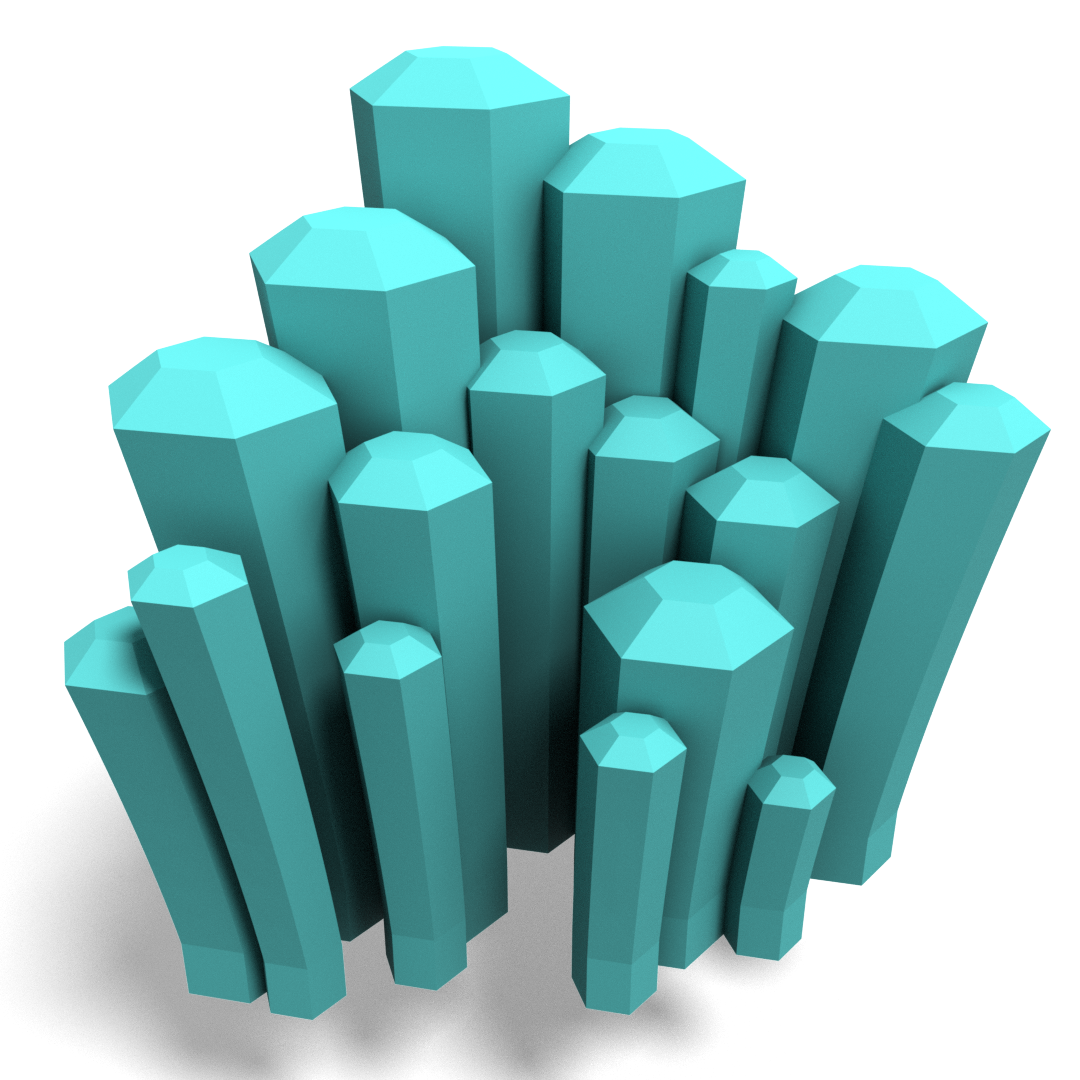}
    \vspace{-1ex} \\
    Mushrooms & Sine Wave Surface & Crystals \\
    \num{14985} vertices & \num{5105} vertices & \num{3305} vertices
    \end{tabular}\\
    \begin{tabular}{ccc}
    \toprule
    $\g_{max} = 10^{-4}$ & $\g_{max} = 10^{-2}$ & $\g_{max} = 1.0$ \\
    \midrule
    \includegraphics[width=\reswidth]{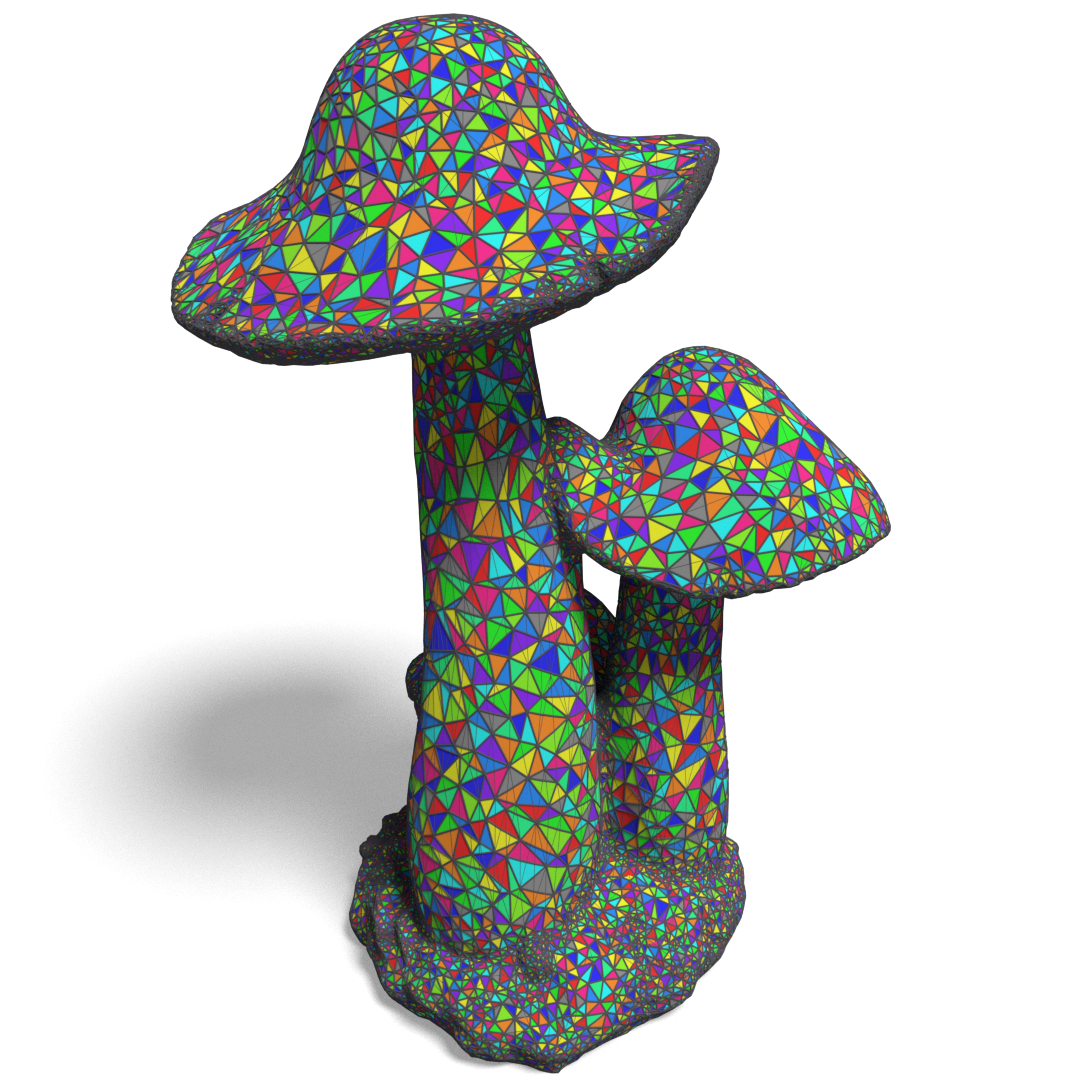} &
    \includegraphics[width=\reswidth]{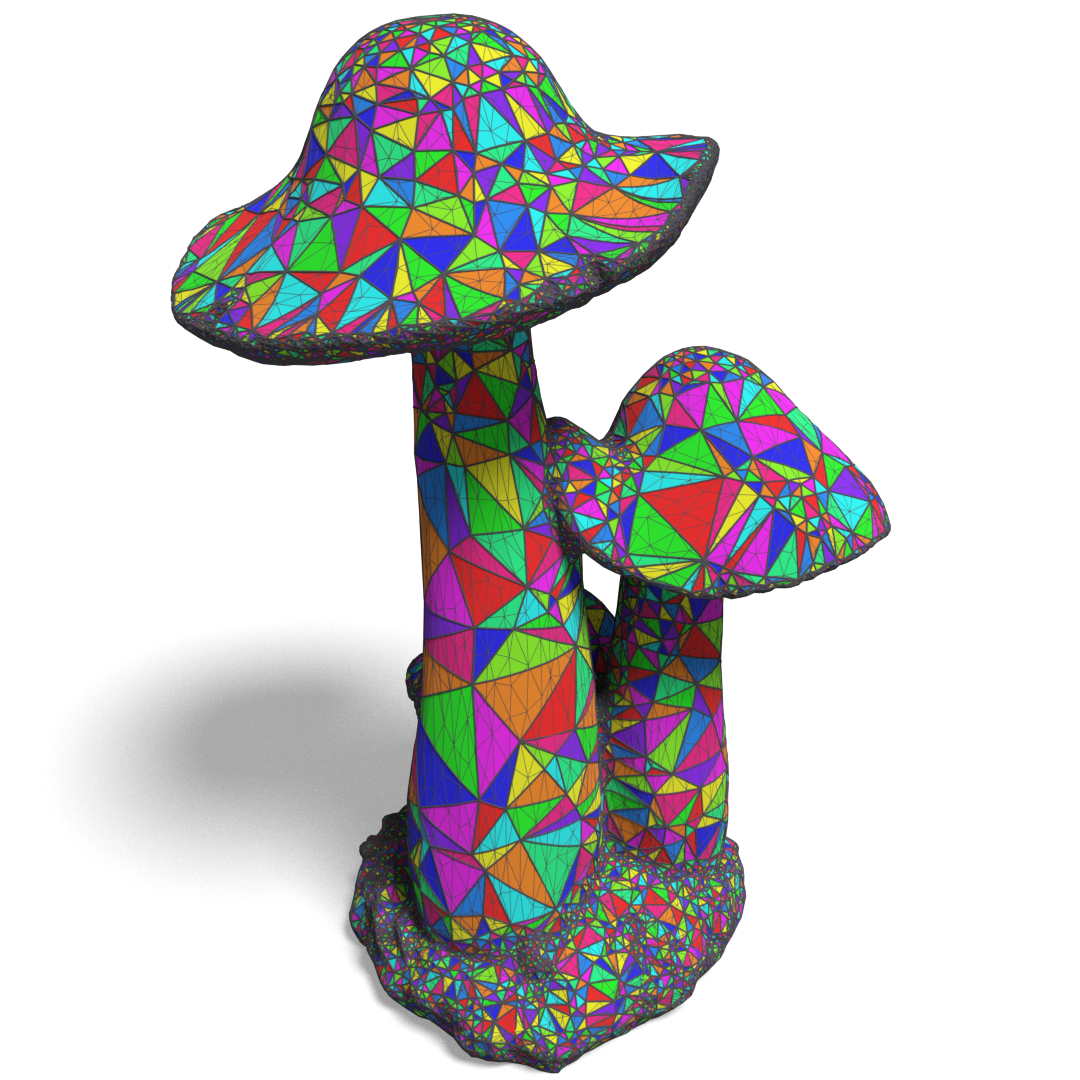} &
    \includegraphics[width=\reswidth]{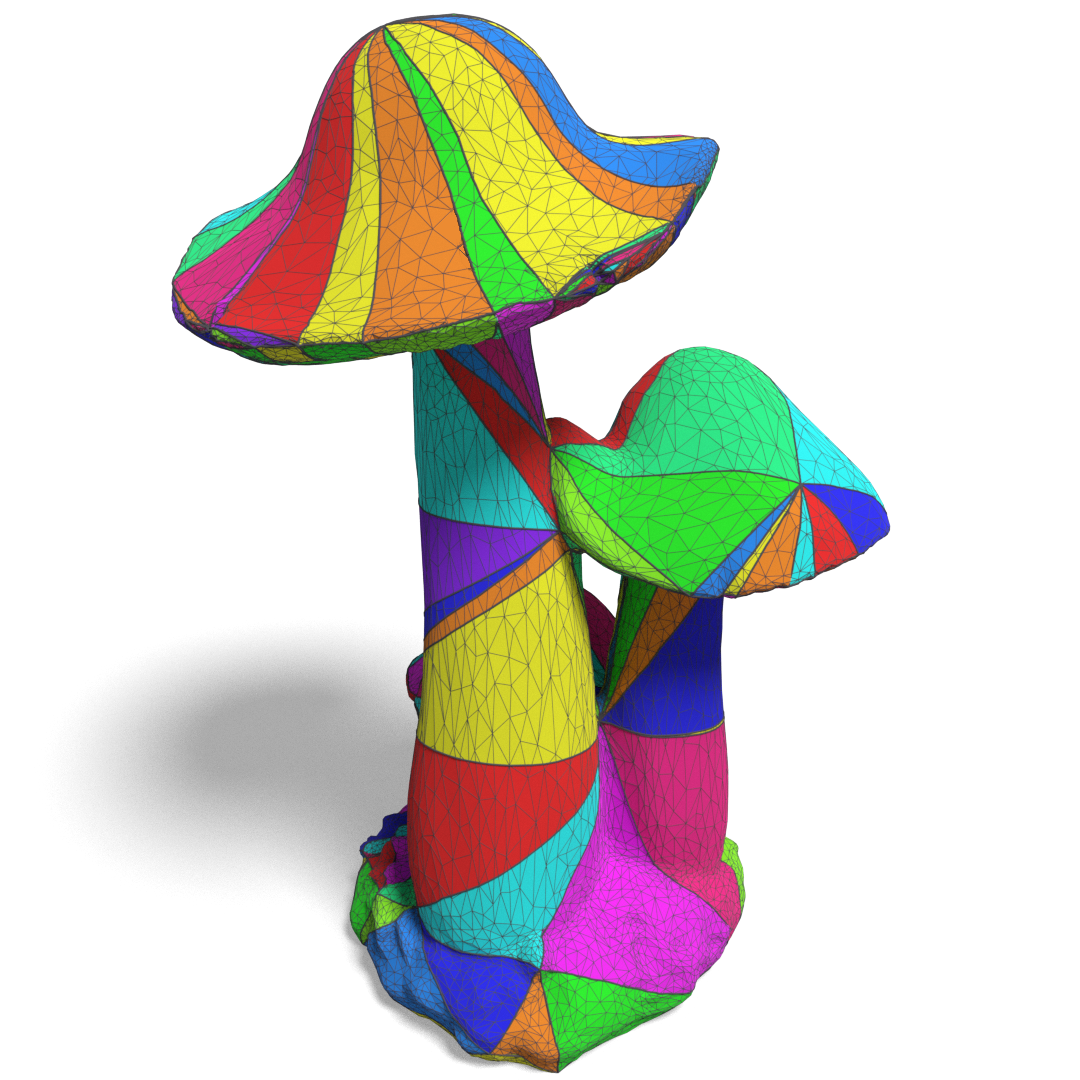}
    \vspace{-1ex} \\
    \num{14698} vertices & \num{11146} vertices & \num{340} vertices \\
    \num{1.92}\% reduction & \num{25.62}\% reduction & \num{97.73}\% reduction
    \vspace{1ex} \\
    \includegraphics[width=\reswidth]{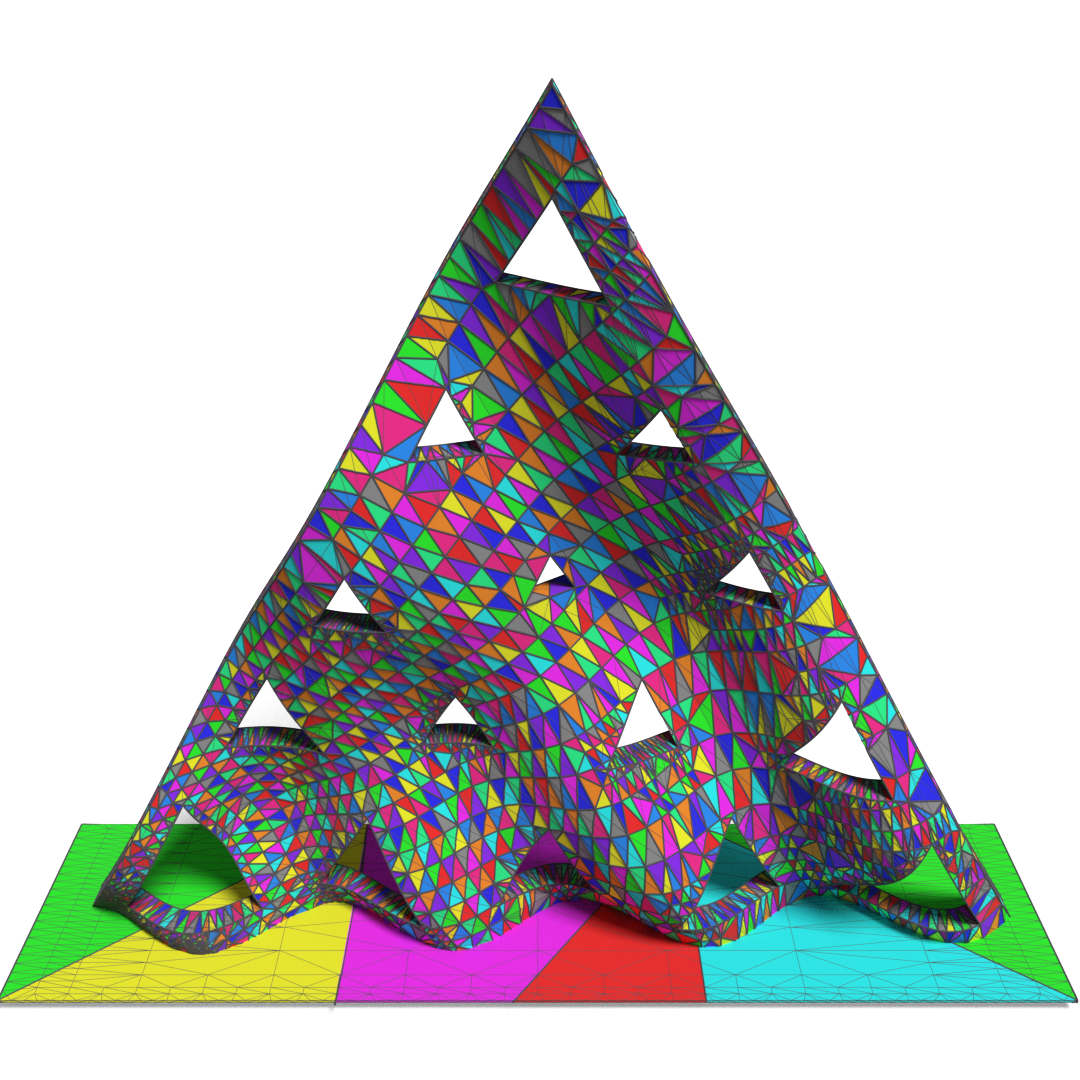} &
    \includegraphics[width=\reswidth]{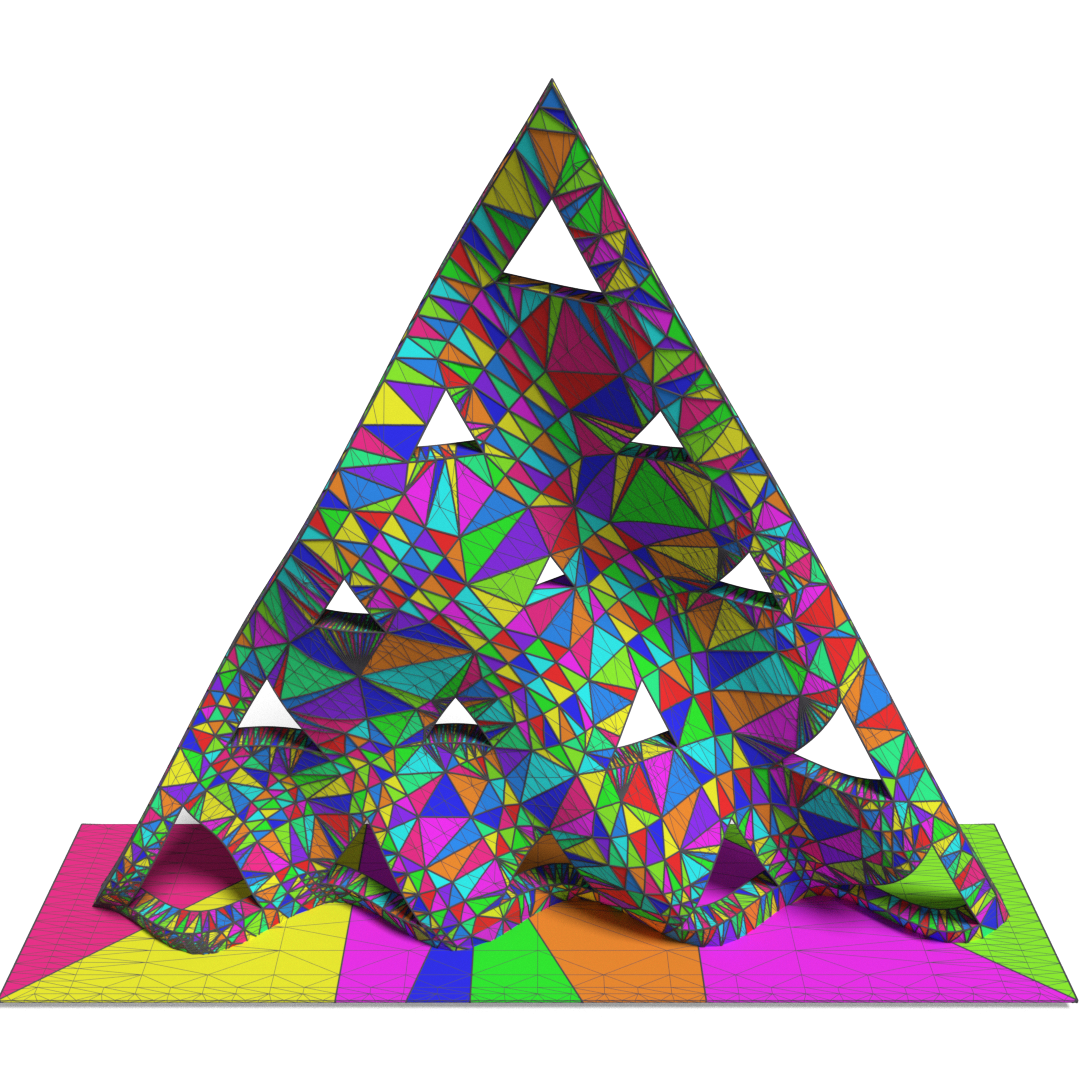} &
    \includegraphics[width=\reswidth]{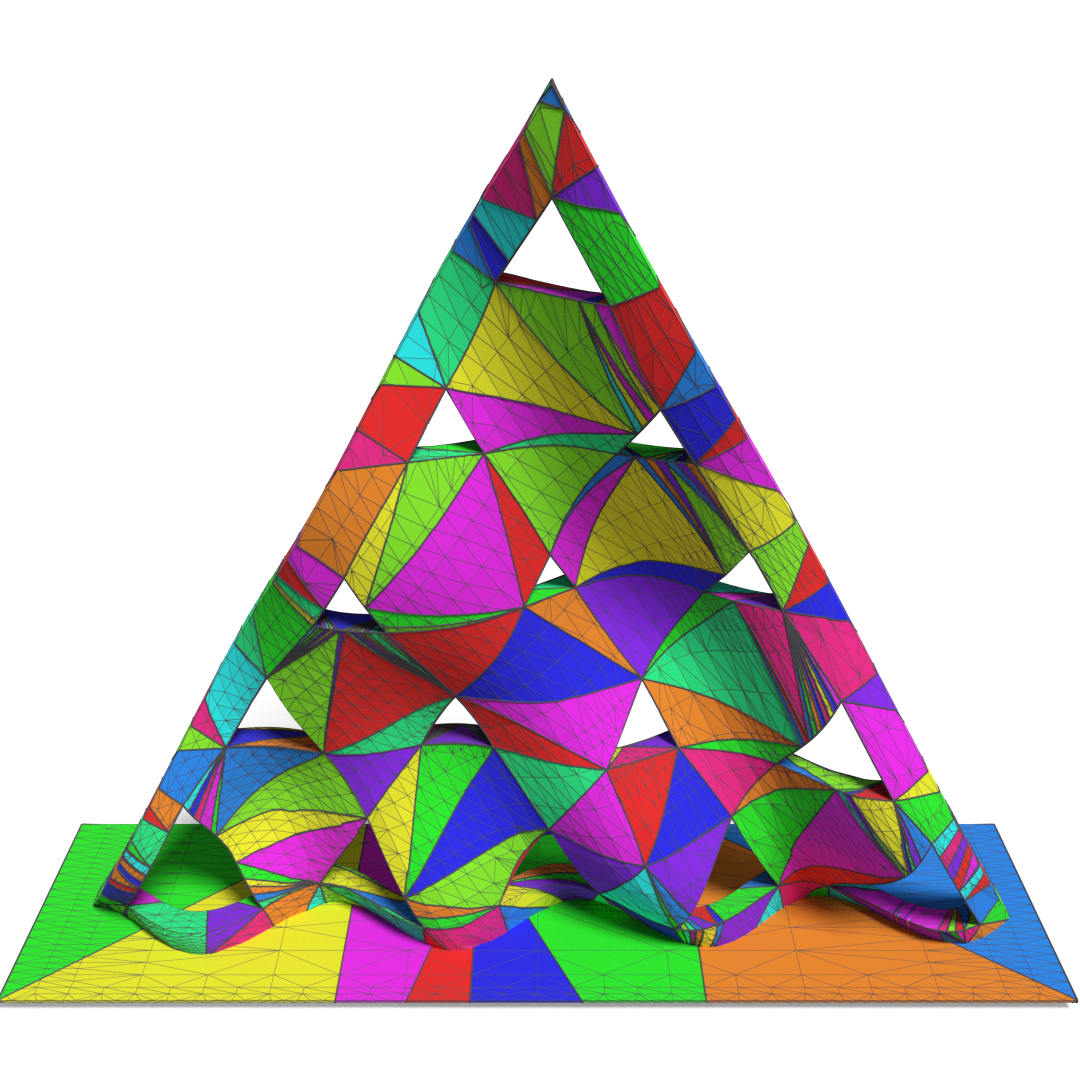}
    \vspace{-1ex} \\
    \num{3447} vertices & \num{2108} vertices & \num{242} vertices \\
    \num{32.48}\% reduction & \num{58.71}\% reduction & \num{96.04}\% reduction
    \vspace{1ex} \\
    \includegraphics[width=\reswidth]{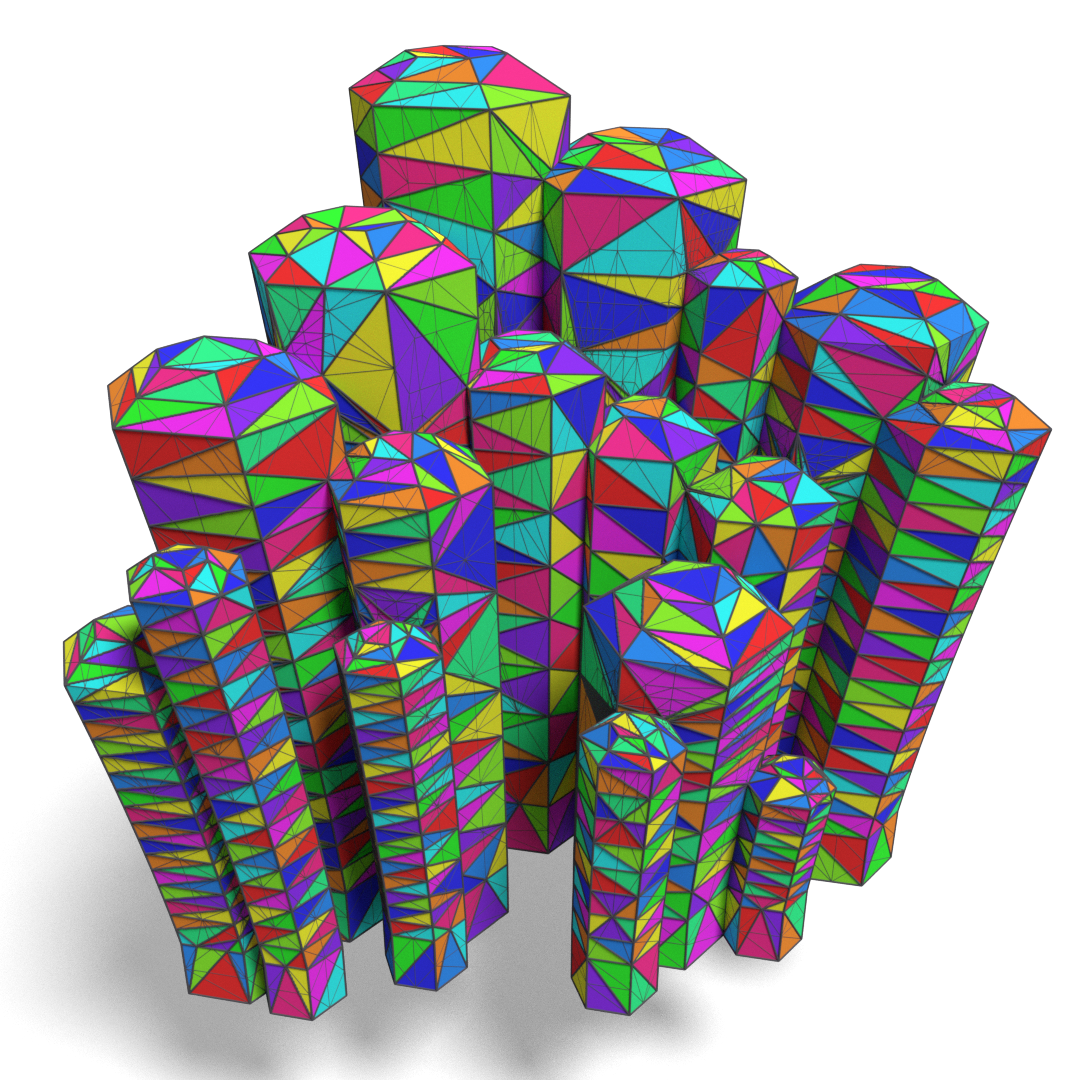} &
    \includegraphics[width=\reswidth]{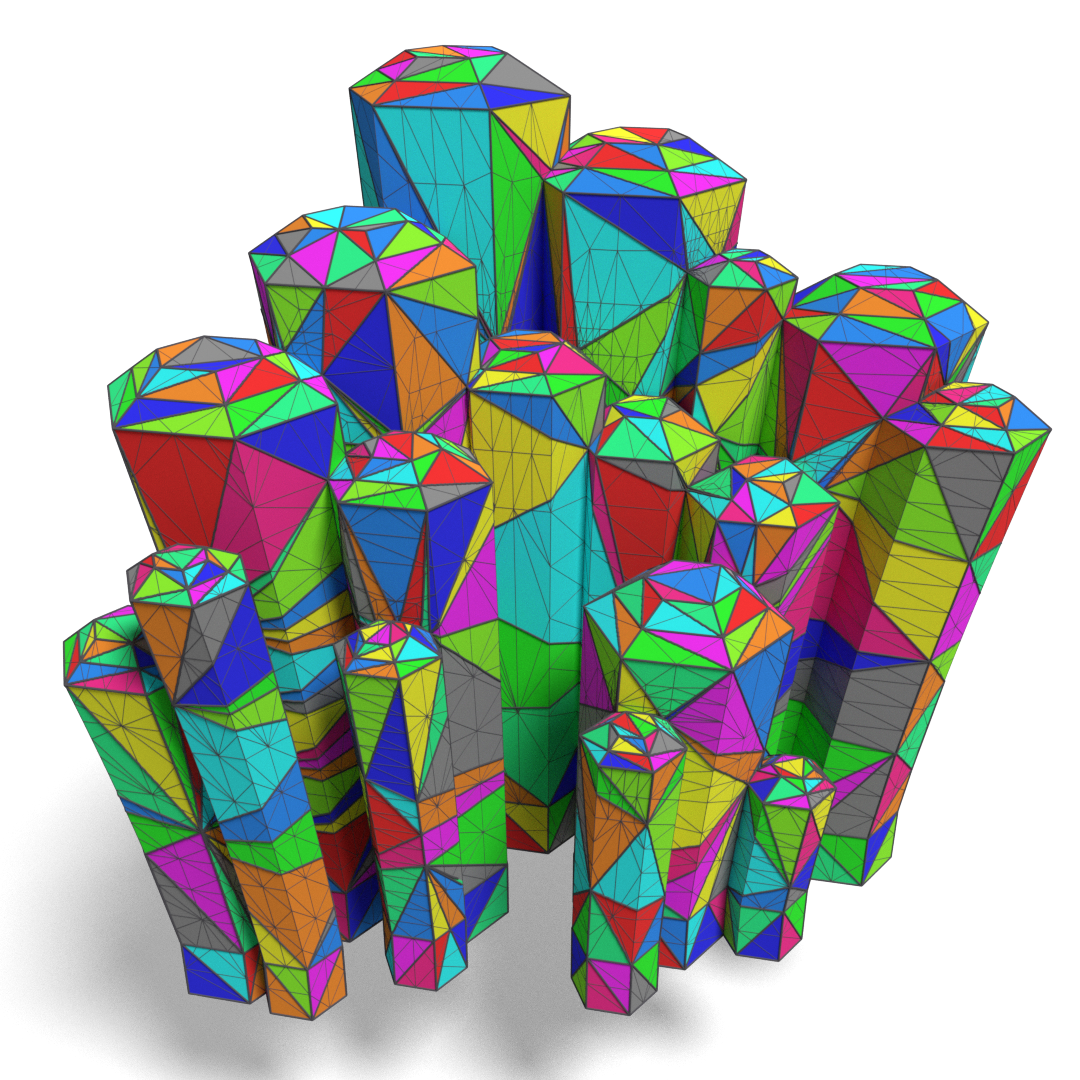} &
    \includegraphics[width=\reswidth]{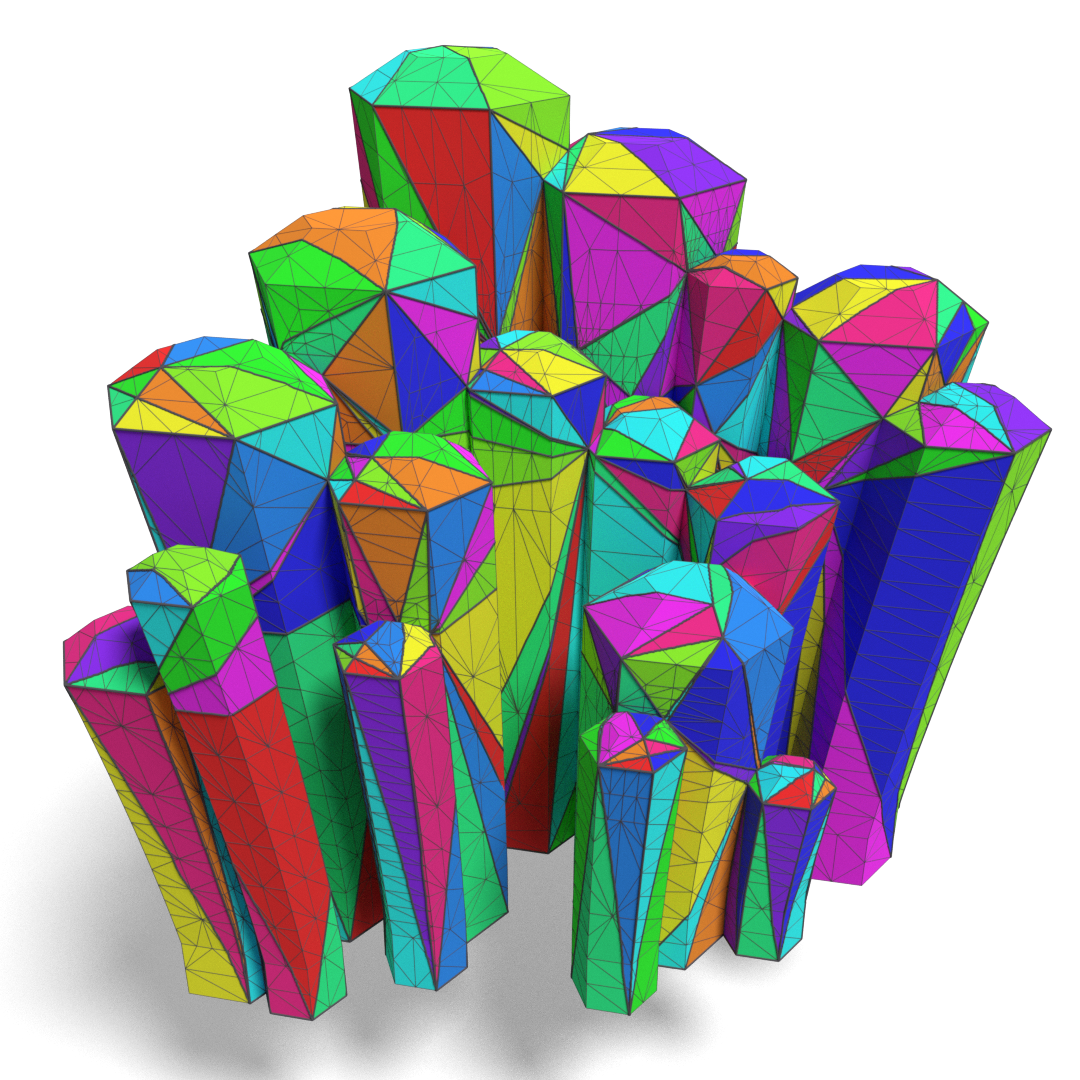}
    \vspace{-1ex} \\
    \num{1882} vertices & \num{718} vertices & \num{307} vertices \\
    \num{43.06}\% reduction & \num{78.28}\% reduction & \num{90.07}\% reduction \\
    \bottomrule
    \end{tabular}
    }
    \caption{Intrinsic mesh simplification results on three meshes for
      different thresholds $\g_{max}$. Note that the simplified intrinsic mesh
    is projected on the original mesh, showing only an approximation of the intrinsic
    triangulation's geometry.}
    \label{fig:results}
\end{figure*}

\begin{figure*}
    \centering
    \def\reswidth{0.5\linewidth}
    {
    \begin{tabular}{cc}
    \includegraphics[width=\reswidth]{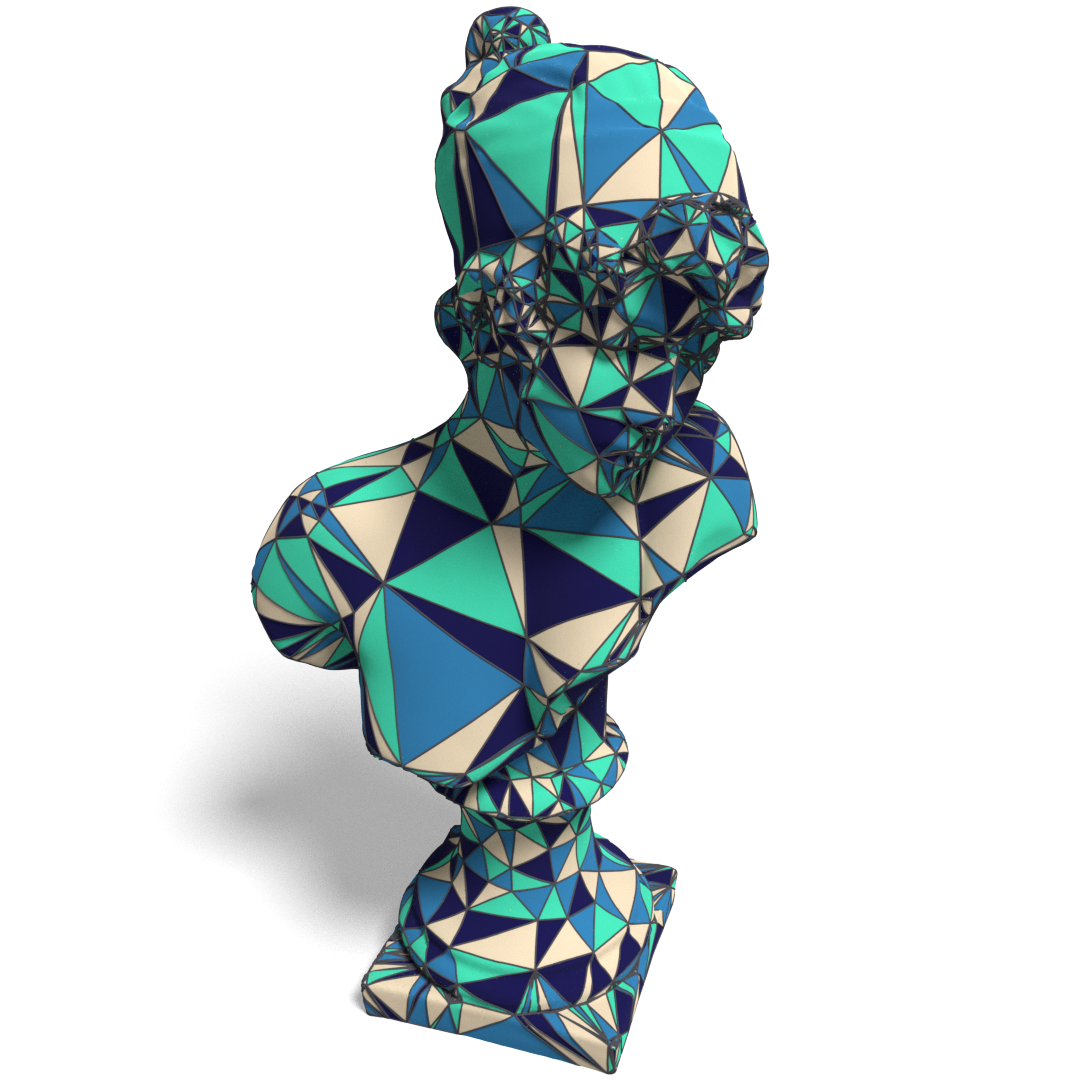} &
    \includegraphics[width=\reswidth]{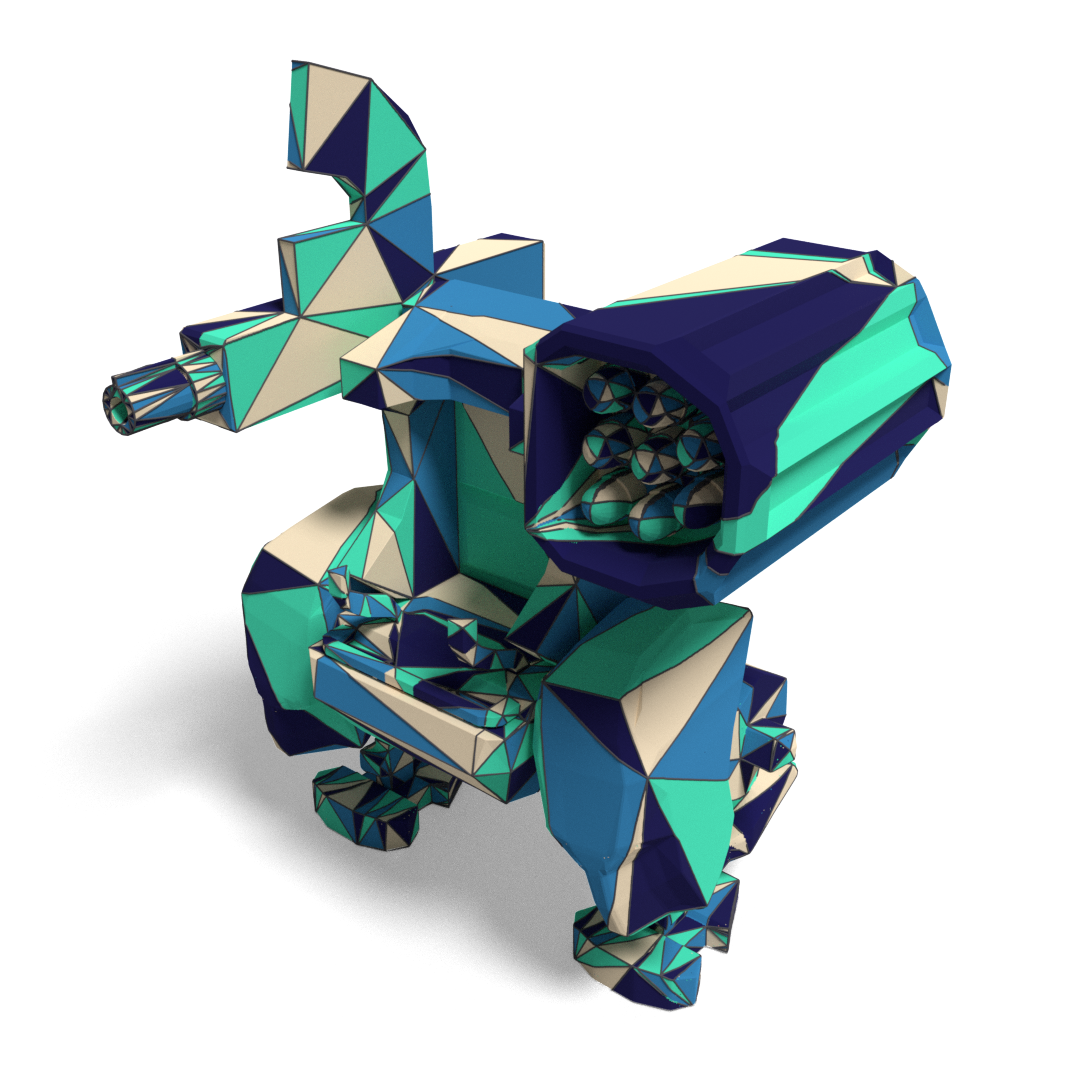} \\
    $\g_{max} = 10^{-1}$ & $\g_{max} = 1.0$ \\
    Original: \num{140864} vertices & Original:  \num{52151} vertices \\
    Simplified: \num{2258} vertices & Simplified: \num{914}  vertices \\
    \num{98.40}\% reduction & \num{98.25}\% reduction \\
    \end{tabular}
    }
    \caption{Our algorithm applied to various meshes from \emph{Thingi10k} with large $\g_{max}$,
    achieving reduction in excess of \num{98}\%.}
    \label{fig:extreme}
\end{figure*}

\autoref{fig:results} shows a selection of \num{3} meshes from the test set. For
each mesh we show the original mesh, and \num{3} thresholds
$\g_{max} = \{10^{-4}, 10^{-2}, 1.0\}$.  For each visualization the
simplified mesh is projected on the original extrinsic mesh, and thus shows
an approximation of the intrinsic geometry of the simplified mesh. Typically, a
common subdivision is utilized to provide a visualization of an intrinsic
triangulation~\cite{Sharp:2019:NIT, Gillespie:2021:ICI} by making use of the
fact that both triangulations have the same intrinsic geometry. When vertices with 
non-zero Gaussian curvature are deleted the correspondence between the two triangulations
breaks down and these methods fail. To visualize a simplified mesh we first perform
a refinement of the original extrinsic mesh. We then track the elements of
the refined mesh during simplification. After simplification we color the
elements of the refined mesh according to the intrinsic face they map to
post-simplification. By tracking the elements of a heavily refined mesh we
are able to produce a visualization that approximates the intrinsic triangulation.
A more robust method of visualizing a simplified intrinsic mesh, perhaps by
producing a (pseudo) common subdivision is an interesting avenue for future research.
For each example in~\autoref{fig:results} we list $\g_{max}$ and the number of vertices
in the simplified mesh.  Note how our method simplifies most in regions that
are (approximately) developable. For example, in the \emph{Sine Wave Surface}
example, we can see that significant simplification takes place on the
base of the model and similarly for the sharp edges of the \emph{Crystals} model. Note how some
of the simplified intrinsic triangles form a curved surface in $\R{3}$. For the lowest
threshold $\g_{max}=10^{-4}$ there is virtually no simplification in the doubly-curved
regions of \emph{Sine Wave Surface} and \emph{Mushrooms} (along the cap). In general there 
is little simplification in \emph{Mushrooms} for $\g_{max}=10^{-4}$, even in the stalk region,
due to noisy vertex placement in the model. However, as the threshold increases, our algorithm
is able to achieve a large reduction in the noisy model.  As expected, for
each example, a higher threshold removes much move vertices; particularly at the highest
threshold $\g_{max}=10^{1.0}$ where our algorithm achieves a reduction of \num{97.73}\%, 
\num{96.04}\%, and \num{90.07}\% for \emph{Mushrooms}, \emph{Sine Wave Surface}, and
\emph{Crystals} respectively. Notice that the doubly curved regions of \emph{Sine Wave Surface}
and \emph{Mushrooms} (on the mushroom caps) are approximated by larger developable patches as
$\g{max}$ increases. \autoref{fig:extreme} shows \num{2} meshes simplified to less than $2$\%
of their original vertex count, approximating each surface with a small number of developable
 patches.

To gain further insight in the efficacy of our intrinsic simplification
method, \autoref{tab:stats} reports the mean percentage and standard deviation
of ideally removable vertices (i.e., those with a Gaussian curvature less than
$\g_{max}$) and the mean percentage (and standard deviation) of actually
removed vertices for seven different thresholds
$\g_{max} = \{10^{-9}, 10^{-6}, 10^{-4}, 10^{-2}, 10^{-1}, 1.0, \pi\}$.  
As expected, with an increasing threshold, more vertices can be removed, 
and in general vertex removal succeeds for almost all candidate vertices.  
In addition, we also report the mean running time (excluding data IO) when
tracking all removed vertices via intrinsic barycentric coordinates. Typically
\num{80}\% of the compute time is spent tracking barycentric coordinates.

\begin{figure*}
    \centering
    \def\reswidth{0.2\linewidth}
    {
    \begin{tabular}{cccc}
    \includegraphics[width=\reswidth]{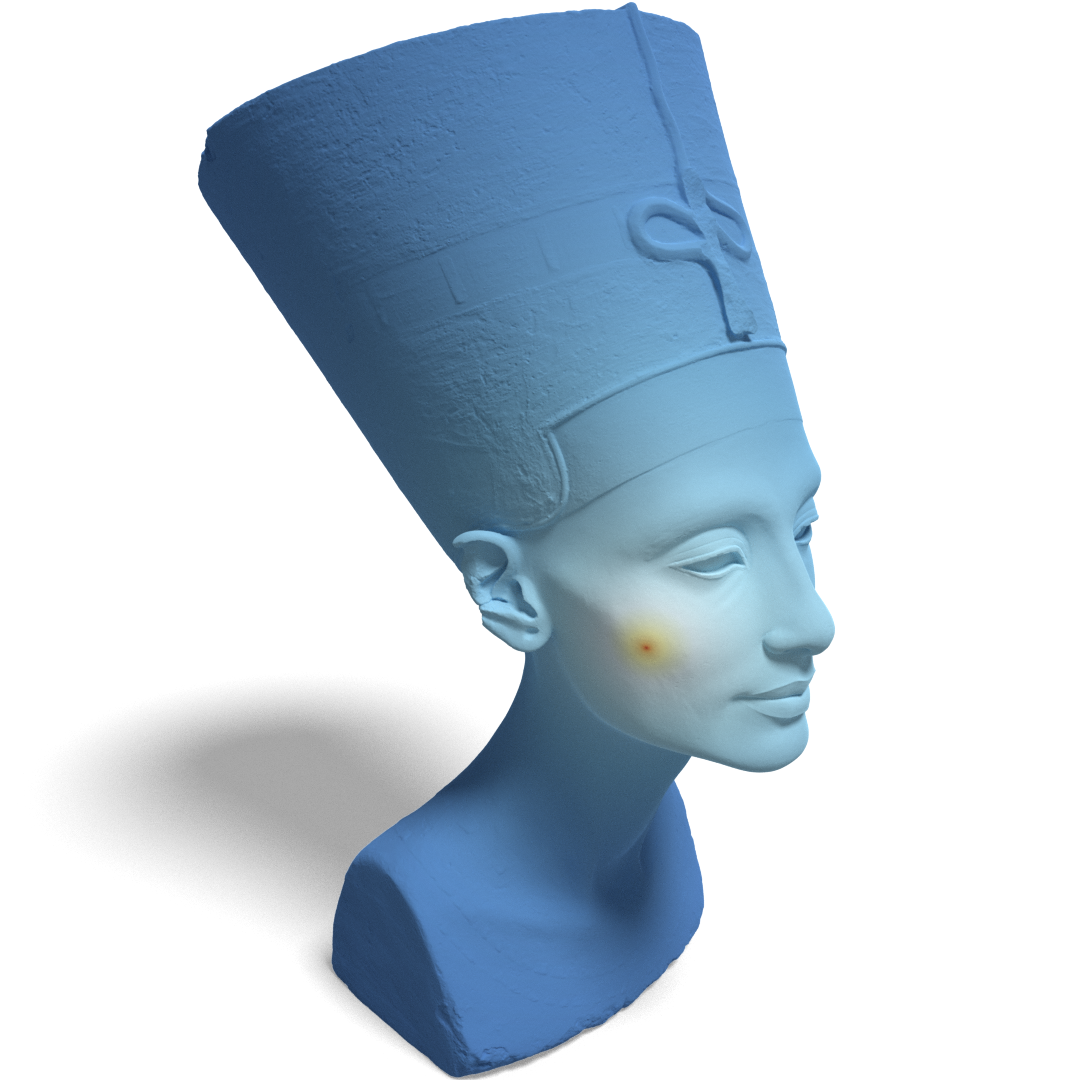} &
    \includegraphics[width=\reswidth]{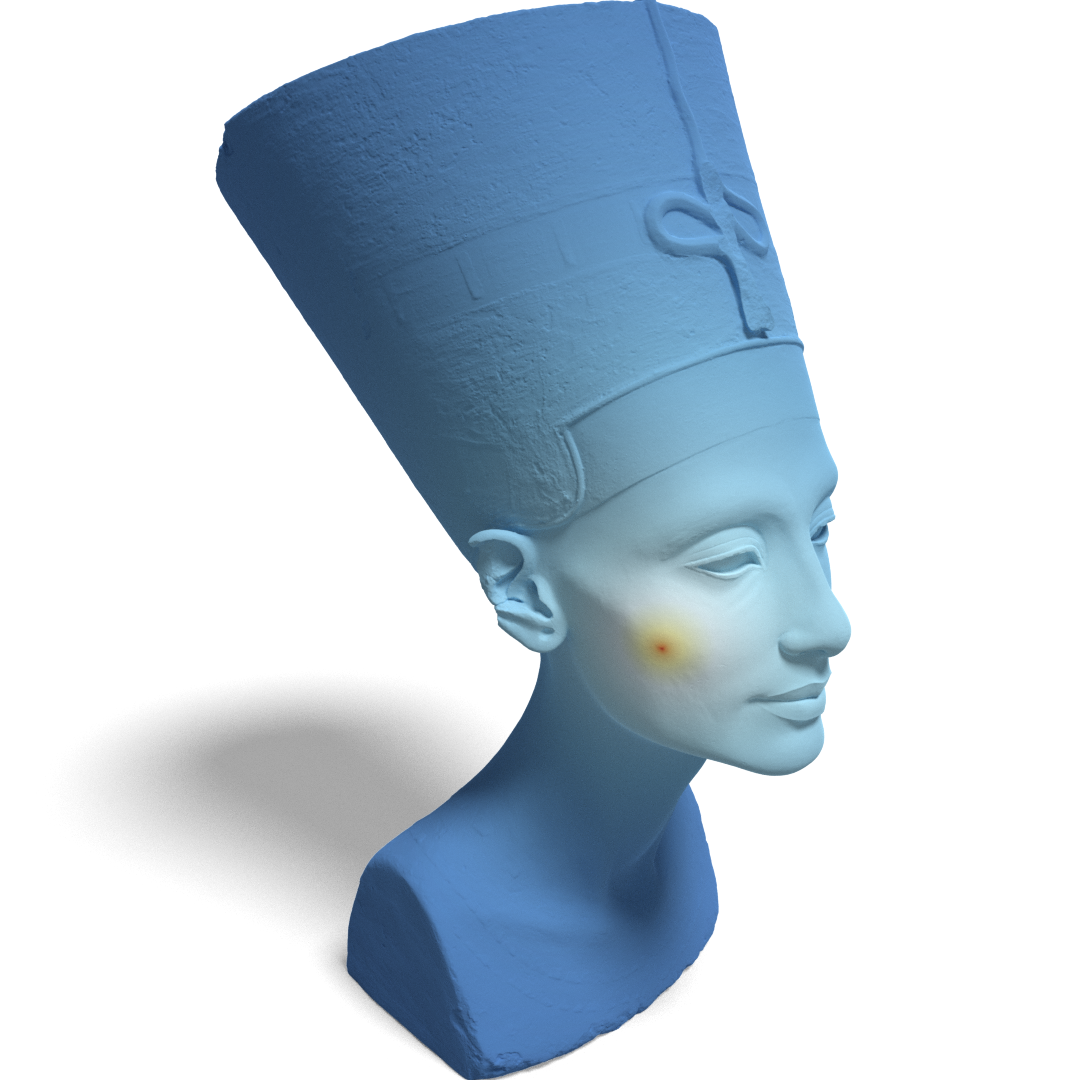} &
    \includegraphics[width=\reswidth]{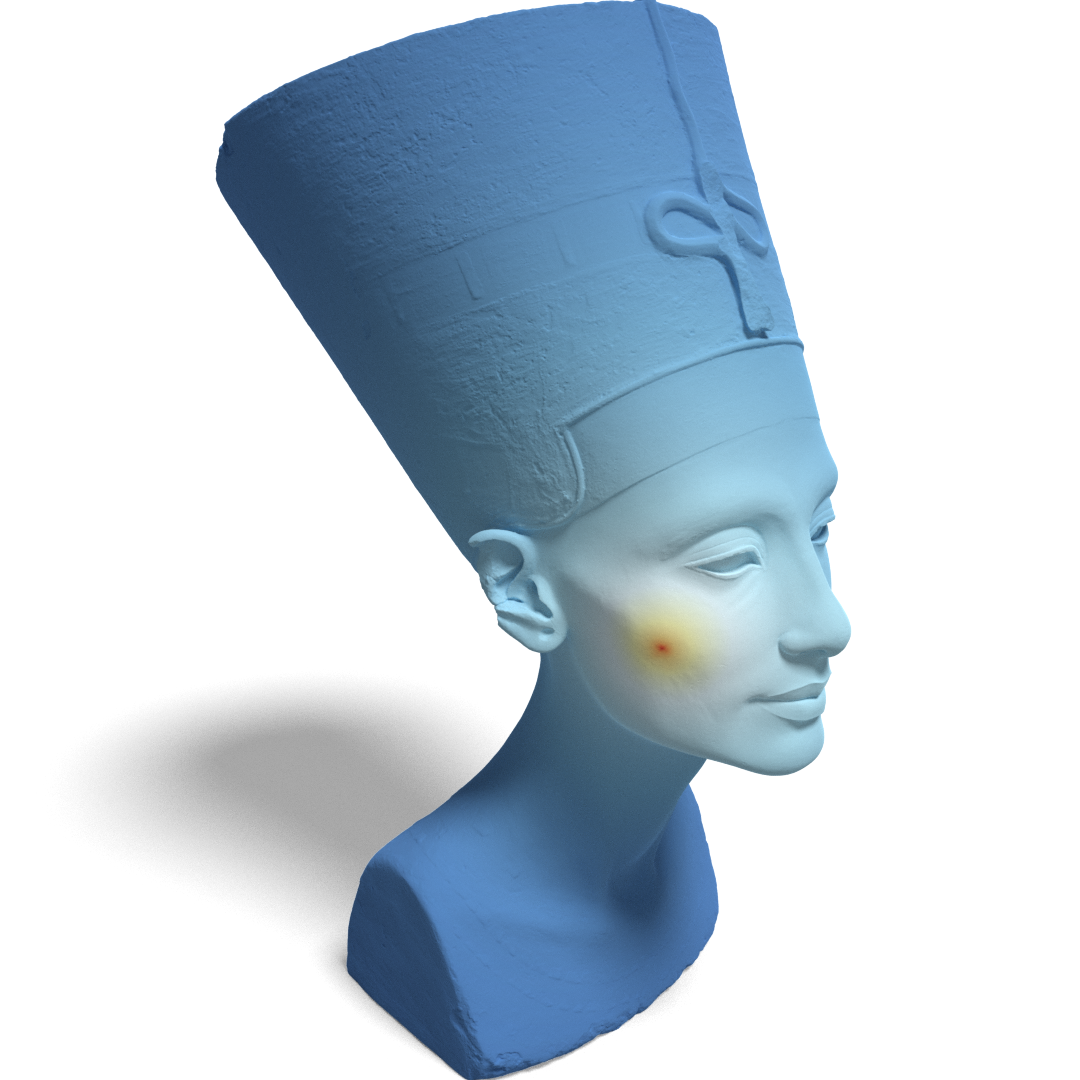} &
    \includegraphics[width=\reswidth]{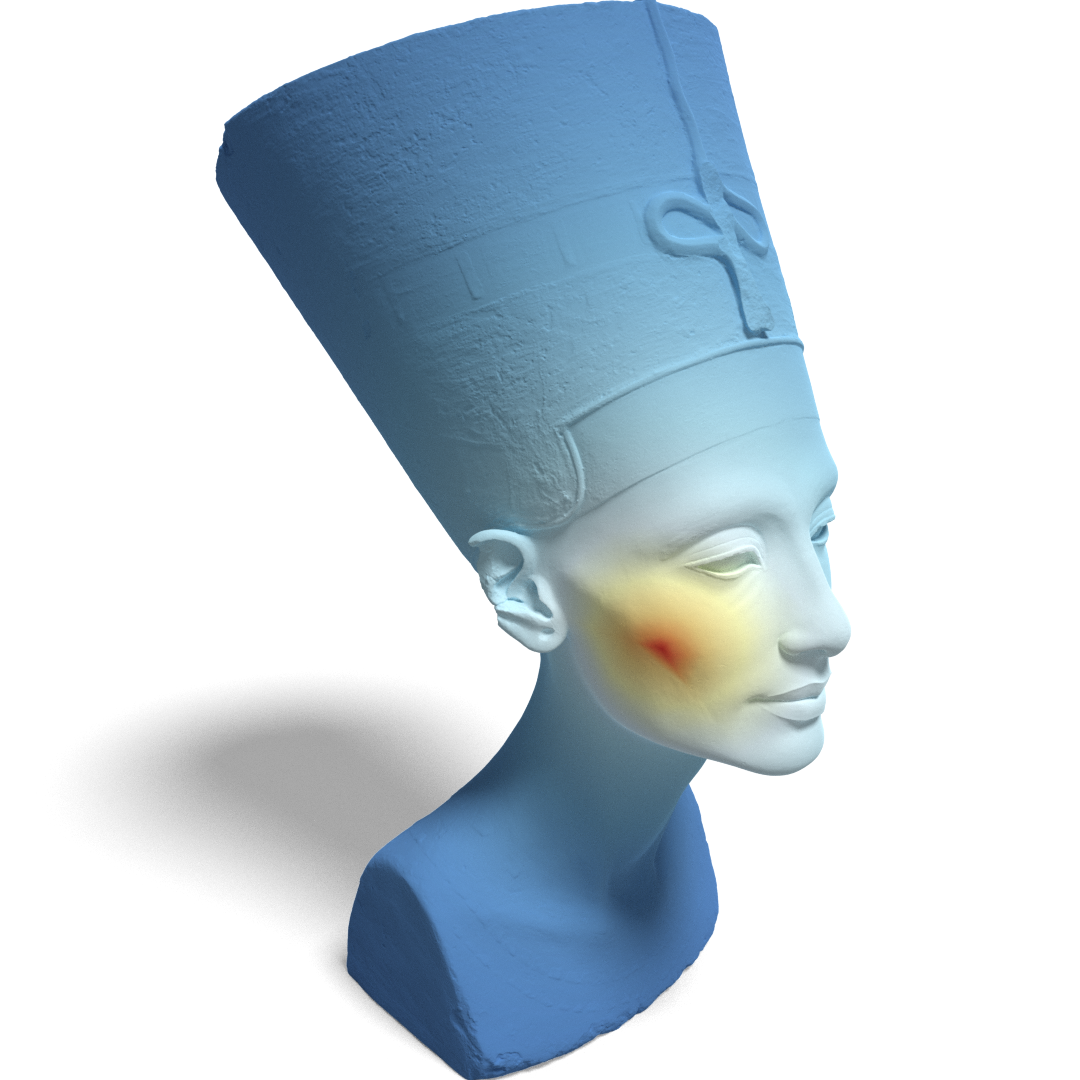}
    \vspace{-0.5ex} \\
    Original & $\g_{max} = 10^{-4}$ & $\g_{max} = 10^{-3}$ & $\g_{max} = 10^{-2}$ \\
    \num{1009118} vertices & \num{963902} vertices & \num{798457} vertices & \num{350793} vertices \\
    & $\text{MSE} = \num{1.77e-05} $ & $\text{MSE} = \num{1.81e-4}$ & $\text{MSE} = \num{4.85e-3}$
    \vspace{1ex} \\
    \includegraphics[width=\reswidth]{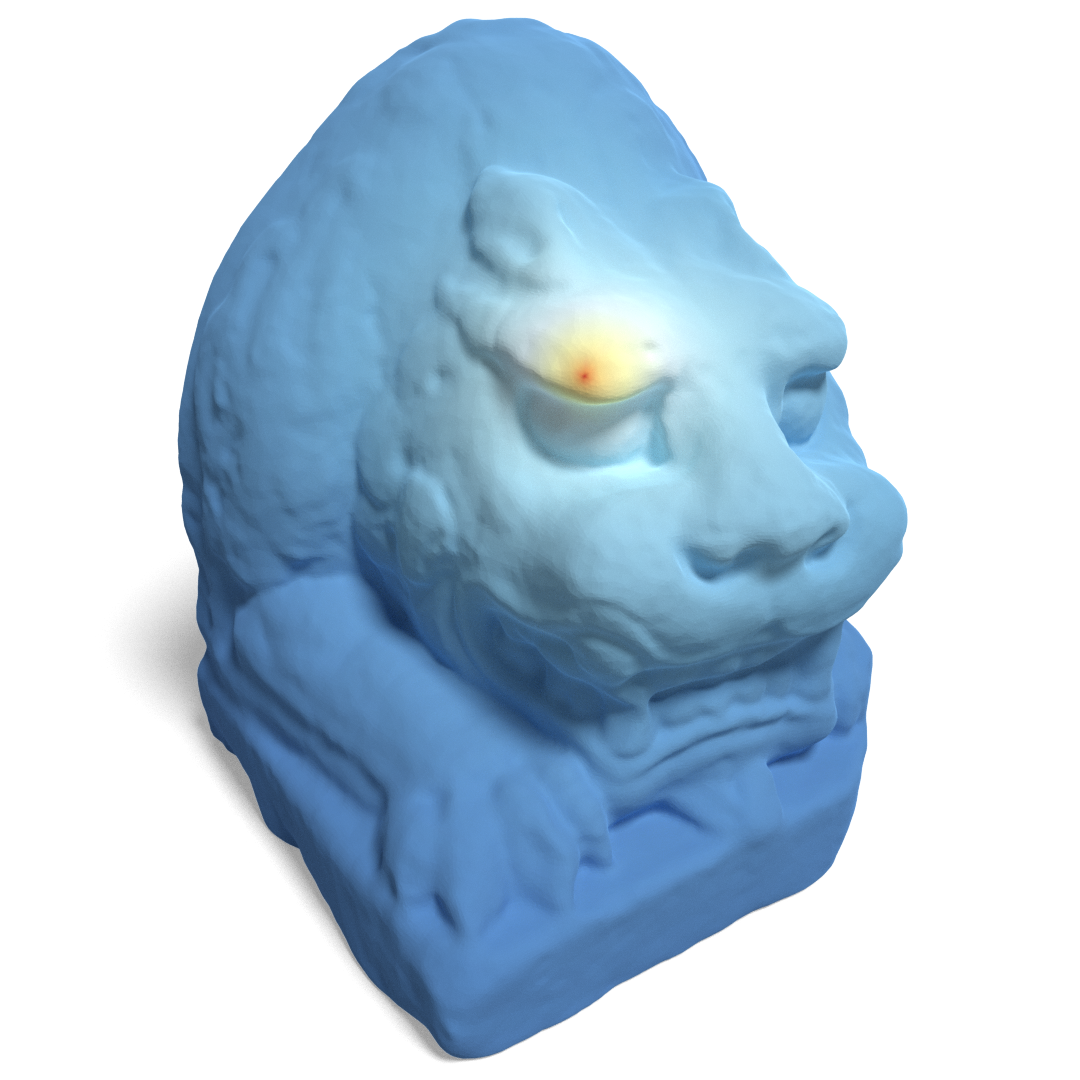} &
    \includegraphics[width=\reswidth]{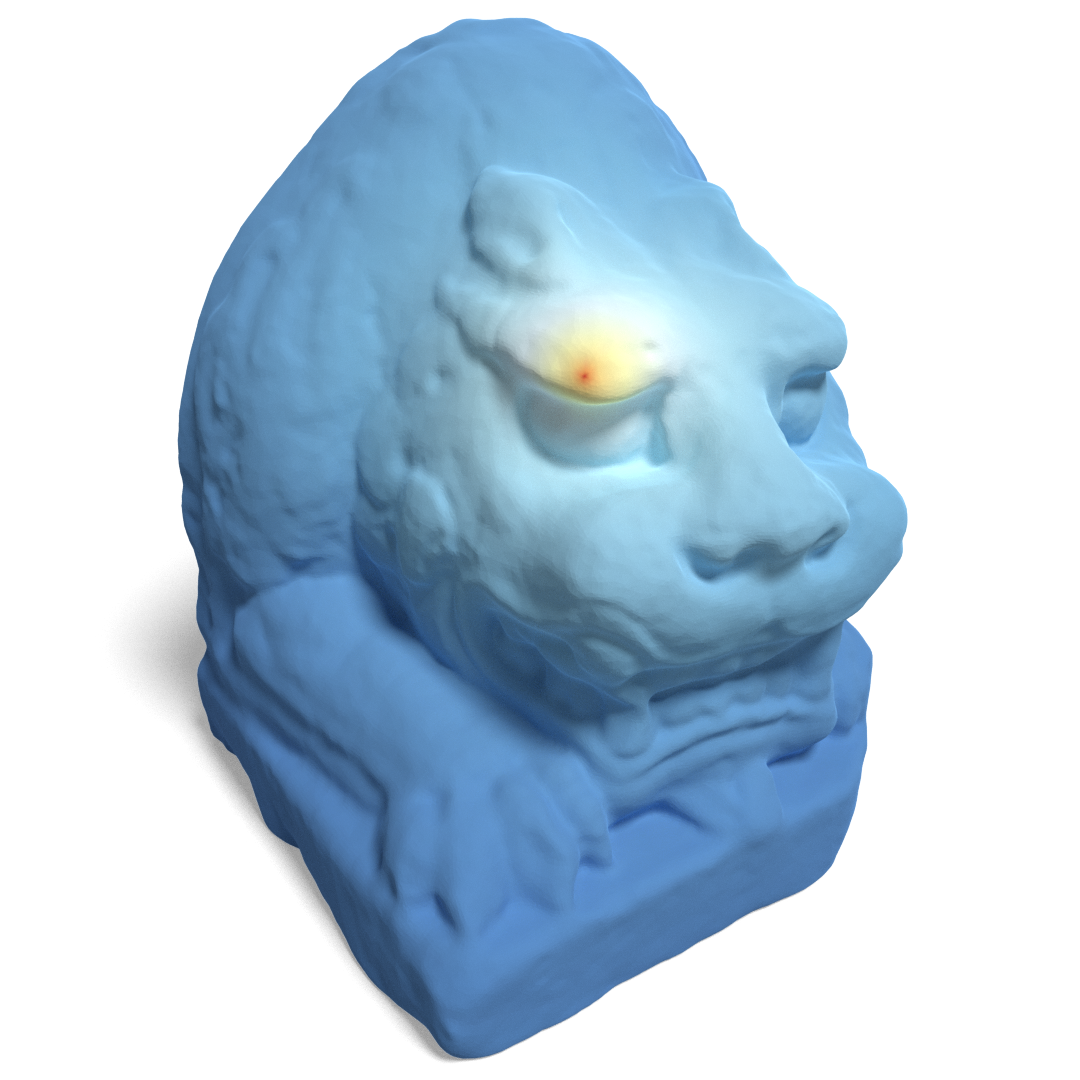} &
    \includegraphics[width=\reswidth]{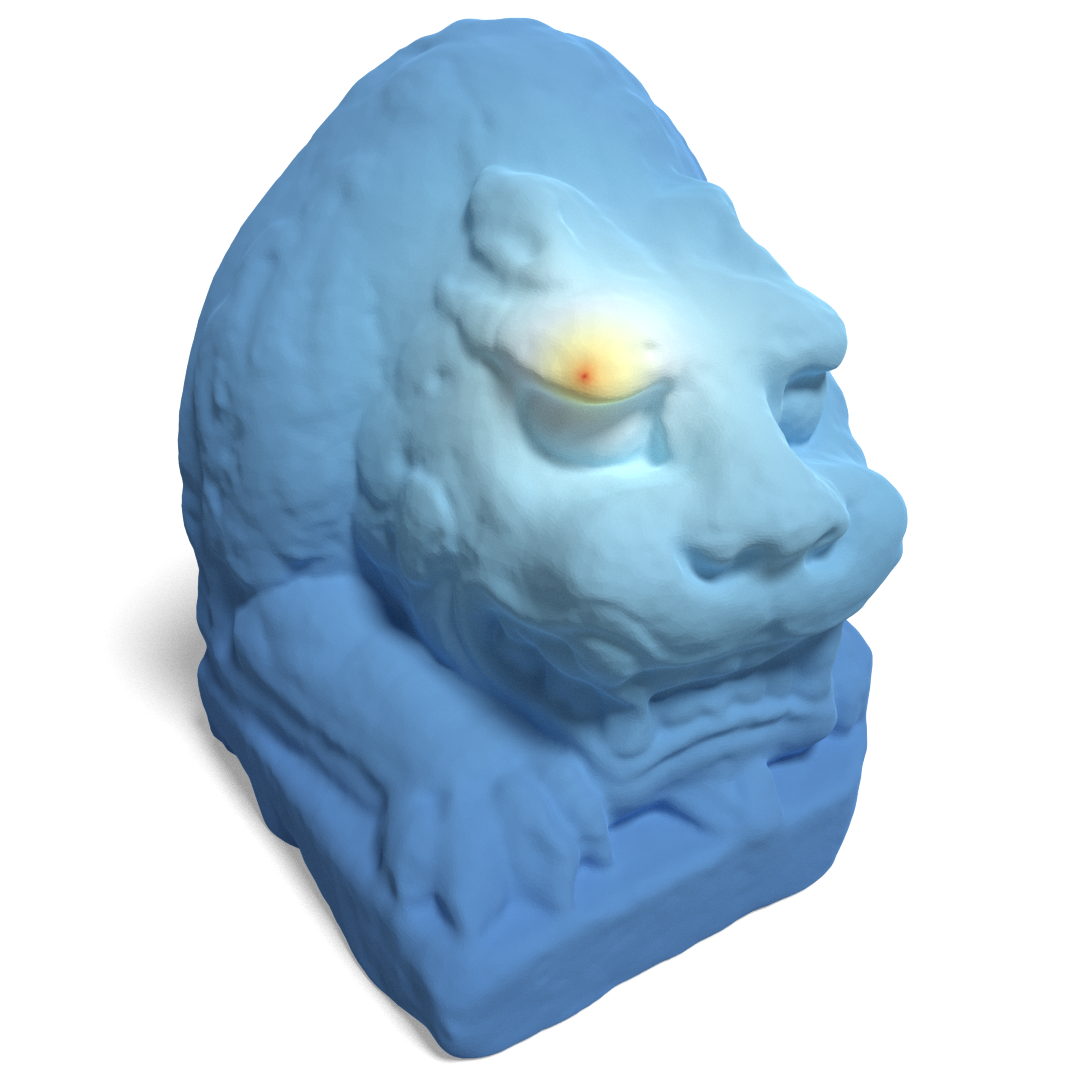} &
    \includegraphics[width=\reswidth]{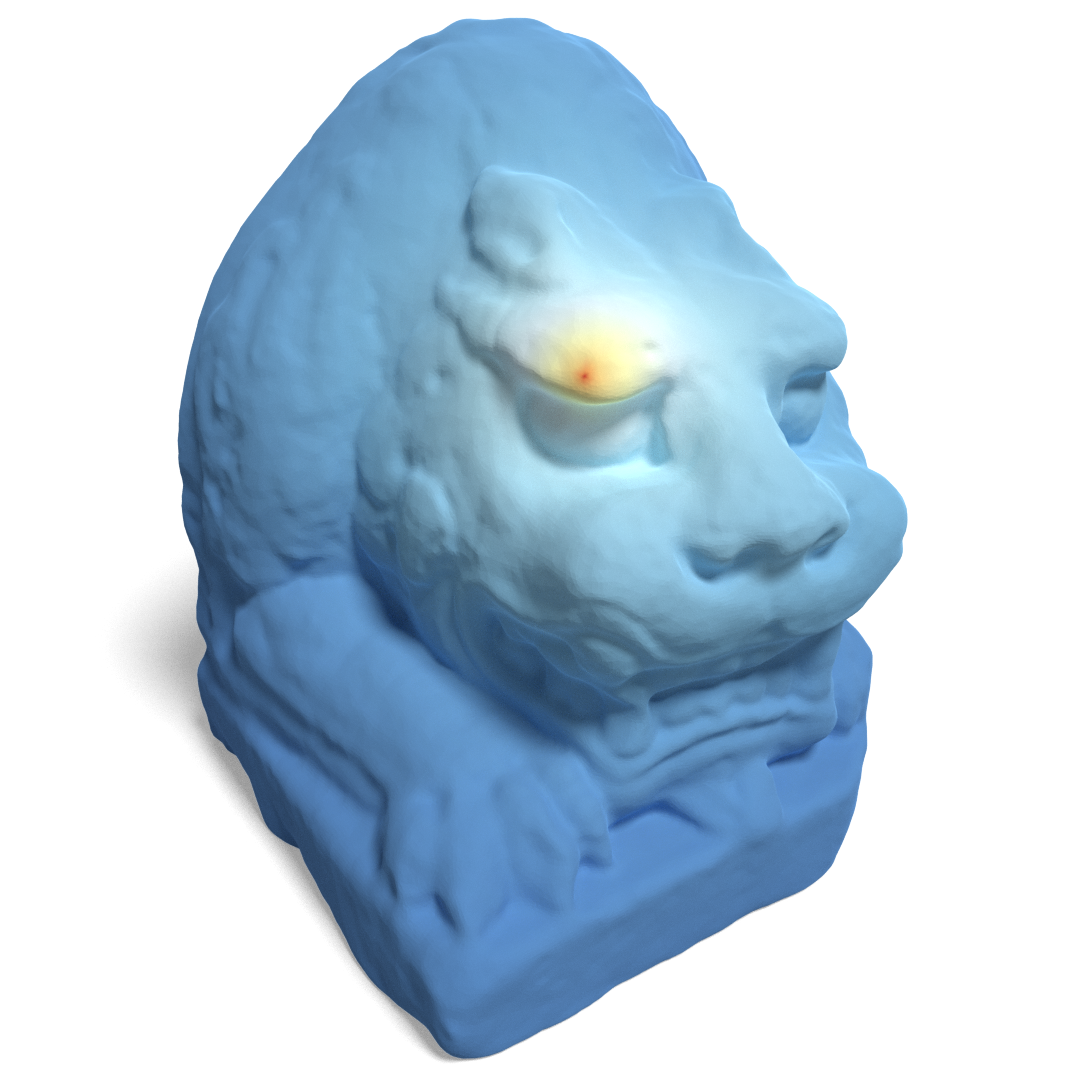}
    \vspace{-1ex} \\
    Original & $\g_{max} = 10^{-4}$ & $\g_{max} = 10^{-3}$ & $\g_{max} = 10^{-2}$ \\
    \num{34999} vertices & \num{34856} vertices & \num{33120} vertices & \num{16038} vertices \\
    & $\text{MSE} = \num{8.28e-8}$ & $\text{MSE} = \num{2.38e-7}$ & $\text{MSE} = \num{1.94e-6}$
    \end{tabular}
    }
    \caption{Visualizations of the solution of a Poisson equation with a
      spike placed on a near-developable vertex (top) or on a vertex with
      high curvature (bottom). The equation is solved directly on the ``raw''
      simplified mesh to better show the impact of simplification.}
    \label{fig:poisson}
\end{figure*}

To demonstrate the impact of removing non-zero Gaussian curvature vertices, we
visualize the solution to a Poisson equation with a spike
centered on a chosen vertex on $2$ selected meshes for $3$ $\g_{max}$
thresholds (\autoref{fig:poisson}); we report the number of vertices and MSE
error over the Poisson solution for each of the vertices in the original mesh, 
where the solution at a removed vertex is interpolated based on the intrinsic
barycentric coordinates. We
deliberately did \emph{not} apply any refinement and instead directly solve the
equation on the simplified mesh to better illustrate the impact of the
simplification.  In the first example, we deliberately placed the source term
on a vertex with low Gaussian curvature. As expected, as the threshold
$\g_{max}$ increases, so does the error.  In the second example, we placed the
source term on a vertex with high Gaussian curvature, which results in a
smaller increase in error because most intrinsic triangles around the source
term will not be simplified.  In a practice, for maximal accuracy, we would
first insert vertices via an intrinsic optimal Delaunay algorithm~\cite{Sharp:2019:NIT}
or adaptive intrinsic mesh refinement~\cite{Sharp:2019:NIT}.  While both methods
will increase the number of intrinsic vertices, neither would be encumbered by
sub-optimally placed intrinsic vertices from the original mesh. Indeed, in the
concurrent work of Liu~\etal~\shortcite{Liu:2023:SSI} they perform refinement
prior to simplification to achieve a better vertex distribution post-simplification.

\begin{figure*}
    \centering
    \def\reswidth{0.22\linewidth}
    {
    \begin{tabular}{cccc}
    \includegraphics[width=\reswidth]{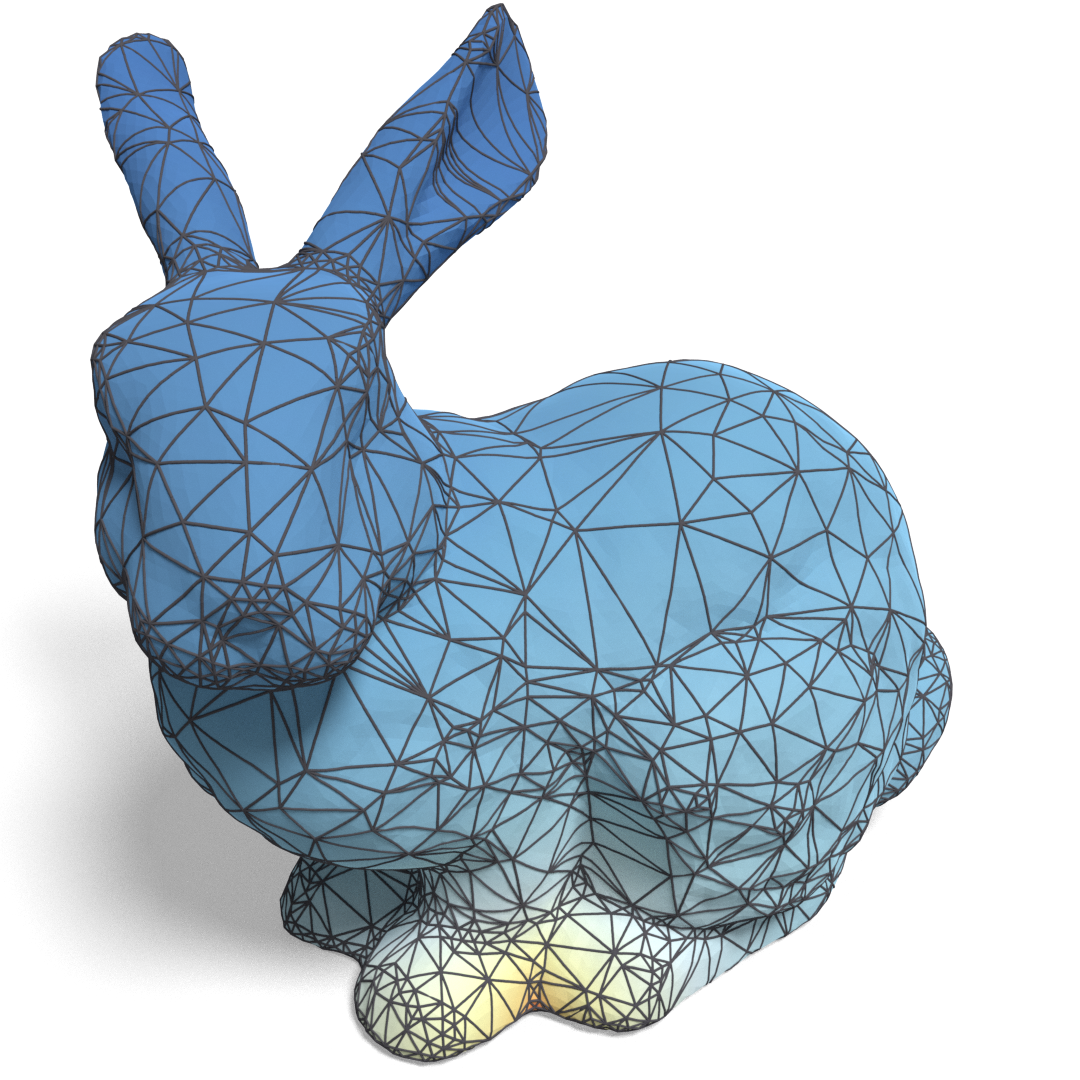} &
    \includegraphics[width=\reswidth]{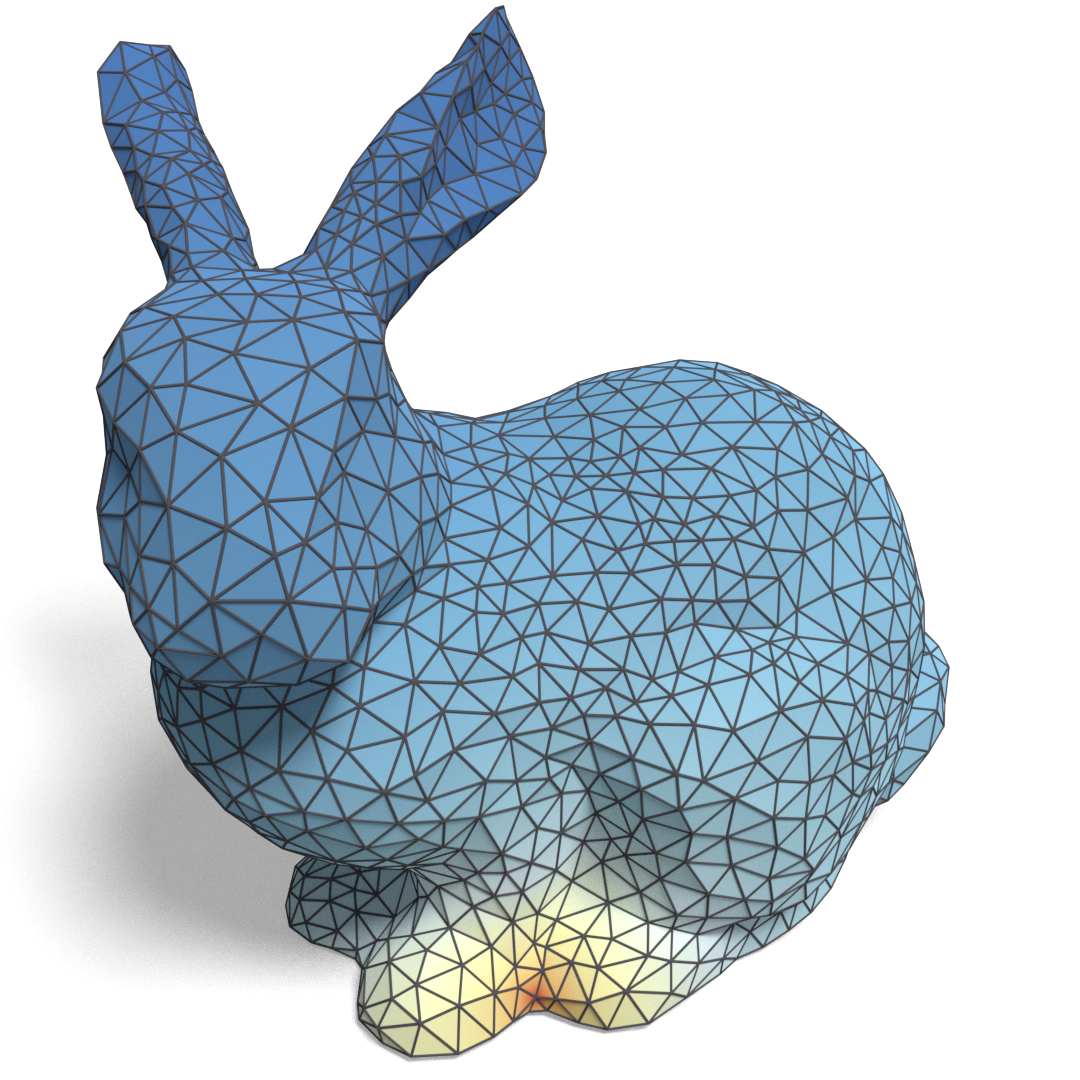}  &
    \includegraphics[width=\reswidth]{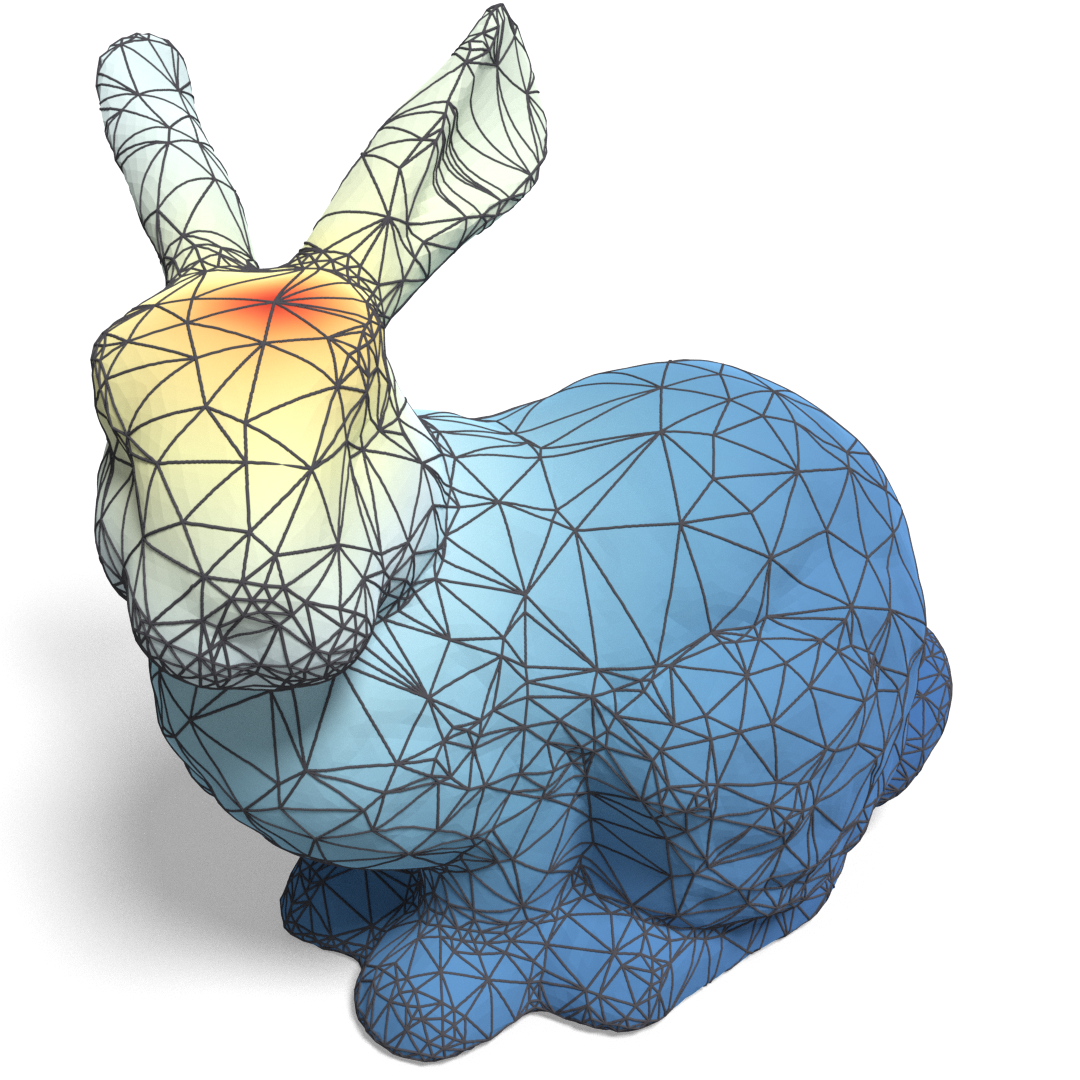} &
    \includegraphics[width=\reswidth]{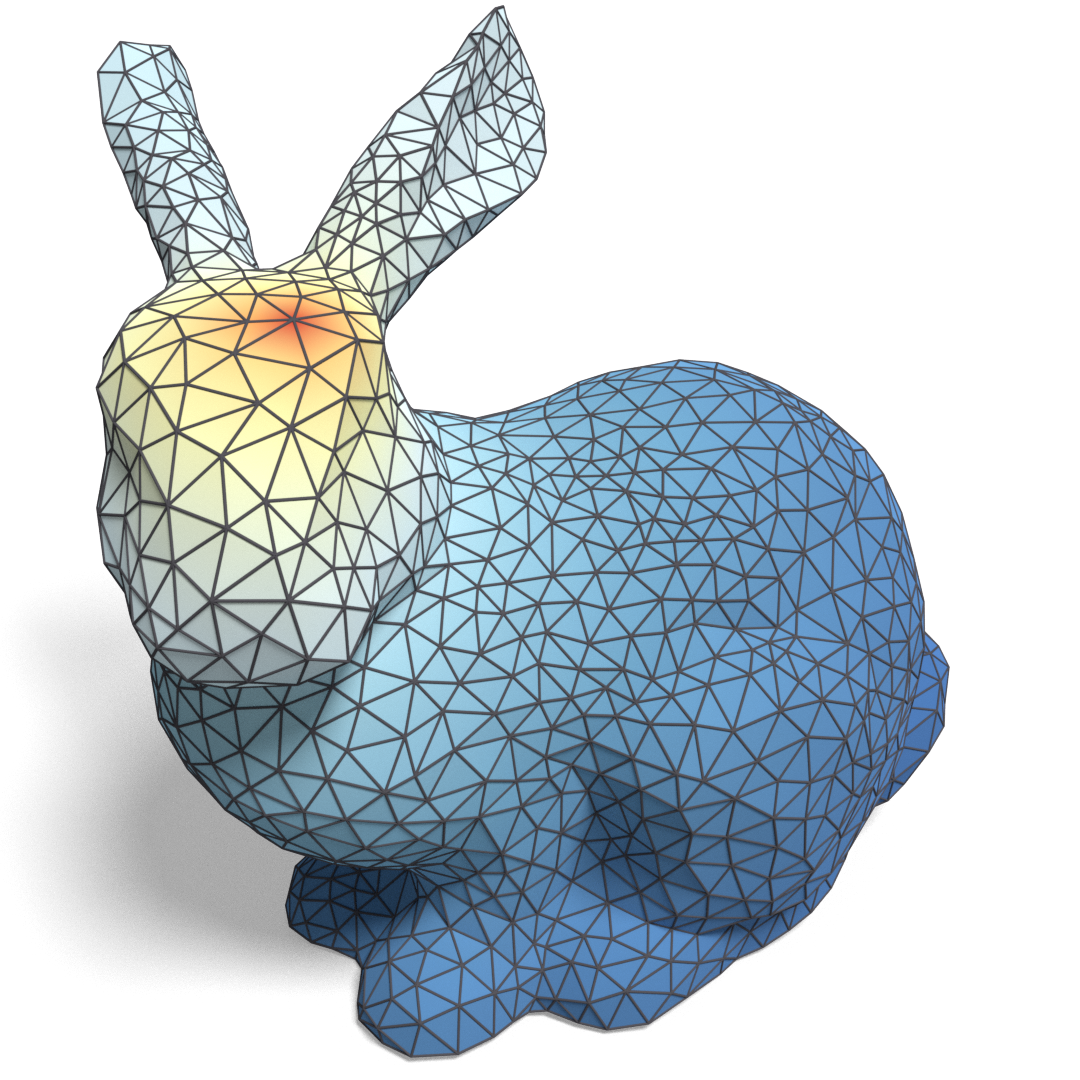}  \\
    Intrinsic Simplification (Ours) & Spectral Simplification &
    Intrinsic Simplification (Ours) & Spectral Simplification \\
    \includegraphics[width=\reswidth]{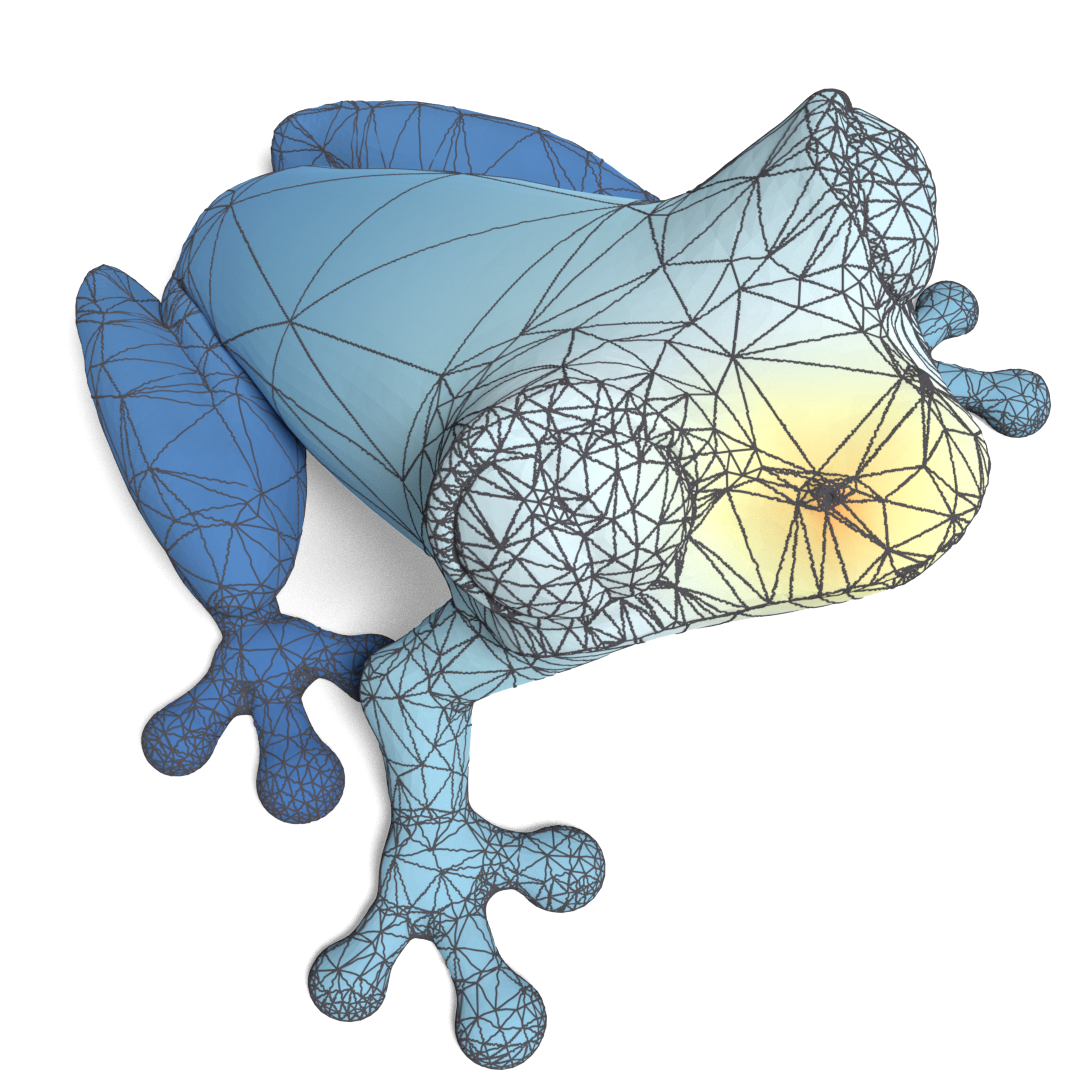} &
    \includegraphics[width=\reswidth]{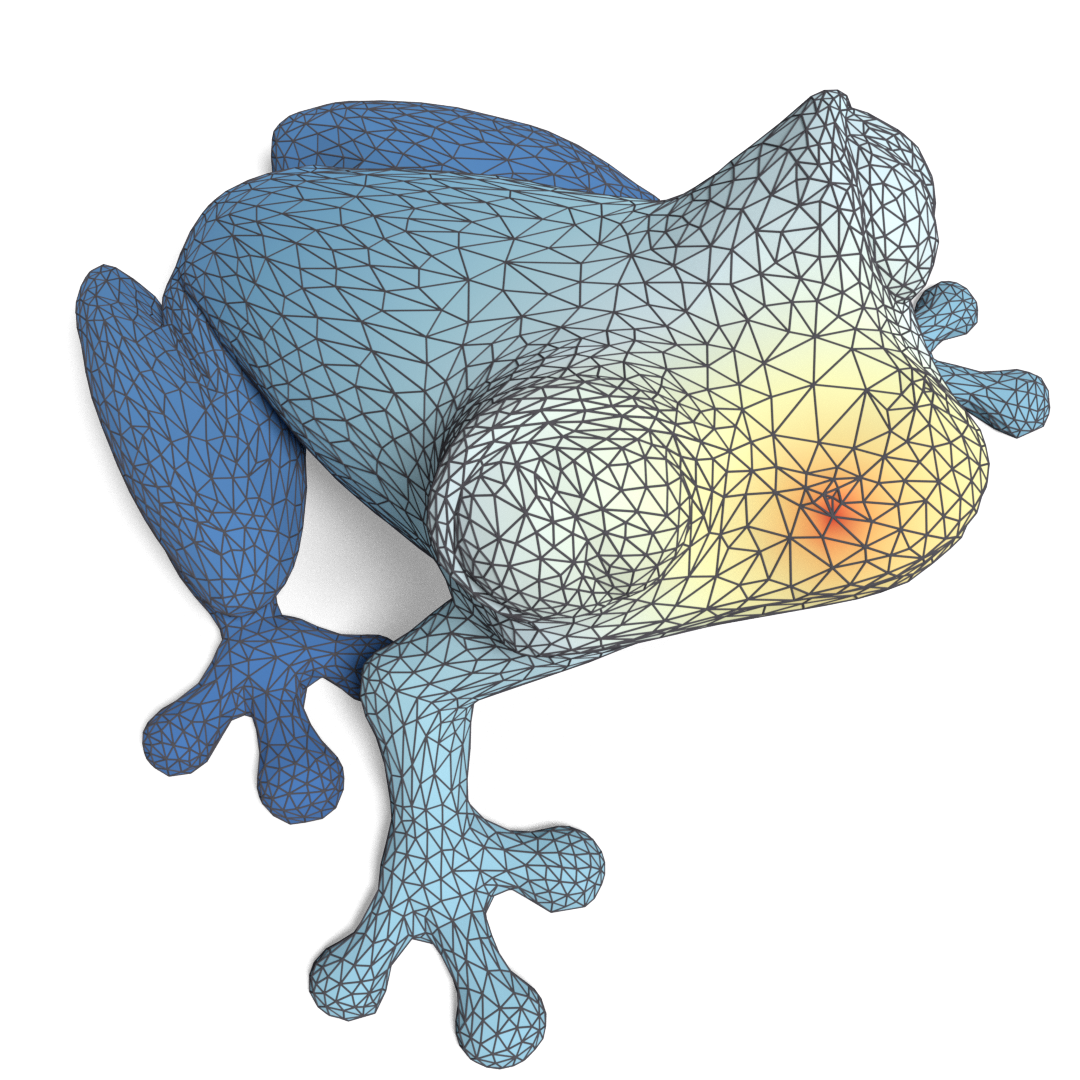} &
    \includegraphics[width=\reswidth]{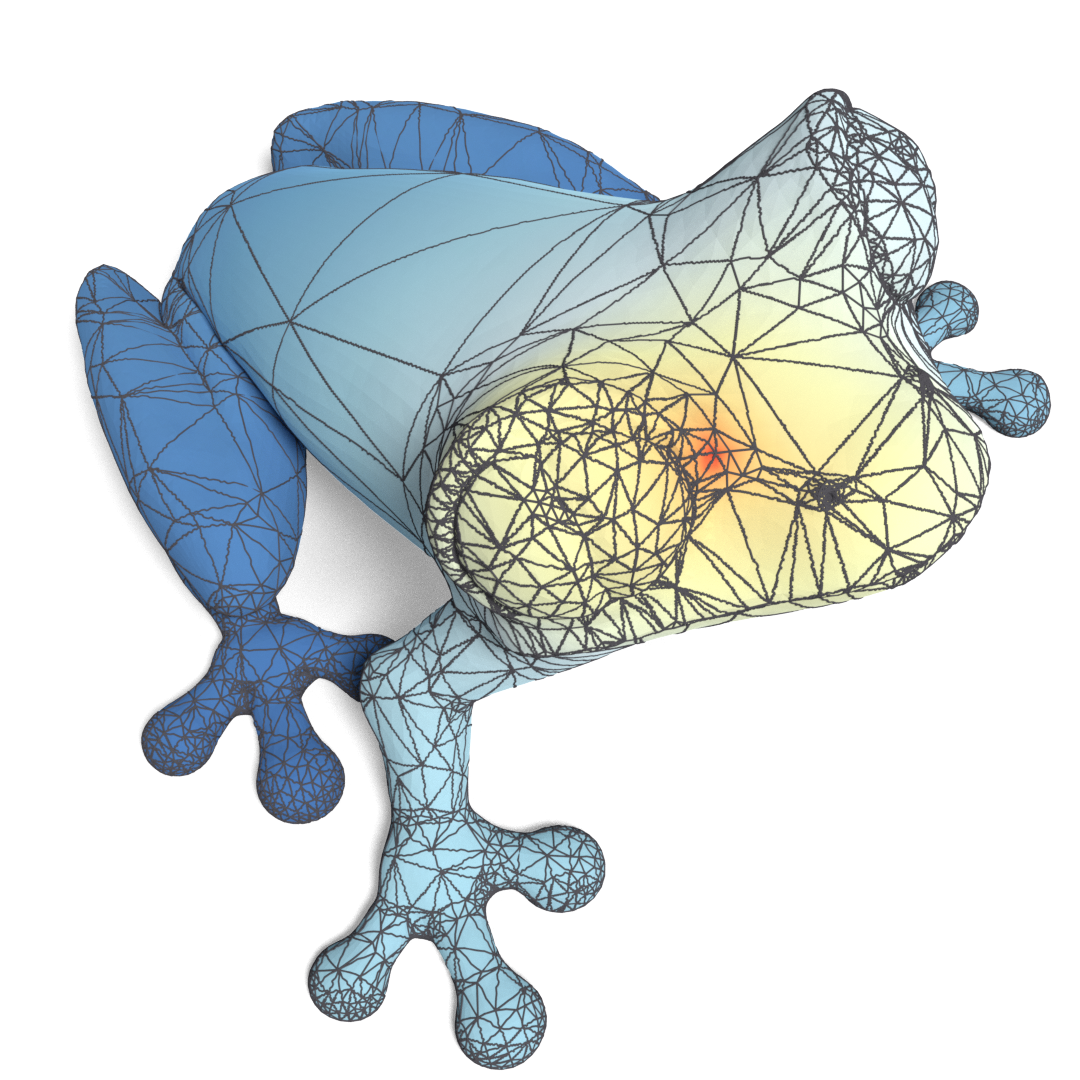} &
    \includegraphics[width=\reswidth]{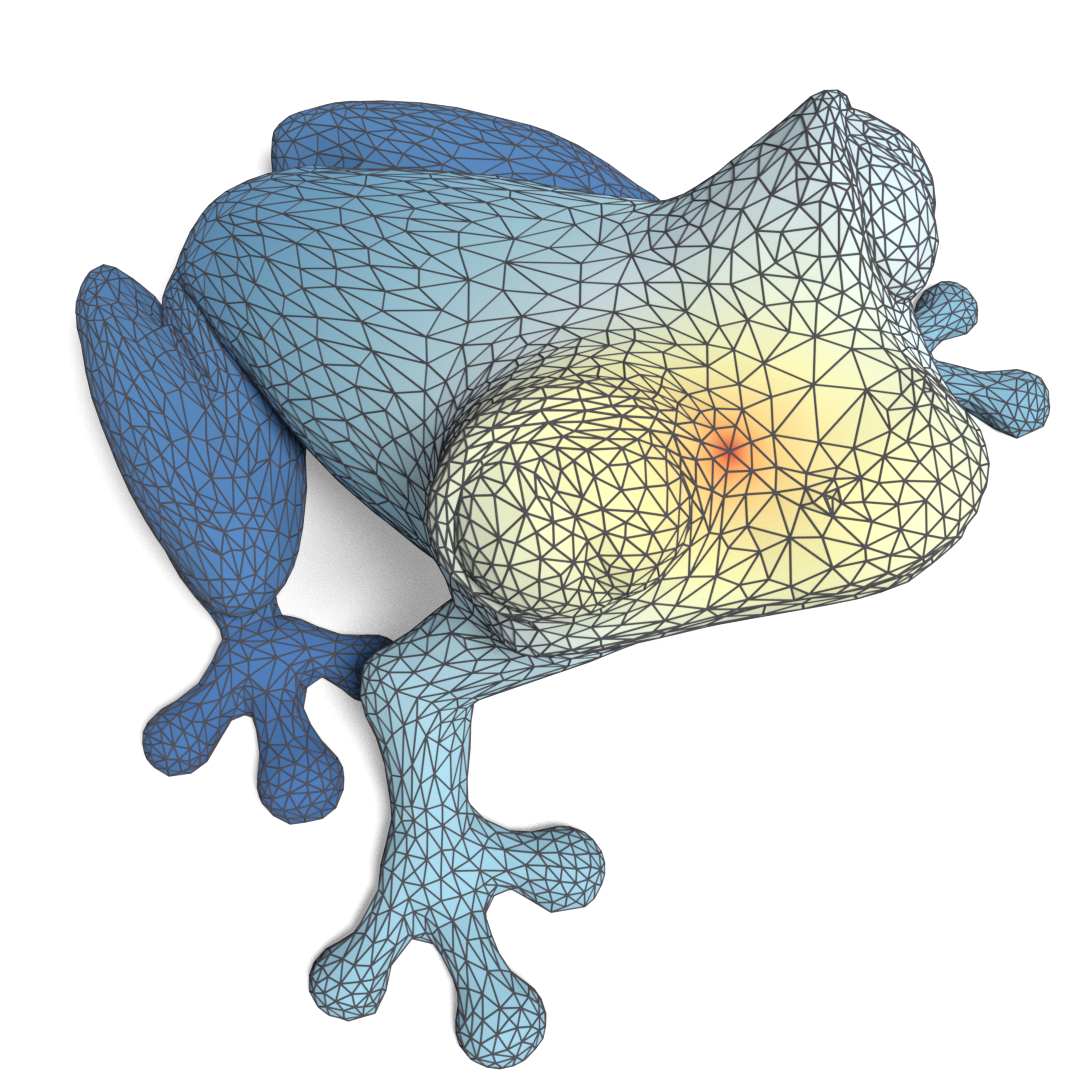} \\
    Intrinsic Simplification (Ours) & QEM Simplification &
    Intrinsic Simplification (Ours) & QEM Simplification
    \end{tabular}
    }
    \caption{Solutions of a Poisson equations computed on simplified meshes
      obtained with Spectral Mesh Simplification (SMS) and QEM. The \emph{Bunny}
      was reduced to \num{1715} vertices (\num{88.00}\% reduction) via our
      intrinsic simplification algorithm and QEM. The \emph{Frog} was reduced
      to \num{4440} vertices (\num{80.99}\% reduction) via our intrinsic
      simplification and SMS. All solutions are computed on the ``raw''
      simplified mesh to better show the impact of simplification.}
    \label{fig:spectral}
\end{figure*}

A key advantage of simplifying the intrinsic representation compared to first
simplifying an extrinsic mesh and then constructing an intrinsic
representation is that we can produce simplified triangles that ride the
underlying surface, and thus better maintain the surface metric.  To better
understand the differences between different simplification methods,
\autoref{fig:spectral} compares intrinsic mesh simplification with two
extrinsic simplifications methods: QEM~\cite{Garland:1997:SSQ} and Spectral
Mesh Simplification~\cite{Lescoat:2020:SMS}.  For each method, we perform an
equal-vertex-count comparison (simplified from $14,\!290$ to $1,\!715$
vertices for the \emph{Bunny}, and from $23,\!356$ to $4,\!440$ vertices for
the \emph{Frog}) to our intrinsically simplified mesh, and highlight the
differences between all methods by visualizing the solution to a Poisson
equation computed directly on the ``raw'' mesh with a source term placed at a
low and high Gaussian curvature vertex. PDE solutions are solved on the intrinsic
mesh and the barycentric coordinates are used to linearly interpolate the solution
at deleted vertices. For QEM we use the implementation provided by 
Lescoat~\etal~\shortcite{Lescoat:2020:SMS}. Note that Spectral Mesh
Simplification optimizes for the intrinsic qualities of the mesh when removing
vertices, and as such on average the solution to the Poisson equation is more
accurate, albeit at a much higher computation cost ($35$ minutes versus $0.62$
seconds for our method on the \emph{Bunny}; for reference QEM took $7.6$
second seconds versus $0.85$ seconds for our method on the \emph{Frog}).
However, as noted before, intrinsic mesh simplification is intended as a
preprocessing step, and in practice we would apply an adaptive refinement or
produce an optimal Delaunay triangulation before solving the Poisson equation.

\section{Limitations and Future Work}
\label{sec:limits}

Our method is not without limitation.  First, as mentioned in ~\autoref{sec:flip}, a potential limitation is imposed
by reducing the valence of a vertex prior to removal. The issue
arises since a vertex $i$ with Gaussian curvature $\g_i$ has a lower bound
on its valence: $\ceil{2-\frac{\g}{\pi}}$. The lower bound exists since the
corner angles $\theta_{1},...,\theta_{n}$ around $i$ must each satisfy
$\theta_{j} < \pi$ since each $\theta_{j}$ belongs to a triangle. Therefore
the total angle around $i$ must satisfy $\alpha_i = \sum_{j=1}^{n} < n\pi$;
and since $\alpha_i = 2\pi - \g_i$, the valence $n$ must satisfy 
$n > \ceil{2-\frac{\g_i}{\pi}}$. The lower bound means that our method is
unable to remove vertices with $\g \leq -\pi$. However, in practice, a typical
mesh only has a few, if any, vertices with such extreme negative curvature.
In fact, on average, models in our validation set (of roughly \num{7000} models)
fewer than $0.2\%$ of vertices violate  $\g > -\pi$. In practice the lower bound on
the valence of a vertex does not affect the practical effectiveness of our method 
for the majority of real-world meshes.

Second, maintaining the correspondence between the intrinsic and extrinsic
triangulations with intrinsic barycentric coordinates forces our
algorithm to update barycentric coordinates of vertices in faces
adjacent to a flipped edge and in one of the three faces incident
to a vertex that is about to be removed. These repeated updates can
cause numerical error to compound as the algorithm progresses. Furthermore,
for large models that are highly decimated, e.g. highly tessellated
cylinder, a large number of removed vertices must have their intrinsic
coordinates updated during each edge flip and vertex removal. Exploring alternative
methods of maintaining the correspondence between the two triangulations
could provide an interesting avenue for future research and open the door
for faster and more robust intrinsic mesh simplification.

Finally, our method of visualizing the simplified intrinsic triangulation
does not accurately represent the new intrinsic geometry. A method that provides
the benefits of a common subdivision can enable a better visualization
as well as provide a more robust way to port data between the two triangulations,
such as enabling a robust change of basis when solving PDEs. Producing a `pseudo'-common
subdivision is an interesting avenue for future research. Alternatively one could
produce a visualization by computing an embedding of the simplified intrinsic mesh,
which could aide in visualizing the deformation of the intrinsic geometry of the
original mesh.
\section{Conclusion}
\label{sec:conclusion}

In this paper we presented a method for intrinsic mesh simplification.
Our method leverages the benefits of intrinsic edge flipping and the
large space of intrinsic triangulations to significantly simplify vertex
deletion to two canonical cases (i.e., valence three for an internal vertex,
and valence two for a boundary vertex).  We use Gaussian curvature as the 
deletion criterion, which, together with our removal strategy, effectively
projects vertices onto a locally developable approximation.  We demonstrated 
the robustness and effectiveness of our method on the Thingi10k dataset and
demonstrated the impact of relaxing the deletion threshold on the solutions
of PDEs.

\section{Acknowledgements}

This work was supported in part by a Virginia Space Grant Grant Consortium
Graduate STEM Research Fellowship Grant and NSF grant IIS-1909028.

\bibliographystyle{ACM-Reference-Format}
\bibliography{references}


\begin{thebibliography}{30}


\ifx \showCODEN    \undefined \def \showCODEN     #1{\unskip}     \fi
\ifx \showDOI      \undefined \def \showDOI       #1{#1}\fi
\ifx \showISBNx    \undefined \def \showISBNx     #1{\unskip}     \fi
\ifx \showISBNxiii \undefined \def \showISBNxiii  #1{\unskip}     \fi
\ifx \showISSN     \undefined \def \showISSN      #1{\unskip}     \fi
\ifx \showLCCN     \undefined \def \showLCCN      #1{\unskip}     \fi
\ifx \shownote     \undefined \def \shownote      #1{#1}          \fi
\ifx \showarticletitle \undefined \def \showarticletitle #1{#1}   \fi
\ifx \showURL      \undefined \def \showURL       {\relax}        \fi
\providecommand\bibfield[2]{#2}
\providecommand\bibinfo[2]{#2}
\providecommand\natexlab[1]{#1}
\providecommand\showeprint[2][]{arXiv:#2}

\bibitem[Bommes et~al\mbox{.}(2013)]%
        {Bommes:2013:IGMQ}
\bibfield{author}{\bibinfo{person}{David Bommes}, \bibinfo{person}{Marcel
  Campen}, \bibinfo{person}{Hans-Christian Ebke}, \bibinfo{person}{Pierre
  Alliez}, {and} \bibinfo{person}{Leif Kobbelt}.}
  \bibinfo{year}{2013}\natexlab{}.
\newblock \showarticletitle{Integer-Grid Maps for Reliable Quad Meshing}.
\newblock \bibinfo{journal}{\emph{ACM Trans. Graph.}} \bibinfo{volume}{32},
  \bibinfo{number}{4}, Article \bibinfo{articleno}{98} (\bibinfo{date}{jul}
  \bibinfo{year}{2013}), \bibinfo{numpages}{12}~pages.
\newblock
\showISSN{0730-0301}
\urldef\tempurl%
\url{https://doi.org/10.1145/2461912.2462014}
\showDOI{\tempurl}


\bibitem[Botsch et~al\mbox{.}(2010)]%
        {Botch:2021:PMP}
\bibfield{author}{\bibinfo{person}{M. Botsch}, \bibinfo{person}{L. Kobbelt},
  \bibinfo{person}{M. Pauly}, \bibinfo{person}{P. Alliez}, {and}
  \bibinfo{person}{B. Levy}.} \bibinfo{year}{2010}\natexlab{}.
\newblock \bibinfo{booktitle}{\emph{Polygon Mesh Processing}}.
\newblock
\showISBNx{9781439865316}


\bibitem[Chen and Xu(2004)]%
        {Chen:2004:ODT}
\bibfield{author}{\bibinfo{person}{Long Chen} {and} \bibinfo{person}{Jinchao
  Xu}.} \bibinfo{year}{2004}\natexlab{}.
\newblock \showarticletitle{Optimal Delaunay triangulations}.
\newblock \bibinfo{journal}{\emph{Journal of Computational Mathematics}}
  \bibinfo{volume}{22}, \bibinfo{number}{2} (\bibinfo{date}{1 March}
  \bibinfo{year}{2004}), \bibinfo{pages}{299--308}.
\newblock


\bibitem[Cohen-Steiner et~al\mbox{.}(2004)]%
        {CohenSteiner:2004:VSA}
\bibfield{author}{\bibinfo{person}{David Cohen-Steiner},
  \bibinfo{person}{Pierre Alliez}, {and} \bibinfo{person}{Mathieu Desbrun}.}
  \bibinfo{year}{2004}\natexlab{}.
\newblock \showarticletitle{Variational Shape Approximation}.
\newblock \bibinfo{journal}{\emph{ACM Trans. Graph.}} \bibinfo{volume}{23},
  \bibinfo{number}{3} (\bibinfo{date}{aug} \bibinfo{year}{2004}),
  \bibinfo{pages}{905–914}.
\newblock


\bibitem[Ebke et~al\mbox{.}(2016)]%
        {Ebke:2016:IQR}
\bibfield{author}{\bibinfo{person}{Hans-Christian Ebke},
  \bibinfo{person}{Patrick Schmidt}, \bibinfo{person}{Marcel Campen}, {and}
  \bibinfo{person}{Leif Kobbelt}.} \bibinfo{year}{2016}\natexlab{}.
\newblock \showarticletitle{Interactively Controlled Quad Remeshing of High
  Resolution 3D Models}.
\newblock \bibinfo{journal}{\emph{ACM Trans. Graph.}} \bibinfo{volume}{35},
  \bibinfo{number}{6}, Article \bibinfo{articleno}{218} (\bibinfo{date}{dec}
  \bibinfo{year}{2016}), \bibinfo{numpages}{13}~pages.
\newblock
\showISSN{0730-0301}
\urldef\tempurl%
\url{https://doi.org/10.1145/2980179.2982413}
\showDOI{\tempurl}


\bibitem[Fisher et~al\mbox{.}(2007)]%
        {Fisher:2007:AAC}
\bibfield{author}{\bibinfo{person}{Matthew Fisher}, \bibinfo{person}{Boris
  Springborn}, \bibinfo{person}{Peter Schr{\"o}der}, {and}
  \bibinfo{person}{Alexander~I Bobenko}.} \bibinfo{year}{2007}\natexlab{}.
\newblock \showarticletitle{An algorithm for the construction of intrinsic
  Delaunay triangulations with applications to digital geometry processing}.
\newblock \bibinfo{journal}{\emph{Computing}} \bibinfo{volume}{81},
  \bibinfo{number}{2} (\bibinfo{year}{2007}), \bibinfo{pages}{199--213}.
\newblock


\bibitem[Floater and Hormann(2005)]%
        {Floater:2005:SPT}
\bibfield{author}{\bibinfo{person}{Michael~S Floater} {and}
  \bibinfo{person}{Kai Hormann}.} \bibinfo{year}{2005}\natexlab{}.
\newblock \showarticletitle{Surface parameterization: a tutorial and survey}.
\newblock \bibinfo{journal}{\emph{Advances in multiresolution for geometric
  modelling}} (\bibinfo{year}{2005}), \bibinfo{pages}{157--186}.
\newblock


\bibitem[Garland and Heckbert(1997)]%
        {Garland:1997:SSQ}
\bibfield{author}{\bibinfo{person}{Michael Garland} {and}
  \bibinfo{person}{Paul~S. Heckbert}.} \bibinfo{year}{1997}\natexlab{}.
\newblock \showarticletitle{Surface Simplification Using Quadric Error
  Metrics}. In \bibinfo{booktitle}{\emph{Proceedings of the 24th Annual
  Conference on Computer Graphics and Interactive Techniques}}
  \emph{(\bibinfo{series}{SIGGRAPH '97})}. \bibinfo{pages}{209–216}.
\newblock


\bibitem[Gillespie et~al\mbox{.}(2021a)]%
        {Gillespie:2021:ICI}
\bibfield{author}{\bibinfo{person}{Mark Gillespie}, \bibinfo{person}{Nicholas
  Sharp}, {and} \bibinfo{person}{Keenan Crane}.}
  \bibinfo{year}{2021}\natexlab{a}.
\newblock \showarticletitle{Integer coordinates for intrinsic geometry
  processing}.
\newblock \bibinfo{journal}{\emph{ACM Trans. Graph.}} \bibinfo{volume}{40},
  \bibinfo{number}{6} (\bibinfo{year}{2021}).
\newblock


\bibitem[Gillespie et~al\mbox{.}(2021b)]%
        {Gillespie:2021:DCE}
\bibfield{author}{\bibinfo{person}{Mark Gillespie}, \bibinfo{person}{Boris
  Springborn}, {and} \bibinfo{person}{Keenan Crane}.}
  \bibinfo{year}{2021}\natexlab{b}.
\newblock \showarticletitle{Discrete Conformal Equivalence of Polyhedral
  Surfaces}.
\newblock \bibinfo{journal}{\emph{ACM Trans. Graph.}} \bibinfo{volume}{40},
  \bibinfo{number}{4} (\bibinfo{year}{2021}).
\newblock


\bibitem[Hormann et~al\mbox{.}(2007)]%
        {Hormann:2007:MPTP}
\bibfield{author}{\bibinfo{person}{Kai Hormann}, \bibinfo{person}{Bruno
  L\'{e}vy}, {and} \bibinfo{person}{Alla Sheffer}.}
  \bibinfo{year}{2007}\natexlab{}.
\newblock \showarticletitle{Mesh Parameterization: Theory and Practice}. In
  \bibinfo{booktitle}{\emph{ACM SIGGRAPH 2007 Courses}} (San Diego, California)
  \emph{(\bibinfo{series}{SIGGRAPH '07})}. \bibinfo{publisher}{Association for
  Computing Machinery}, \bibinfo{address}{New York, NY, USA},
  \bibinfo{pages}{1–es}.
\newblock
\showISBNx{9781450318235}
\urldef\tempurl%
\url{https://doi.org/10.1145/1281500.1281510}
\showDOI{\tempurl}


\bibitem[Hu et~al\mbox{.}(2018)]%
        {Hu:2018:TMW}
\bibfield{author}{\bibinfo{person}{Yixin Hu}, \bibinfo{person}{Qingnan Zhou},
  \bibinfo{person}{Xifeng Gao}, \bibinfo{person}{Alec Jacobson},
  \bibinfo{person}{Denis Zorin}, {and} \bibinfo{person}{Daniele Panozzo}.}
  \bibinfo{year}{2018}\natexlab{}.
\newblock \showarticletitle{Tetrahedral Meshing in the Wild}.
\newblock \bibinfo{journal}{\emph{ACM Trans. Graph.}} \bibinfo{volume}{37},
  \bibinfo{number}{4}, Article \bibinfo{articleno}{60} (\bibinfo{date}{jul}
  \bibinfo{year}{2018}).
\newblock


\bibitem[Khan et~al\mbox{.}(2020)]%
        {Khan:2020:SRS}
\bibfield{author}{\bibinfo{person}{Dawar Khan}, \bibinfo{person}{Alexander
  Plopski}, \bibinfo{person}{Yuichiro Fujimoto}, \bibinfo{person}{Masayuki
  Kanbara}, \bibinfo{person}{Gul Jabeen}, \bibinfo{person}{Yongjie~Jessica
  Zhang}, \bibinfo{person}{Xiaopeng Zhang}, {and} \bibinfo{person}{Hirokazu
  Kato}.} \bibinfo{year}{2020}\natexlab{}.
\newblock \showarticletitle{Surface remeshing: A systematic literature review
  of methods and research directions}.
\newblock \bibinfo{journal}{\emph{IEEE TVCG}} \bibinfo{volume}{28},
  \bibinfo{number}{3} (\bibinfo{year}{2020}), \bibinfo{pages}{1680--1713}.
\newblock


\bibitem[Khodakovsky et~al\mbox{.}(2003)]%
        {Khodakovsky:2003:GSP}
\bibfield{author}{\bibinfo{person}{Andrei Khodakovsky}, \bibinfo{person}{Nathan
  Litke}, {and} \bibinfo{person}{Peter Schr\"{o}der}.}
  \bibinfo{year}{2003}\natexlab{}.
\newblock \showarticletitle{Globally Smooth Parameterizations with Low
  Distortion}.
\newblock \bibinfo{journal}{\emph{ACM Trans. Graph.}} \bibinfo{volume}{22},
  \bibinfo{number}{3} (\bibinfo{date}{jul} \bibinfo{year}{2003}),
  \bibinfo{pages}{350–357}.
\newblock
\showISSN{0730-0301}
\urldef\tempurl%
\url{https://doi.org/10.1145/882262.882275}
\showDOI{\tempurl}


\bibitem[Lee et~al\mbox{.}(1998)]%
        {Lee:1998:MAPS}
\bibfield{author}{\bibinfo{person}{Aaron~WF Lee}, \bibinfo{person}{Wim
  Sweldens}, \bibinfo{person}{Peter Schr{\"o}der}, \bibinfo{person}{Lawrence
  Cowsar}, {and} \bibinfo{person}{David Dobkin}.}
  \bibinfo{year}{1998}\natexlab{}.
\newblock \showarticletitle{MAPS: Multiresolution adaptive parameterization of
  surfaces}. In \bibinfo{booktitle}{\emph{Proceedings of the 25th annual
  conference on Computer graphics and interactive techniques}}.
  \bibinfo{pages}{95--104}.
\newblock


\bibitem[Lescoat et~al\mbox{.}(2020)]%
        {Lescoat:2020:SMS}
\bibfield{author}{\bibinfo{person}{Thibault Lescoat},
  \bibinfo{person}{Hsueh-Ti~Derek Liu}, \bibinfo{person}{Jean-Marc Thiery},
  \bibinfo{person}{Alec Jacobson}, \bibinfo{person}{Tamy Boubekeur}, {and}
  \bibinfo{person}{Maks Ovsjanikov}.} \bibinfo{year}{2020}\natexlab{}.
\newblock \showarticletitle{Spectral mesh simplification}. In
  \bibinfo{booktitle}{\emph{Comp. Graph. Forum}}, Vol.~\bibinfo{volume}{39}.
  \bibinfo{pages}{315--324}.
\newblock


\bibitem[Liu et~al\mbox{.}(2023)]%
        {Liu:2023:SSI}
\bibfield{author}{\bibinfo{person}{Derek Liu}, \bibinfo{person}{Mark
  Gillespie}, \bibinfo{person}{Benjamin Chislett}, \bibinfo{person}{Nicholas
  Sharp}, \bibinfo{person}{Alec Jacobson}, {and} \bibinfo{person}{Keenan
  Crane}.} \bibinfo{year}{2023}\natexlab{}.
\newblock \showarticletitle{Surface Simplification using Intrinsic Error
  Metrics}.
\newblock \bibinfo{journal}{\emph{ACM Trans. Graph.}} \bibinfo{volume}{XX},
  \bibinfo{number}{X} (\bibinfo{year}{2023}).
\newblock


\bibitem[Popovi\'{c} and Hoppe(1997)]%
        {Popovic:1997:PSC}
\bibfield{author}{\bibinfo{person}{Jovan Popovi\'{c}} {and}
  \bibinfo{person}{Hugues Hoppe}.} \bibinfo{year}{1997}\natexlab{}.
\newblock \showarticletitle{Progressive Simplicial Complexes}. In
  \bibinfo{booktitle}{\emph{Proceedings of the 24th Annual Conference on
  Computer Graphics and Interactive Techniques}}
  \emph{(\bibinfo{series}{SIGGRAPH '97})}. \bibinfo{pages}{217–224}.
\newblock


\bibitem[Qi et~al\mbox{.}(2022)]%
        {Qi:2022:BFW}
\bibfield{author}{\bibinfo{person}{Yang Qi}, \bibinfo{person}{Dario Seyb},
  \bibinfo{person}{Benedikt Bitterli}, {and} \bibinfo{person}{Wojciech
  Jarosz}.} \bibinfo{year}{2022}\natexlab{}.
\newblock \showarticletitle{A bidirectional formulation for Walk on Spheres}.
\newblock \bibinfo{journal}{\emph{Comp. Graph. Forum}} \bibinfo{volume}{41},
  \bibinfo{number}{4} (\bibinfo{year}{2022}), \bibinfo{pages}{51--62}.
\newblock


\bibitem[Rossignac and Borrel(1993)]%
        {Rossignac:1993:MR3}
\bibfield{author}{\bibinfo{person}{Jarek Rossignac} {and} \bibinfo{person}{Paul
  Borrel}.} \bibinfo{year}{1993}\natexlab{}.
\newblock \showarticletitle{Multi-resolution 3D approximations for rendering
  complex scenes}.
\newblock In \bibinfo{booktitle}{\emph{Modeling in computer graphics}}.
  \bibinfo{publisher}{Springer}, \bibinfo{pages}{455--465}.
\newblock


\bibitem[Sawhney and Crane(2020)]%
        {Sawhney:2020:MCG}
\bibfield{author}{\bibinfo{person}{Rohan Sawhney} {and} \bibinfo{person}{Keenan
  Crane}.} \bibinfo{year}{2020}\natexlab{}.
\newblock \showarticletitle{Monte Carlo Geometry Processing: A Grid-Free
  Approach to PDE-Based Methods on Volumetric Domains}.
\newblock \bibinfo{journal}{\emph{ACM Trans. Graph.}} \bibinfo{volume}{39},
  \bibinfo{number}{4}, Article \bibinfo{articleno}{123} (\bibinfo{date}{aug}
  \bibinfo{year}{2020}).
\newblock


\bibitem[Schroeder et~al\mbox{.}(1992)]%
        {Schroeder:1992:DTM}
\bibfield{author}{\bibinfo{person}{William~J Schroeder},
  \bibinfo{person}{Jonathan~A Zarge}, {and} \bibinfo{person}{William~E
  Lorensen}.} \bibinfo{year}{1992}\natexlab{}.
\newblock \showarticletitle{Decimation of triangle meshes}. In
  \bibinfo{booktitle}{\emph{Proceedings of the 19th annual conference on
  Computer graphics and interactive techniques}}. \bibinfo{pages}{65--70}.
\newblock


\bibitem[Sell\'en et~al\mbox{.}(2019)]%
        {Sellan:2019:SGP}
\bibfield{author}{\bibinfo{person}{Silvia Sell\'en}, \bibinfo{person}{Herng~Yi
  Cheng}, \bibinfo{person}{Yuming Ma}, \bibinfo{person}{Mitchell Dembowski},
  {and} \bibinfo{person}{Alec Jacobson}.} \bibinfo{year}{2019}\natexlab{}.
\newblock \showarticletitle{Solid Geometry Processing on Deconstructed
  Domains}.
\newblock \bibinfo{journal}{\emph{Comp. Graph. Forum}} \bibinfo{volume}{38},
  \bibinfo{number}{1} (\bibinfo{year}{2019}), \bibinfo{pages}{564--579}.
\newblock


\bibitem[Sharp and Crane(2020)]%
        {Sharp:2020:YCF}
\bibfield{author}{\bibinfo{person}{Nicholas Sharp} {and}
  \bibinfo{person}{Keenan Crane}.} \bibinfo{year}{2020}\natexlab{}.
\newblock \showarticletitle{You Can Find Geodesic Paths in Triangle Meshes by
  Just Flipping Edges}.
\newblock \bibinfo{journal}{\emph{ACM Trans. Graph.}} \bibinfo{volume}{39},
  \bibinfo{number}{6}, Article \bibinfo{articleno}{249} (\bibinfo{date}{nov}
  \bibinfo{year}{2020}).
\newblock


\bibitem[Sharp et~al\mbox{.}(2019)]%
        {Sharp:2019:NIT}
\bibfield{author}{\bibinfo{person}{Nicholas Sharp}, \bibinfo{person}{Yousuf
  Soliman}, {and} \bibinfo{person}{Keenan Crane}.}
  \bibinfo{year}{2019}\natexlab{}.
\newblock \showarticletitle{Navigating intrinsic triangulations}.
\newblock \bibinfo{journal}{\emph{ACM Trans. Graph.}} \bibinfo{volume}{38},
  \bibinfo{number}{4} (\bibinfo{year}{2019}).
\newblock


\bibitem[Sheffer et~al\mbox{.}(2006)]%
        {Sheffer:2006:MPM}
\bibfield{author}{\bibinfo{person}{Alla Sheffer}, \bibinfo{person}{Emil Praun},
  {and} \bibinfo{person}{Kenneth Rose}.} \bibinfo{year}{2006}\natexlab{}.
\newblock \showarticletitle{Mesh Parameterization Methods and Their
  Applications}.
\newblock \bibinfo{journal}{\emph{Found. Trends. Comput. Graph. Vis.}}
  \bibinfo{volume}{2}, \bibinfo{number}{2} (\bibinfo{date}{jan}
  \bibinfo{year}{2006}), \bibinfo{pages}{105–171}.
\newblock
\showISSN{1572-2740}
\urldef\tempurl%
\url{https://doi.org/10.1561/0600000011}
\showDOI{\tempurl}


\bibitem[Springborn et~al\mbox{.}(2008)]%
        {Springborn:2008:CET}
\bibfield{author}{\bibinfo{person}{Boris Springborn}, \bibinfo{person}{Peter
  Schr\"{o}der}, {and} \bibinfo{person}{Ulrich Pinkall}.}
  \bibinfo{year}{2008}\natexlab{}.
\newblock \showarticletitle{Conformal Equivalence of Triangle Meshes}.
\newblock \bibinfo{journal}{\emph{ACM Trans. Graph.}} \bibinfo{volume}{27},
  \bibinfo{number}{3} (\bibinfo{date}{aug} \bibinfo{year}{2008}),
  \bibinfo{pages}{1–11}.
\newblock
\showISSN{0730-0301}
\urldef\tempurl%
\url{https://doi.org/10.1145/1360612.1360676}
\showDOI{\tempurl}


\bibitem[Takayama(2022)]%
        {Takayama:2022:CIT}
\bibfield{author}{\bibinfo{person}{Kenshi Takayama}.}
  \bibinfo{year}{2022}\natexlab{}.
\newblock \showarticletitle{Compatible Intrinsic Triangulations}.
\newblock \bibinfo{journal}{\emph{ACM Trans. Graph.}} \bibinfo{volume}{41},
  \bibinfo{number}{4}, Article \bibinfo{articleno}{57} (\bibinfo{date}{jul}
  \bibinfo{year}{2022}).
\newblock


\bibitem[Zhou et~al\mbox{.}(2016)]%
        {Zhou:2016:MAS}
\bibfield{author}{\bibinfo{person}{Qingnan Zhou}, \bibinfo{person}{Eitan
  Grinspun}, \bibinfo{person}{Denis Zorin}, {and} \bibinfo{person}{Alec
  Jacobson}.} \bibinfo{year}{2016}\natexlab{}.
\newblock \showarticletitle{Mesh Arrangements for Solid Geometry}.
\newblock \bibinfo{journal}{\emph{ACM Trans. Graph.}} \bibinfo{volume}{35},
  \bibinfo{number}{4}, Article \bibinfo{articleno}{39} (\bibinfo{date}{jul}
  \bibinfo{year}{2016}).
\newblock


\bibitem[Zhou and Jacobson(2016)]%
        {Zhou:2016:TTK}
\bibfield{author}{\bibinfo{person}{Qingnan Zhou} {and} \bibinfo{person}{Alec
  Jacobson}.} \bibinfo{year}{2016}\natexlab{}.
\newblock \showarticletitle{Thingi10K: A Dataset of 10,000 3D-Printing Models}.
\newblock \bibinfo{journal}{\emph{arXiv preprint arXiv:1605.04797}}
  (\bibinfo{year}{2016}).
\newblock


\end{thebibliography}

\end{document}